\documentclass[twoside,11pt,final]{entics} 
\usepackage{enticsmacro}
\usepackage{graphicx}
\usepackage{amsmath, amssymb}
\usepackage{quiver}
\usepackage{mathpartir}
\usepackage[all]{xy}

\usepackage{xcolor}
\usepackage{color}
\usepackage{fancyvrb}

\DefineVerbatimEnvironment{Highlighting}{Verbatim}{commandchars=\\\{\}}
\newenvironment{Shaded}{}{}

\newcommand{\DataTypeTok}[1]{\textcolor[rgb]{0.56,0.13,0.00}{#1}}

\newcommand{\KeywordTok}[1]{\textcolor[rgb]{0.00,0.44,0.13}{\textbf{#1}}}
\newcommand{\NormalTok}[1]{#1}

\newcommand{\OtherTok}[1]{\textcolor[rgb]{0.00,0.44,0.13}{#1}}
\newcommand{\PreprocessorTok}[1]{\textcolor[rgb]{0.74,0.48,0.00}{#1}}

\usepackage[utf8]{inputenc}
\usepackage{newunicodechar}
\newunicodechar{λ}{\ensuremath{\mathnormal\lambda}}
\newunicodechar{∀}{\ensuremath{\mathnormal\forall}}
\newunicodechar{≡}{\ensuremath{\mathnormal\equiv}}
\newunicodechar{ℓ}{\ensuremath{\mathnormal\ell}}
\newunicodechar{κ}{\ensuremath{\mathnormal\kappa}}
\newunicodechar{Σ}{\ensuremath{\mathnormal\Sigma}}
\newunicodechar{⊔}{\ensuremath{\mathnormal\sqcup}}
\newunicodechar{♭}{\ensuremath{\mathnormal\flat}}
\newunicodechar{ε}{\ensuremath{\mathnormal\epsilon}}
\newunicodechar{₀}{\ensuremath{\mathnormal{_0}}}
\newunicodechar{⊥}{\ensuremath{\mathnormal\bot}}
\newunicodechar{⊤}{\ensuremath{\mathnormal\top}}
\newunicodechar{α}{\ensuremath{\mathnormal\alpha}}
\newunicodechar{β}{\ensuremath{\mathnormal\beta}}
\newunicodechar{η}{\ensuremath{\mathnormal\eta}}
\newunicodechar{⁻}{\ensuremath{\mathnormal{^-}}}
\newunicodechar{¹}{\ensuremath{\mathnormal{^1}}}
\newunicodechar{ℕ}{\ensuremath{\mathbb{N}}}
\newunicodechar{ω}{\ensuremath{\mathnormal\omega}}
\newunicodechar{∘}{\ensuremath{\mathnormal\circ}}
\newunicodechar{◃}{\ensuremath{\mathnormal\triangleleft}}
\newunicodechar{⊗}{\ensuremath{\mathnormal\otimes}}
\newunicodechar{□}{\ensuremath{\mathnormal\Box}}
\newunicodechar{∥}{\ensuremath{\mathnormal\Vert}}
\newunicodechar{⇆}{\ensuremath{\mathnormal\leftrightarrows}}
\newunicodechar{𝓤}{\ensuremath{\mathnormal\mathcal{U}}}
\newunicodechar{𝔲}{\ensuremath{\mathnormal\mathfrak{u}}}
\newunicodechar{♯}{\ensuremath{\mathnormal\sharp}}
\newunicodechar{σ}{\ensuremath{\mathnormal\sigma}}
\newunicodechar{Π}{\ensuremath{\mathnormal\Pi}}
\newunicodechar{𝕪}{\ensuremath{\mathnormal y}}
\newunicodechar{≃}{\ensuremath{\mathnormal\simeq}}
\newunicodechar{μ}{\ensuremath{\mathnormal\mu}}
\newunicodechar{ρ}{\ensuremath{\mathnormal\rho}}
\newunicodechar{⇈}{\ensuremath{\mathnormal\upuparrows}}
\newunicodechar{π}{\ensuremath{\mathnormal\pi}}
\newunicodechar{γ}{\ensuremath{\mathnormal\gamma}}
\newunicodechar{δ}{\ensuremath{\mathnormal\delta}}
\newunicodechar{Ϝ}{\ensuremath{\mathnormal\digamma}}
\newunicodechar{𝔽}{\ensuremath{\mathnormal\mathbb{F}}}
\newunicodechar{ℙ}{\ensuremath{\mathnormal\mathbb{P}}}
\newunicodechar{↔}{\ensuremath{\mathnormal\leftrightarrow}}
\newunicodechar{≺}{\ensuremath{\mathnormal\prec}}
\newunicodechar{ₘ}{\ensuremath{\mathnormal{_m}}}
\newunicodechar{ₙ}{\ensuremath{\mathnormal{_m}}}
\newunicodechar{𝓥}{\ensuremath{\mathnormal\mathcal{V}}}
\newunicodechar{𝔳}{\ensuremath{\mathnormal\mathfrak{v}}}
\newunicodechar{₁}{\ensuremath{\mathnormal{_1}}}
\newunicodechar{⁼}{\ensuremath{\mathnormal{^=}}}

\newcommand{\tri}{\mathbin{\triangleleft}}
\def\conf{MFPS 2025} 	%%Fill in the acronym for your conference (with year)
\volume{5}			%Fill in the ENTICS volume number here
\def\volu{5}			% and here
\def\papno{1}			%Fill in your paper number here
\def\mydoi{16885}
\def\lastname{Aberlé, Spivak} %Lastnames appear in the running header 
\def\runtitle{Polynomial Universes in Homotopy Type Theory} %Short title appears in the running header on even pages. 
\def\copynames{C. Aberlé, D. Spivak}    %% Fill in the first initial and last name of the authors
\begin{document}
%%%Note the beginning and end of the frontmatter section that starts here%%%%%
\begin{frontmatter}
  \title{Polynomial Universes in Homotopy Type Theory} 						%%Title here and the
 \thanks[ALL]{This material is based upon work supported by the Air Force Office of
 Scientific Research under award numbers FA9550-20-1-0348 and FA9550-23-1-0376.}   %%Text of \thanks[ALL} here..
 %%%%%%%%%%%%%%%%%%%%%%%%%%%%			This Thanks is optional.
  %%%%Now the author(s) names(s)%%%%%
  \author{C.B. Aberlé \thanksref{a}\thanksref{myemail}}	%%Note NO SPACE between 
   \author{David I. Spivak\thanksref{b}\thanksref{coemail}}		%last name and \thanksref{...} 
    %%%Next come the addresses%%%%
   \address[a]{Department of Computer Science\\ Carnegie Mellon University\\				%or between \thanksrefs...
    Pittsburgh, PA, USA}  							
   \thanks[myemail]{Email: \href{mailto:caberle@andrew.cmu.edu} {\texttt{\normalshape
        caberle@andrew.cmu.edu}}} 
   %%%Note: if both authors share same institution, only list the address once, after the second 
   %%%author. 
   %%%There also is a link from the first author to the co-author's address to show how to list 
   %%%affiliations to more than one institution, when needed. 
  \address[b]{Topos Institute, Berkeley, CA, USA} 
  \thanks[coemail]{Email:  \href{mailto:dspivak@gmail.com} {\texttt{\normalshape
  dspivak@gmail.com}}}
\begin{abstract} 
Awodey, later with Newstead, showed how polynomial functors with extra structure (termed ``natural models'') hold within them the categorical semantics for dependent type theory. Their work presented these ideas clearly but ultimately led them outside of the usual category of polynomial functors to a particular \emph{tricategory} of polynomials in order to explain all of the structure possessed by such models. This paper builds off that work---explicating the categorical semantics of dependent type theory by axiomatizing them entirely in terms of the usual category of polynomial functors. In order to handle the higher-categorical coherences required for such an explanation, we work with polynomial functors in the language of Homotopy Type Theory (HoTT), which allows for higher-dimensional structures to be expressed purely within this category. The move to HoTT moreover enables us to express a key additional condition on polynomial functors---\emph{univalence}---which is sufficient to guarantee that models of type theory expressed as univalent polynomials satisfy all higher coherences of their corresponding algebraic structures, purely in virtue of being closed under the usual constructors of dependent type theory. We call polynomial functors satisfying this condition \emph{polynomial universes}. As an example of the simplification to the theory of natural models this enables, we highlight the fact that a polynomial universe being closed under dependent product types implies the existence of a distributive law of monads, which witnesses the usual distributivity of dependent products over dependent sums.
\end{abstract}
\end{frontmatter}
\section{Introduction}\label{intro}

The \emph{dependency} of types-of-things on values-of-things is fundamental to our ability to express complex mathematical ideas and build up sophisticated abstractions. By taking this essential idea to heart, dependent type theory \cite{MLTT} provides both of the following: \begin{itemize}
  \item An elegant \textbf{syntax} for expressing mathematical ideas, that can moreover be \textbf{computably} realized.
  \item A robust \textbf{categorical semantics} that allow type theoretic syntax to be used in order to work in the internal languages of well-structured categories, and even more complex structures, such as $\infty$-categories.
  \end{itemize} 
  
Of these, it is the \textbf{categorical semantics} of dependent type theory that shall be our focus in this paper. Specifically, although these categorical semantics are evidently quite powerful, they are also notoriously subtle, owing to the issue of \emph{strictness}, whereby various identities that typically hold only up to isomorphism in arbitrary categories must hold \emph{strictly} in order to soundly model the type-theoretic syntax.
  
A key device in resolving this difficulty turns out to be the categorical machinery of \emph{polynomial functors}. Awodey \cite{AwodeyNatMod}---and later Newstead \cite{NewsteadNatMod}---have shown that there is a strong connection between dependent type theory and polynomial functors, via the concept of \emph{natural models,} which cleanly solves the strictness problem via the type-theoretic concept of a \emph{type universe,} and such universes turn out to naturally be regarded as certain polynomial endofunctors on a suitably-chosen category of presheaves.

Although the elementary structure of natural models is thus
straightforwardly described by considering them as objects in a category
of polynomial functors, Awodey and Newstead were ultimately led to construct a rather complicated \emph{tricategory} of polynomial functors in order to make sense of those parts of natural
models that require identities to hold only \emph{up to isomorphism},
rather than strictly. There is thus an evident tension between
\emph{strict} and \emph{weak} identities that has not yet been fully
resolved in the story of natural models. In the present work, we build
on Awodey and Newstead's work to resolve this impasse by showing
how polynomial universes can be fully axiomatized in terms of the ordinary category of polynomial
functors, by defining this category internally in the language
of \emph{Homotopy Type Theory} (HoTT) \cite{hottbook}.

HoTT thus provides a \emph{synthetic} setting in which to work with polynomial functors and their higher-categorical coherences without leaving the usual category of polynomial functors. As we shall see, this has a great simplifying effect upon the resultant theory, and reveals many additional structures, both of polynomial universes, and of the category of polynomial functors as a whole. As an illustration of this, we show how every polynomial universe \(u\) closed under the usual type formers of dependent type theory, regarded as a polynomial pseudomonad, gives rise to distributive law of $u$ over itself, which in particular witnesses the usual distributive law of dependent products over dependent sums. \footnote{A precursor to this story was attempted by the second-named author in a \href{https://topos.site/blog/2021-07-01-jump-monads/}{blog post}, which may be useful for readers seeking intuition. There, a proposed self-distributive law of the list monad was claimed to witness dependent products. However, the author only later noticed that one of the four equations for distributive laws was doomed to fail \cite{zwart2022no} due to the 1-dimensionality of types in the category of sets. The present project was inspired by this.}

Additionally, the move from set theory to HoTT as a setting for developing the theory of polynomial universes makes it well-suited to formalization in a proof assistant. We have formalized the main results of this paper in Agda, with the code of this formalization given in the appendix.

\section{The Trouble with Dependent Types}

In what follows, we work in an informal setting of Martin-Löf type theory \cite{MLTT} and its categorical semantics \cite{seely1984}, which we assume familiarity with on the part of the reader.

In the typical (naïve) categorical semantics of dependent type theory \cite{seely1984}, one
considers a category \(\mathcal{C}\) whose objects are considered as corresponding to \emph{contexts} $\Gamma$, such that morphisms $f : \Gamma \to \Delta$ correspond to \emph{substitutions between contexts}, with a type $\Gamma \vdash A ~ \mathsf{type}$ dependent upon $\Gamma$ being represented as the substitution $\Gamma, A \to \Gamma$ that forgets the type $A$ from the extended context $\Gamma, A$ (commonly called a \emph{display map}). Application of a substitution $f : \Gamma \to \Delta$ to a type $\Delta \vdash A$ is then represented as the pullback $$
\begin{tikzcd}
	{\Gamma, A[f]} & {\Delta, A} \\
	\Gamma & \Delta
	\arrow[from=1-1, to=1-2]
	\arrow[from=1-1, to=2-1]
	\arrow["\lrcorner"{anchor=center, pos=0.125}, draw=none, from=1-1, to=2-2]
	\arrow[from=1-2, to=2-2]
	\arrow["f"', from=2-1, to=2-2]
\end{tikzcd}
$$ 

In particular, any display map \(A : \Gamma, A \to \Gamma\)
induces a functor \(\mathcal{C}/\Gamma \to \mathcal{C}/\Gamma, A\)
by substitution along \(A\), which corresponds to \emph{weakening} a context by adding a variable of type $A$. The left and right adjoints to the weakening functor
(if they exist) then correspond to $\Sigma$ and $\Pi$
types, respectively. $$
\begin{tikzcd}
	{\mathcal{C}/\Gamma, A} \\
	\\
	{\mathcal{C}/\Gamma}
	\arrow[""{name=0, anchor=center, inner sep=0}, "{\Sigma_A}"', curve={height=30pt}, from=1-1, to=3-1]
	\arrow[""{name=1, anchor=center, inner sep=0}, "{\Pi_A}", curve={height=-30pt}, from=1-1, to=3-1]
	\arrow[""{name=2, anchor=center, inner sep=0}, "{A^*}"', from=3-1, to=1-1]
	\arrow["\dashv"{anchor=center, rotate=3}, draw=none, from=0, to=2]
	\arrow["\dashv"{anchor=center, rotate=-3}, draw=none, from=2, to=1]
\end{tikzcd}
$$ 

In order for the operation of forming $\Sigma$ and $\Pi$ types to be stable under substitution (i.e. pullback), these must additionally satisfy the \emph{Beck-Chevalley} condition: $$
\small \begin{tikzcd}
	{\Gamma, x : A[f]} & {\Delta, x : A} \\
	\Gamma & \Delta
	\arrow["g", from=1-1, to=1-2]
	\arrow["{A[f]}"', from=1-1, to=2-1]
	\arrow["\lrcorner"{anchor=center, pos=0.125}, draw=none, from=1-1, to=2-2]
	\arrow["A", from=1-2, to=2-2]
	\arrow["f", from=2-1, to=2-2]
\end{tikzcd} \implies (\Sigma_A(B))[f] \xleftarrow{\simeq} \Sigma_{A[f]} (B[g])
$$ 

Unfortunately, this pleasingly simple story of categorical dependent types is a fantasy, and the interpretation of
type-theoretic syntax into categorical semantics sketched above is
unsound, as it stands. The problem in essentials is that, in the syntax
of type theory, substitution is \emph{strictly} associative, and \emph{strictly} satisfies the Beck-Chevalley condition. However, in the
above categorical semantics, substitution is computed by pullback, which is in general only associative up to isomorphism, and likewise for the Beck-Chevalley condition. It is precisely this problem which natural models exist to solve.

\subsection{Natural Models}

As mentioned previously, the key insight (due originally to Voevodsky) in formulating the notion of a natural model is that the problem of strictness in the semantics of type theory has, in a sense, already been solved by the
notion of a \emph{type universe}, i.e. a type $\mathcal{U}$ whose elements can themselves be regarded as types. In categorical semantics, we interpret such a universe as a display map $u : \mathcal{U}_\bullet \to \mathcal{U}$. A type in context $\Gamma$ may then be represented as a morphism $A : \Gamma \to \mathcal{U}$, rather than a display map $\Gamma, A \to \Gamma$, which we may recover as the pullback: $$
\begin{tikzcd}
	{\Gamma, A} & {\mathcal{U}_\bullet} \\
	\Gamma & {\mathcal{U}}
	\arrow[from=1-1, to=1-2]
	\arrow[from=1-1, to=2-1]
	\arrow["\lrcorner"{anchor=center, pos=0.125}, draw=none, from=1-1, to=2-2]
	\arrow["u", from=1-2, to=2-2]
	\arrow["A"', from=2-1, to=2-2]
\end{tikzcd}
$$ We say that a display map $\Gamma, A \to \Gamma$ is \emph{classified} by $\mathcal{U}$ if there is a pullback square as above.

Hence given a universe of types \(\mathcal{U}\), rather than
representing substitution as pullback, we can simply represent the action of applying a substitution $f : \Gamma \to \Delta$ to a family of types $A : \Delta \to \mathcal{U}$ as the precomposition $A \circ f : \Gamma \to \mathcal{U}$, which is automatically strictly associative, and strictly satisfies the Beck-Chevalley condition for $\Sigma$ types / $\Pi$ types if $\mathcal{U}$ is closed under $\Sigma$ types / $\Pi$ types, respectively (although what it means for $\mathcal{U}$ to be closed under $\Sigma$ and $\Pi$ types is rather nontrivial---indeed, this topic forms the main subject of this paper).

To interpret the syntax of dependent type theory in a category $\mathcal{C}$ of contexts and substitutions, it therefore suffices to \emph{embed} $\mathcal{C}$ into a category whose type-theoretic internal language possesses such a \emph{universe} whose types correspond to those of $\mathcal{C}$. A natural candidate for such an embedding is the \emph{Yoneda embedding} $\mathbf{y} : \mathcal{C} \to \mathbf{Set}^{\mathcal{C}^{\textnormal{op}}}$.

Hence we can work in the category of presheaves $\mathbf{Set}^{\mathcal{C}^{\textnormal{op}}}$ to study the type-theoretic language of $\mathcal{C}$. The universe $\mathcal{U}$ is then given by: \begin{enumerate} \item an object of $\mathbf{Set}^{\mathcal{C}^{\textnormal{op}}}$, i.e. a contravariant functorial assignment, to each context $\Gamma$, of a set $\mathcal{U}(\Gamma)$ of \emph{types in context} $\Gamma$, together with \item an object $u \in \mathbf{Set}^{\mathcal{C}^{\textnormal{op}}}/\mathcal{U}$, i.e. a natural transformation $u : \mathcal{U}_\bullet \to \mathcal{U}$, where for each context $\Gamma$, $\mathcal{U}_\bullet(\Gamma)$ is the set of terms in context $\Gamma$, and $u_\Gamma : \mathcal{U}_\bullet(\Gamma) \to \mathcal{U}(\Gamma)$ assigns each term to its type. \end{enumerate}

The condition that all types in $\mathcal{U}$ ``belong to $\mathcal{C}$'' can then be expressed by requiring $u$ to be \emph{representable} in the following sense: for any representable $\gamma \in \mathbf{Set}^{\mathcal{C}^{\textnormal{op}}}$ with $\alpha : \gamma \to \mathcal{U}$, the pullback $$
\begin{tikzcd}
	{\gamma . \alpha} & {\mathcal{U}_\bullet} \\
	\gamma & {\mathcal{U}}
	\arrow[from=1-1, to=1-2]
	\arrow[from=1-1, to=2-1]
	\arrow["\lrcorner"{anchor=center, pos=0.125}, draw=none, from=1-1, to=2-2]
	\arrow["u", from=1-2, to=2-2]
	\arrow["\alpha"', from=2-1, to=2-2]
\end{tikzcd}
$$ is representable. In particular, this says that, given a context $\Gamma$ and a type $A \in \mathcal{U}[\Gamma]$, there is a context $\Gamma, A$ together with a substitution $\Gamma, A \to A$ that corresponds to the above pullback under the Yoneda embedding. Type-theoretically, this corresponds to the operation of \emph{context extension}.

The question, then, is how to express that $\mathcal{C}$ has $\Sigma$ types, $\Pi$ types, etc., in terms of the structure of $u$. Toward answering this question, we may note that $u$ gives rise to an endofunctor (indeed, a \emph{polynomial endofunctor}) $P_u : \mathbf{Set}^{\mathcal{C}^{\textnormal{op}}} \to \mathbf{Set}^{\mathcal{C}^{\textnormal{op}}}$, defined by $$
P_u(X) = \sum_{A : \mathcal{U}} X^{\mathcal{U}_\bullet[A]}
$$ (This notation will be made precise momentarily). As it turns out, much of the type-theoretic structure of $u$ (and by extension $\mathcal{C}$) can be accounted for in terms of this functor. For instance, $u$ is closed under unit and $\Sigma$ types if and only if $P_u$ carries the structure of a \emph{Cartesian pseudomonad} on $\mathbf{Set}^{\mathcal{C}^{\textnormal{op}}}$ (c.f. Theorem 2.3 of \cite{NewsteadNatMod}). To see why this is the case, we could, on the one hand, proceed as in past developments of the theory of natural models and define a certain tricategory of polynomial functors in which such pseudomonads can be defined. However, here we diverge from these past approaches, and instead develop the theory of polynomial functors in the language of \emph{Homotopy Type Theory}.

\section{Polynomial Functors in HoTT}

In order to understand the higher-dimensional identities and coherences of type formers in natural models, we now change our setting from the \emph{extensional} type-theoretic language of 1-topoi such as $\mathbf{Set}^{\mathcal{C}^{\textnormal{op}}}$ to the \emph{intensional} type-theoretic language of $\infty$-topoi, which is a form of HoTT \cite{shulman2019}. All the constructions on natural models we considered previously carry over to this setting---mutatis mutandis---by replacing the category of presheaves $\mathbf{Set}^{\mathcal{C}^{\textnormal{op}}}$ with the $\infty$-category of presheaves $\infty\mathbf{Grpd}^{\mathcal{C}^{\textnormal{op}}}$.\footnote{Since essentially all of the categorical structures considered henceforth in this paper shall be infinite dimensional, we shall generally omit the prefix ``$\infty$'' from our descriptions of these structures. Hence hereafter ``category'' means $\infty$-category, ``functor'' means $\infty$-functor, ``limit'' means homotopy limit, and so on, unless otherwise specified.}

Working internally in the language of $\infty\mathbf{Grpd}^{\mathcal{C}^{\textnormal{op}}}$, let $\texttt{Type}$ denote the \emph{type} of types \footnote{Technically, we assume an infinite hierarchy of stratified type universes in order to avoid inconsistency. However, for the purpose of high-level explanation, we shall generally omit universe levels and speak generically of the ``type'' of all types and similarly the category of all such, even though, strictly speaking, neither of these exist.} and let $\mathbf{Type}$ be the \emph{category} of types and functions between them. For any type $A$, let $y^A$ denote the corresponding co-representable functor $\mathbf{Type} \to \mathbf{Type}$ that maps a type $y$ to the function type \texttt{A → y}. A \emph{polynomial endofunctor} $P : \mathbf{Type} \to \mathbf{Type}$ is an endofunctor on $\mathbf{Type}$ equivalent to a sum of such co-representables $$
P(y) = \sum_{x : A} y^{B[x]}
$$ for some type $A$ and a family of types $B[x]$ indexed by $x : A$. The data of a polynomial functor is thus uniquely determined by the choice of $A$ and $B$. \footnote{A similar treatment to ours of universes via polynomial functors, there referred to as ``containers,'' is given in \cite{MonadicContainers}, including the treatment of $\Sigma$ types via monadic structure. In our case, we leverage some additional properties of the category of polynomial enfodunctors, such as the Vertical-Cartesian factorization system in particular, to additionally characterize $\Pi$ types in terms of distributive laws.} Hence we may represent the \emph{type} of such functors as that of pairs $(A , B)$ of this form. In Agda, this type can be expressed as follows:

\begin{Shaded}
\begin{Highlighting}[]
\NormalTok{Poly }\OtherTok{:} \OtherTok{(}\NormalTok{ℓ κ }\OtherTok{:}\NormalTok{ Level}\OtherTok{)} \OtherTok{→}\NormalTok{ Type }\OtherTok{((}\NormalTok{lsuc ℓ}\OtherTok{)}\NormalTok{ ⊔ }\OtherTok{(}\NormalTok{lsuc κ}\OtherTok{))}
\NormalTok{Poly ℓ κ }\OtherTok{=}\NormalTok{ Σ }\OtherTok{(}\NormalTok{Type ℓ}\OtherTok{)} \OtherTok{(λ}\NormalTok{ A }\OtherTok{→}\NormalTok{ A }\OtherTok{→}\NormalTok{ Type κ}\OtherTok{)}
\end{Highlighting}
\end{Shaded}

The observant reader may note the striking similarity of the above-given
formula for a polynomial functor and the endofunctor on
\(\mathbf{Set}^{\mathcal{C}^{op}}\) defined in the previous section on
natural models. This is no accident---given a type \texttt{𝓤} and
a function \texttt{u\ :\ 𝓤\ →\ Type} corresponding to a natural model as
described previously, we obtain the corresponding polynomial
\texttt{𝔲\ :\ Poly} as the pair \texttt{(𝓤\ ,\ u)}. Hence we can study
the structure of \texttt{𝓤} and \texttt{u} in terms of \texttt{𝔲}, allowing for an elegant treatment of the theory
of natural models.

For \(p = \sum_{a : A} y^{B(a)}\),
\(q = \sum_{c : C} y^{D(c)} \in \mathbf{Poly}\), a natural transformation
\(p \to q\) is equivalently given by: \begin{itemize} \item a
\emph{forward} map \texttt{f\ :\ A\ →\ B}, and \item a \emph{backward} map
\texttt{g\ :\ (a\ :\ A)\ →\ D\ (f\ a)\ →\ B\ a} \end{itemize} as can be seen from the
following argument via Yoneda: \[
\begin{array}{rl}
& \int_{y \in \mathbf{Type}} \left( \sum_{a : A} y^{B(a)}  \right) \to \sum_{c : C} y^{D(c)}\\
\simeq & \prod_{a : A} \int_{y \in \mathbf{Type}} y^{B(a)} \to \sum_{c : C} y^{D(c)}\\
\simeq & \prod_{a : A} \sum_{c : C} B(a)^{D(c)}\\
\simeq & \sum_{f : A \to C} \prod_{a : A} B(a)^{D(f(c))}
\end{array}
\] In Agda, we use the notation \(p \leftrightarrows q\) to the latter type (AKA the type of \emph{dependent lenses}---or just \emph{lenses}---from
\(p\) to \(q\)), which may be written as follows:

\begin{Shaded}
\begin{Highlighting}[]
\OtherTok{\_}\NormalTok{⇆}\OtherTok{\_} \OtherTok{:} \OtherTok{∀} \OtherTok{\{}\NormalTok{ℓ0 ℓ1 κ0 κ1}\OtherTok{\}} \OtherTok{→}\NormalTok{ Poly ℓ0 κ0 }\OtherTok{→}\NormalTok{ Poly ℓ1 κ1 }\OtherTok{→}\NormalTok{ Type }\OtherTok{(}\NormalTok{ℓ0 ⊔ ℓ1 ⊔ κ0 ⊔ κ1}\OtherTok{)}
\OtherTok{(}\NormalTok{A , B}\OtherTok{)}\NormalTok{ ⇆ }\OtherTok{(}\NormalTok{C , D}\OtherTok{)} \OtherTok{=}\NormalTok{ Σ }\OtherTok{(}\NormalTok{A }\OtherTok{→}\NormalTok{ C}\OtherTok{)} \OtherTok{(λ}\NormalTok{ f }\OtherTok{→} \OtherTok{(}\NormalTok{a }\OtherTok{:}\NormalTok{ A}\OtherTok{)} \OtherTok{→}\NormalTok{ D }\OtherTok{(}\NormalTok{f a}\OtherTok{)} \OtherTok{→}\NormalTok{ B a}\OtherTok{)}
\end{Highlighting}
\end{Shaded}

Hence we have a category \(\mathbf{Poly}\) of polynomial functors and
lenses between them. Our goal, then, is to show how the type-theoretic
structure of a natural model naturally arises from the structure of this
category. In fact, \(\mathbf{Poly}\) is replete with categorical
structures of all kinds, of which we now mention but a few of particular importance to us.

We say that a lens \texttt{(f\ ,\ f♯)\ :\ (A\ ,\ B)\ ⇆\ (C\ ,\ D)} is
\emph{Cartesian} if for every \texttt{a\ :\ A}, the map
\texttt{f♯\ a\ :\ D{[}f\ a{]}\ →\ B\ a} is an
equivalence.\footnote{For a proof that this notion of Cartesian morphism between polynomials is equivalent to the ordinary definition of Cartesian natural transformations between polynomial functors, see Chapter 5.5 of \cite{SpivakNiu}}

\begin{Shaded}
  \begin{Highlighting}[]
  \KeywordTok{module}\NormalTok{ Cart }\OtherTok{\{}\NormalTok{ℓ0 ℓ1 κ0 κ1}\OtherTok{\}} \OtherTok{\{}\NormalTok{p }\OtherTok{:}\NormalTok{ Poly ℓ0 κ0}\OtherTok{\}} 
                   \OtherTok{(}\NormalTok{q }\OtherTok{:}\NormalTok{ Poly ℓ1 κ1}\OtherTok{)} \OtherTok{(}\NormalTok{f }\OtherTok{:}\NormalTok{ p ⇆ q}\OtherTok{)} \KeywordTok{where}
  
  \NormalTok{    isCartesian }\OtherTok{:} \DataTypeTok{Set} \OtherTok{(}\NormalTok{ℓ0 ⊔ κ0 ⊔ κ1}\OtherTok{)}
  \NormalTok{    isCartesian }\OtherTok{=} \OtherTok{(}\NormalTok{a }\OtherTok{:}\NormalTok{ fst p}\OtherTok{)} \OtherTok{→}\NormalTok{ isEquiv }\OtherTok{(}\NormalTok{snd f a}\OtherTok{)}
  
  \KeywordTok{open}\NormalTok{ Cart }\KeywordTok{public}
  \end{Highlighting}
  \end{Shaded}

  In what follows, Cartesian lenses shall be of special interest to us, owing to the fact that the existence of a Cartesian lens $p \to u$ in a certain sense expresses that $u$ is closed under all the types encoded by $p$. Specifically, if we view \texttt{p\ =\ (A\ ,\ B)} as an \texttt{A}-indexed
  family of types, given by \texttt{B}, then the existence of a Cartesian lens \texttt{(f\ ,\ f♯)\ :\ p\ ⇆\ 𝔲} essentially shows that, for each \texttt{a : A} there is an element \texttt{f a : 𝓤} such that $B[a] \simeq u[f ~ a]$, i.e. every type in the family $B[a]$ is equivalent to one for which there exists a ``code'' in 𝓤. To show that 𝔲 is closed under $\Sigma$ types, $\Pi$ types, etc., we therefore need only find polynomials that suitably represent these types, and ask that there be Cartesian morphisms from these to $u$.

  Cartesian lenses are also closed under composition, so there is a wide subcategory $\mathbf{Poly^{Cart}}$ of the category of polynomial functors, whose morphisms are Cartesian lenses. By studying which categorical properties of $\mathbf{Poly}$ are inherited by $\mathbf{Poly^{Cart}}$, we can then deduce the corresponding properties of polynomial universes/natural models. This shall be our method of choice in what follows.

\subsection{Composition of Polynomial Functors and $\Sigma$ Types}\label{sec:composition}

As endofunctors on \(\mathbf{Type}\), polynomial functors may
straightforwardly be composed. To show that the resulting composite is
itself (equivalent to) a polynomial functor, we can reason via the
following chain of equivalences: given polynomials \texttt{(A\ ,\ B)}
and \texttt{(C\ ,\ D)}, their composite, evaluated at a type \texttt{y}
is \[
\begin{array}{rl}
& \sum_{a : A} \left( \sum_{c : C} y^{D(c)} \right)^{B(a)}\\
\simeq & \sum_{a : A} \sum_{f : B(a) \to C} \prod_{b : B(a)} y^{D(f(b))}\\
\simeq & \sum_{(a , f) : \sum_{a : A} C^{B(a)}} y^{\sum_{b : B(a)} D(f(b))}
\end{array}
\] This then defines a monoidal product \(◃\) on \(\mathbf{Poly}\) with
monoidal unit given by the identity functor \texttt{𝕪}:

\begin{Shaded}
  \begin{Highlighting}[]
  \NormalTok{𝕪 }\OtherTok{:}\NormalTok{ Poly lzero lzero}
  \NormalTok{𝕪 }\OtherTok{=} \OtherTok{(}\NormalTok{⊤ , }\OtherTok{λ} \OtherTok{\_} \OtherTok{→}\NormalTok{ ⊤}\OtherTok{)}
  
  \OtherTok{\_}\NormalTok{◃}\OtherTok{\_} \OtherTok{:} \OtherTok{∀} \OtherTok{\{}\NormalTok{ℓ0 ℓ1 κ0 κ1}\OtherTok{\}} \OtherTok{→}\NormalTok{ Poly ℓ0 κ0 }\OtherTok{→}\NormalTok{ Poly ℓ1 κ1 }
        \OtherTok{→}\NormalTok{ Poly }\OtherTok{(}\NormalTok{ℓ0 ⊔ κ0 ⊔ ℓ1}\OtherTok{)} \OtherTok{(}\NormalTok{κ0 ⊔ κ1}\OtherTok{)}
  \OtherTok{(}\NormalTok{A , B}\OtherTok{)}\NormalTok{ ◃ }\OtherTok{(}\NormalTok{C , D}\OtherTok{)} \OtherTok{=} 
      \OtherTok{(}\NormalTok{ Σ A }\OtherTok{(λ}\NormalTok{ a }\OtherTok{→}\NormalTok{ B a }\OtherTok{→}\NormalTok{ C}\OtherTok{)} 
  \NormalTok{    , }\OtherTok{λ} \OtherTok{(}\NormalTok{a , f}\OtherTok{)} \OtherTok{→}\NormalTok{ Σ }\OtherTok{(}\NormalTok{B a}\OtherTok{)} \OtherTok{(λ}\NormalTok{ b }\OtherTok{→}\NormalTok{ D }\OtherTok{(}\NormalTok{f b}\OtherTok{))} \OtherTok{)}
  \end{Highlighting}
  \end{Shaded}

  By construction, the existence of a Cartesian lens
  \texttt{(σ\ ,\ σ♯)\ :\ 𝔲\ ◃\ 𝔲\ ⇆\ 𝔲} effectively shows that \texttt{𝔲}
  is closed under \texttt{Σ} types, since:
  
  \begin{itemize}
  \item
    \texttt{σ} maps a pair (A , B) consisting of \texttt{A\ :\ 𝓤} and
    \texttt{B\ :\ (u\ A)\ →\ 𝓤} to a term \texttt{σ(A,B)} representing the
    \texttt{Σ} type. This corresponds to the type formation rule
    \[ \inferrule{\Gamma \vdash A : \mathsf{Type}\\ \Gamma, x : A \vdash B[x] ~ \mathsf{Type}}{\Gamma \vdash \Sigma x : A . B[x] ~ \mathsf{Type}} \]
  \item
    For all \texttt{(A\ ,\ B)} as above, \texttt{σ♯\ (A\ ,\ B)} takes a
    term of type \texttt{σ\ (A\ ,\ B)} and yields a term
    \texttt{fst\ (σ♯\ (A\ ,\ B))\ :\ A} along with a term
    \texttt{snd\ (σ♯\ (A\ ,\ B))\ :\ B\ (fst\ (σ♯\ (A\ ,\ B)))},
    corresponding to the elimination rules \[
    \inferrule{\Gamma \vdash p : \Sigma x : A . B[x]}{\Gamma \vdash \pi_1(p) : A} \quad \inferrule{\Gamma \vdash p : \Sigma x : A . B[x]}{\Gamma \vdash \pi_2(p) : B[\pi_1(p)]} \]
  \item
    The fact that \texttt{σ♯\ (A\ ,\ B)} has is an equivalence implies it
    has an inverse
    \texttt{σ♯⁻¹\ (A\ ,\ B)\ :\ Σ\ (u\ A)\ (λ\ x\ →\ u\ (B\ x))\ →\ u\ (σ\ (A\ ,\ B))},
    which takes a pair of terms to a term of the corresponding pair type,
    and thus corresponds to the introduction rule
    \[ \inferrule{\Gamma \vdash a : A\\ \Gamma \vdash b : B[a]}{\Gamma \vdash (a , b) : \Sigma x : A . B[x]} \]
  \item
    The fact that \(σ♯⁻¹ (A , B)\) is both a left and a right inverse to
    \(σ♯\) then implies the usual \(β\) and \(η\) laws for dependent pair
    types
    \[ \pi_1(a , b) = a \quad \pi_2(a , b) = b \quad p = (\pi_1(p) , \pi_2(p)) \]
  \end{itemize} Similarly, the existence of a Cartesian lens \((η , η♯) : 𝕪 ⇆ 𝔲\)
  implies that \(𝔲\) classifies the unit type, in
  that:
  
  \begin{itemize}
  \item
    There is an element \texttt{η\ tt\ :\ 𝓤} which represents the unit
    type. This corresponds to the type formation rule
    \[ \inferrule{~}{\Gamma \vdash \top : \mathsf{Type}}\]
  \item
    The ``elimination rule'' \texttt{η♯\ tt\ :\ u(η\ tt)\ →\ ⊤}, applied
    to an element \texttt{x\ :\ u(η\ tt)} is trivial, in that it simply
    discards \texttt{x}. This corresponds to the fact that, in the
    ordinary type-theoretic presentation, \(\top\) does not have an
    elimination rule.
  \item
    However, since this trivial elimination rule has an inverse
    \texttt{η♯⁻¹\ tt\ :\ ⊤\ →\ u\ (η\ tt)}, it follows that there is a
    (unique) element \texttt{η♯⁻¹\ tt\ tt\ :\ u\ (η\ tt)}, which
    corresponds to the introduction rule for \(\top\):
    \[\inferrule{~}{\Gamma \vdash \mathsf{tt} : \top}\]
  \item
    Moreover, the uniqueness of this element corresponds to the
    \(\eta\)-law for \(\top\):
    \[\frac{\Gamma \vdash x : \top}{\Gamma \vdash x = \mathsf{tt}}\]
  \end{itemize}
  
  But then, what sorts of laws can we expect Cartesian lenses as above to
  obey, and is the existence of such a lens all that is needed to ensure
  that the natural model \(𝔲\) has dependent pair types in the original
  sense of Awodey \& Newstead's definition in terms of Cartesian
  (pseudo)monads \cite{AwodeyNatMod,NewsteadNatMod}, or is some
  further data required? And what about \texttt{Π} types, or other type
  formers? To answer these questions, we will need to study the structure
  of \texttt{◃}, along with some closely related functors, in a bit more
  detail. In particular, we shall see that the structure of \texttt{◃} as
  a monoidal product on \(\mathbf{Poly}\) reflects many of the basic
  identities one expects to hold of \texttt{Σ} types.

  For instance, the associativity of \texttt{◃} corresponds to the
  associativity of \texttt{Σ} types.
  
  \begin{Shaded}
  \begin{Highlighting}[]
  \KeywordTok{module}\NormalTok{ ◃Assoc }\OtherTok{\{}\NormalTok{ℓ0 ℓ1 ℓ2 κ0 κ1 κ2}\OtherTok{\}} \OtherTok{(}\NormalTok{p }\OtherTok{:}\NormalTok{ Poly ℓ0 κ0}\OtherTok{)} 
                \OtherTok{(}\NormalTok{q }\OtherTok{:}\NormalTok{ Poly ℓ1 κ1}\OtherTok{)} \OtherTok{(}\NormalTok{r }\OtherTok{:}\NormalTok{ Poly ℓ2 κ2}\OtherTok{)} \KeywordTok{where}
  
  \NormalTok{    ◃assoc }\OtherTok{:} \OtherTok{((}\NormalTok{p ◃ q}\OtherTok{)}\NormalTok{ ◃ r}\OtherTok{)}\NormalTok{ ⇆ }\OtherTok{(}\NormalTok{p ◃ }\OtherTok{(}\NormalTok{q ◃ r}\OtherTok{))}
  \NormalTok{    ◃assoc }\OtherTok{=} \OtherTok{(} \OtherTok{(λ} \OtherTok{((}\NormalTok{a , γ}\OtherTok{)}\NormalTok{ , δ}\OtherTok{)} 
                    \OtherTok{→} \OtherTok{(}\NormalTok{a , }\OtherTok{(λ}\NormalTok{ b }\OtherTok{→} \OtherTok{(}\NormalTok{γ b , }\OtherTok{λ}\NormalTok{ d }\OtherTok{→}\NormalTok{ δ }\OtherTok{(}\NormalTok{b , d}\OtherTok{)))))} 
  \NormalTok{             , }\OtherTok{(λ} \OtherTok{\_} \OtherTok{(}\NormalTok{b , }\OtherTok{(}\NormalTok{d , x}\OtherTok{))} \OtherTok{→} \OtherTok{((}\NormalTok{b , d}\OtherTok{)}\NormalTok{ , x}\OtherTok{))} \OtherTok{)}
  
  \KeywordTok{open}\NormalTok{ ◃Assoc }\KeywordTok{public}
  \end{Highlighting}
  \end{Shaded}
  
  \noindent while the left and right unitors of \texttt{◃} correspond to the fact
  that \texttt{⊤} is both a left and a right unit for \texttt{Σ}.
  
  \begin{Shaded}
  \begin{Highlighting}[]
  \KeywordTok{module}\NormalTok{ ◃LRUnit }\OtherTok{\{}\NormalTok{ℓ κ}\OtherTok{\}} \OtherTok{(}\NormalTok{p }\OtherTok{:}\NormalTok{ Poly ℓ κ}\OtherTok{)} \KeywordTok{where}
  
  \NormalTok{    ◃unitl }\OtherTok{:} \OtherTok{(}\NormalTok{𝕪 ◃ p}\OtherTok{)}\NormalTok{ ⇆ p}
  \NormalTok{    ◃unitl }\OtherTok{=} \OtherTok{(} \OtherTok{(λ} \OtherTok{(\_}\NormalTok{ , a}\OtherTok{)} \OtherTok{→}\NormalTok{ a tt}\OtherTok{)}\NormalTok{ , }\OtherTok{λ} \OtherTok{(\_}\NormalTok{ , a}\OtherTok{)}\NormalTok{ x }\OtherTok{→} \OtherTok{(}\NormalTok{tt , x}\OtherTok{)} \OtherTok{)}
  
  \NormalTok{    ◃unitr }\OtherTok{:} \OtherTok{(}\NormalTok{p ◃ 𝕪}\OtherTok{)}\NormalTok{ ⇆ p}
  \NormalTok{    ◃unitr }\OtherTok{=} \OtherTok{(} \OtherTok{(λ} \OtherTok{(}\NormalTok{a , γ}\OtherTok{)} \OtherTok{→}\NormalTok{ a}\OtherTok{)}\NormalTok{ , }\OtherTok{(λ} \OtherTok{(}\NormalTok{a , γ}\OtherTok{)}\NormalTok{ b }\OtherTok{→} \OtherTok{(}\NormalTok{b , tt}\OtherTok{))} \OtherTok{)}
  
  \KeywordTok{open}\NormalTok{ ◃LRUnit }\KeywordTok{public}
  \end{Highlighting}
  \end{Shaded}
  
  In fact, \texttt{◃} restricts to a monoidal product on
  \(\mathbf{Poly^{Cart}}\), since the functorial action of \texttt{◃} on
  lenses preserves Cartesian lenses, and all of the above-defined structure morphisms for \texttt{◃} are
  Cartesian. We should expect, then, for these equivalences to be somehow reflected
  in the structure of Cartesian lenses \texttt{η\ :\ 𝕪\ ⇆\ 𝔲} and
  \texttt{μ\ :\ 𝔲\ ◃\ 𝔲\ ⇆\ 𝔲} witnessing the closure of $u$ under $\top$ and $\Sigma$. This would be the case, e.g., if the
  following diagrams were to commute: \[
  \begin{tikzcd}
      {y \triangleleft u} & {u \triangleleft u } & {u \triangleleft y} \\
      & {u}
      \arrow["{\eta \triangleleft u}", from=1-1, to=1-2]
      \arrow["{\mathsf{\triangleleft unitl}}"{description}, from=1-1, to=2-2]
      \arrow["\mu", from=1-2, to=2-2]
      \arrow["{u \triangleleft \eta}"', from=1-3, to=1-2]
      \arrow["{\mathsf{\triangleleft unitr}}"{description}, from=1-3, to=2-2]
  \end{tikzcd} \quad (M1) \qquad \begin{tikzcd}
      {(u \triangleleft u) \triangleleft u} & {u \triangleleft (u \triangleleft u)} & {u \triangleleft u} \\
      {u \triangleleft u} && {u}
      \arrow["{\mathsf{\triangleleft assoc}}", from=1-1, to=1-2]
      \arrow["{\mu \triangleleft u}"', from=1-1, to=2-1]
      \arrow["{u \triangleleft \mu}", from=1-2, to=1-3]
      \arrow["\mu", from=1-3, to=2-3]
      \arrow["\mu"', from=2-1, to=2-3]
  \end{tikzcd} \quad (M2)
  \] One may recognize these as the usual diagrams for a monoid in a monoidal
  category, hence (since \texttt{◃} corresponds to endofunctor composition) for a \emph{monad} as typically defined.
  However, because of the higher-categorical structure of types in HoTT,
  we should not only ask for these diagrams to commute, but for the cells
  exhibiting that these diagrams commute to themselves be subject to
  higher coherences, and so on, giving \texttt{𝔲} not the structure of a monad, but rather of an \emph{\(\infty\)-monad}.
  
  Yet demonstrating that \(𝔲\) is an \(\infty\)-monad involves specifying
  a potentially infinite amount of coherence data. Have we therefore
  traded both the Scylla of equality-up-to-isomorphism and the Charybdis
  of strictness for an even worse fate of higher coherence hell? The
  answer to this question, surprisingly, is negative, as there is a way to
  implicitly derive all of this data from a single axiom, which
  corresponds to the characteristic axiom of HoTT itself: univalence, as we shall show in the following section. For now, however, we turn to the other main type former of dependent type theory which we have not yet considered: the dependent function type, or \texttt{Π} type.

\subsection{The $\upuparrows$ Functor \& $\Pi$ Types}

We have so far considered how polynomial universes may be equipped with
structure to interpret the unit type and dependent pair types. We have
not yet, however, said much in the way of \emph{dependent function
types.} In order to rectify this omission, it will first be prudent to
consider some additional structure on the category of polynomial
functors -- specifically a new functor
\({\upuparrows}\)
that plays a similar role for \texttt{Π} types as the composition
\(\triangleleft : \mathbf{Poly} \times \mathbf{Poly} \to \mathbf{Poly}\)
played for \texttt{Σ} types, and which in turn bears a close connection
to \emph{distributive laws} in \(\mathbf{Poly}\).

The \(\upuparrows\) functor can be loosely defined as the solution to
the following problem: given a polynomial \texttt{𝔲}, find
\texttt{𝔲\ ⇈\ 𝔲} such that \texttt{𝔲} has a Cartesian morphism from \texttt{𝔲\ ⇈\ 𝔲} if and
only if \texttt{𝔲} has the structure to interpret \texttt{Π} types (in
the same way that \texttt{𝔲} has a Cartesian morphism from \texttt{𝔲\ ◃\ 𝔲} if and only if
\texttt{𝔲} has the structure to interpret \texttt{Σ} types).
Generalizing this to arbitrary pairs of polynomials
\(p = (A , B), ~ q = (C , D)\) then yields the following formula for
\(p \upuparrows q\): \[
p \upuparrows q = \sum_{(a , f) : \sum_{a : A} C^{B(a)}} y^{\prod_{b : B(a)} D(f(b))}
\] and the following definition in Agda:

\begin{Shaded}
\begin{Highlighting}[]
\OtherTok{\_}\NormalTok{⇈}\OtherTok{\_} \OtherTok{:} \OtherTok{∀} \OtherTok{\{}\NormalTok{ℓ0 ℓ1 κ0 κ1}\OtherTok{\}} \OtherTok{→}\NormalTok{ Poly ℓ0 κ0 }\OtherTok{→}\NormalTok{ Poly ℓ1 κ1 }
      \OtherTok{→}\NormalTok{ Poly }\OtherTok{(}\NormalTok{ℓ0 ⊔ κ0 ⊔ ℓ1}\OtherTok{)} \OtherTok{(}\NormalTok{κ0 ⊔ κ1}\OtherTok{)}
\OtherTok{(}\NormalTok{A , B}\OtherTok{)}\NormalTok{ ⇈ }\OtherTok{(}\NormalTok{C , D}\OtherTok{)} \OtherTok{=} 
    \OtherTok{(}\NormalTok{ Σ A }\OtherTok{(λ}\NormalTok{ a }\OtherTok{→}\NormalTok{ B a }\OtherTok{→}\NormalTok{ C}\OtherTok{)} 
\NormalTok{    , }\OtherTok{(λ} \OtherTok{(}\NormalTok{a , f}\OtherTok{)} \OtherTok{→} \OtherTok{(}\NormalTok{b }\OtherTok{:}\NormalTok{ B a}\OtherTok{)} \OtherTok{→}\NormalTok{ D }\OtherTok{(}\NormalTok{f b}\OtherTok{)))}
\end{Highlighting}
\end{Shaded}

Note that this construction is straightforwardly functorial with respect
to arbitrary lenses in its 2nd argument. Functoriality of the 1st
argument is trickier, however. For reasons that will become apparent
momentarily, we define the functorial action
\(p \upuparrows q \to p' \upuparrows q\) of \(\upuparrows\)
on a lens \(f : p \to p'\) equipped with a left inverse
\(f' : p' \to p\), i.e.~such that
\(f' \circ f = \text{id}_p\).
\footnote{To see why this is the right choice of morphism for which ⇈ is functorial in its first argument, we note that pairs consisting of a morphism and a left inverse for it are equivalently the morphisms between identity morphisms in the \emph{twisted arrow category} of $\mathbf{Poly}$, i.e. diagrams of the following form: $$
\begin{array}{ccc}
p & \to & q\\
= & & =\\
p & \leftarrow & q
\end{array}
$$}

\begin{Shaded}
\begin{Highlighting}[]
\NormalTok{⇈Lens }\OtherTok{:} \OtherTok{∀} \OtherTok{\{}\NormalTok{ℓ0 ℓ1 ℓ2 ℓ3 κ0 κ1 κ2 κ3}\OtherTok{\}}
        \OtherTok{→} \OtherTok{\{}\NormalTok{p }\OtherTok{:}\NormalTok{ Poly ℓ0 κ0}\OtherTok{\}} \OtherTok{(}\NormalTok{r }\OtherTok{:}\NormalTok{ Poly ℓ2 κ2}\OtherTok{)}
        \OtherTok{→} \OtherTok{\{}\NormalTok{q }\OtherTok{:}\NormalTok{ Poly ℓ1 κ1}\OtherTok{\}} \OtherTok{(}\NormalTok{s }\OtherTok{:}\NormalTok{ Poly ℓ3 κ3}\OtherTok{)}
        \OtherTok{→} \OtherTok{(}\NormalTok{f }\OtherTok{:}\NormalTok{ p ⇆ r}\OtherTok{)} \OtherTok{(}\NormalTok{f\textquotesingle{} }\OtherTok{:}\NormalTok{ r ⇆ p}\OtherTok{)} 
        \OtherTok{→}\NormalTok{ EqLens p }\OtherTok{(}\NormalTok{id p}\OtherTok{)} \OtherTok{(}\NormalTok{comp p f f\textquotesingle{}}\OtherTok{)}
        \OtherTok{→} \OtherTok{(}\NormalTok{g }\OtherTok{:}\NormalTok{ q ⇆ s}\OtherTok{)} \OtherTok{→} \OtherTok{(}\NormalTok{p ⇈ q}\OtherTok{)}\NormalTok{ ⇆ }\OtherTok{(}\NormalTok{r ⇈ s}\OtherTok{)}
\NormalTok{⇈Lens }\OtherTok{\{}\NormalTok{p }\OtherTok{=}\NormalTok{ p}\OtherTok{\}}\NormalTok{ r s }\OtherTok{(}\NormalTok{f , f♯}\OtherTok{)} \OtherTok{(}\NormalTok{f\textquotesingle{} , f\textquotesingle{}♯}\OtherTok{)} \OtherTok{(}\NormalTok{e , e♯}\OtherTok{)} \OtherTok{(}\NormalTok{g , g♯}\OtherTok{)} \OtherTok{=} 
    \OtherTok{(} \OtherTok{(λ} \OtherTok{(}\NormalTok{a , γ}\OtherTok{)} \OtherTok{→} \OtherTok{(}\NormalTok{f a , }\OtherTok{(λ}\NormalTok{ x }\OtherTok{→}\NormalTok{ g }\OtherTok{(}\NormalTok{γ }\OtherTok{(}\NormalTok{f♯ a x}\OtherTok{)))))}
\NormalTok{    , }\OtherTok{(λ} \OtherTok{(}\NormalTok{a , γ}\OtherTok{)}\NormalTok{ Ϝ x }\OtherTok{→} 
\NormalTok{         g♯ }\OtherTok{(}\NormalTok{γ x}\OtherTok{)} 
            \OtherTok{(}\NormalTok{transp }\OtherTok{(λ}\NormalTok{ y }\OtherTok{→}\NormalTok{ snd s }\OtherTok{(}\NormalTok{g }\OtherTok{(}\NormalTok{γ y}\OtherTok{)))} 
                    \OtherTok{(}\NormalTok{sym }\OtherTok{(}\NormalTok{e♯ a x}\OtherTok{))} 
                    \OtherTok{(}\NormalTok{Ϝ }\OtherTok{(}\NormalTok{f\textquotesingle{}♯ }\OtherTok{(}\NormalTok{f a}\OtherTok{)} \OtherTok{(}\NormalTok{transp }\OtherTok{(}\NormalTok{snd p}\OtherTok{)} \OtherTok{(}\NormalTok{e a}\OtherTok{)}\NormalTok{ x}\OtherTok{)))))} \OtherTok{)}
\end{Highlighting}
\end{Shaded}

\noindent A Cartesian lens
\texttt{(π\ ,\ π♯)\ :\ 𝔲\ ⇈\ 𝔲\ ⇆\ 𝔲} then effectively shows that \texttt{𝔲}
is closed under \texttt{Π}-types, since:

\begin{itemize}
\item
  \texttt{π} maps a pair \texttt{(A\ ,\ B)} consisting of
  \texttt{A\ :\ 𝓤} and \texttt{B\ :\ u(A)\ →\ 𝓤} to a term
  \texttt{π(A,B)} representing the corresponding \texttt{Π} type. This
  corresponds to the type formation rule
  \[ \inferrule{\Gamma \vdash A : \mathsf{Type}\\ \Gamma, x : A \vdash B[x] ~ \mathsf{Type}}{\Gamma \vdash \Pi x : A . B[x] ~ \mathsf{Type}} \]
\item
  The ``elimination rule'' \texttt{π♯\ (A\ ,\ B)}, for any pair
  \texttt{(A\ ,\ B)} as above, maps an element \texttt{f\ :\ π(A,B)} to
  a function \texttt{π♯\ (A\ ,\ B)\ f\ :\ (a\ :\ u(A))\ →\ u\ (B\ x)}
  which takes an element \texttt{x} of \texttt{A} and yields an element
  of \texttt{B\ x}. This corresponds to the rule for function
  application: \[
  \inferrule{\Gamma \vdash f : \Pi x : A . B[x]\\ \Gamma \vdash a : A}{\Gamma \vdash f ~ a : B[a]}
  \]
\item
  Moreover, for all \texttt{(A\ ,\ B)}, the inverse map
  \texttt{π♯⁻¹\ (A\ ,\ B)\ :\ (x\ :\ u(A))\ →\ u(B(x))\ →\ u(π(A,B))} to \texttt{π♯\ (A\ ,\ B)} corresponds to \(\lambda\)-abstraction: \[
  \inferrule{\Gamma, x : A \vdash f[x] : B[x]}{\Gamma \vdash \lambda x . f[x] : \Pi x : A . B[x]}
  \]
\item
  The fact that \texttt{π♯⁻¹\ (A\ ,\ B)} is both a left and a right
  inverse to \texttt{π♯} then corresponds to the \(\beta\) and \(\eta\)
  laws for \texttt{Π} types. \[
  (\lambda x . f[x]) ~ a = f[a] \qquad f = \lambda x . f ~ x
  \]
\end{itemize}

Although it is clear enough that \(\upuparrows\) serves its
intended purpose of characterizing \texttt{Π} types polynomially, its construction seems somewhat more ad hoc than that of
\(\triangleleft\), which similarly characterized \texttt{Σ} types in
polynomial universes while arising quite naturally from composition of
polynomial functors. We would like to better understand what additional
properties \(\upuparrows\) must satisfy, and how these in turn are
reflected as properties of polynomial universes with \texttt{Π} types.
In fact, we will ultimately show that this construction is intimately
linked with a quite simple structure on polynomial universes \texttt{𝔲},
namely a \emph{distributive law} of \texttt{𝔲} (viewed as a monad) over
itself. Before that, however, we note some other key properties of
\(\upuparrows\).

Specifically, let \(\mathbf{Poly}_{R}\) be the category whose objects
are polynomials and whose morphisms are lenses equipped with left
inverses. Straightforwardly, \(\triangleleft\) restricts to a monoidal
product on \(\mathbf{Poly}_R\), since it is functorial in both arguments
and must preserve left/right inverses. Hence \(\upuparrows\) can be
viewed as a functor
\(\mathbf{Poly}_R \times \mathbf{Poly} \to \mathbf{Poly}\). Then
\(\upuparrows\) moreover naturally carries the structure of an
\emph{action} on \(\mathbf{Poly}\) of the monoidal category
\(\mathbf{Poly}_R\) equipped with \(\triangleleft\), in that there are
equivalences \[
y \upuparrows p \simeq p \quad \text{and} \quad
(p \triangleleft q) \upuparrows r \simeq p \upuparrows (q \upuparrows r)
\] defined as follows:

\begin{Shaded}
\begin{Highlighting}[]
\KeywordTok{module}\NormalTok{ Unit⇈ }\OtherTok{\{}\NormalTok{ℓ κ}\OtherTok{\}} \OtherTok{(}\NormalTok{p }\OtherTok{:}\NormalTok{ Poly ℓ κ}\OtherTok{)} \KeywordTok{where}

\NormalTok{    𝕪⇈ }\OtherTok{:} \OtherTok{(}\NormalTok{𝕪 ⇈ p}\OtherTok{)}\NormalTok{ ⇆ p}
\NormalTok{    𝕪⇈ }\OtherTok{=} \OtherTok{(} \OtherTok{(λ} \OtherTok{(\_}\NormalTok{ , a}\OtherTok{)} \OtherTok{→}\NormalTok{ a tt}\OtherTok{)}\NormalTok{ , }\OtherTok{λ} \OtherTok{(\_}\NormalTok{ , a}\OtherTok{)}\NormalTok{ b tt }\OtherTok{→}\NormalTok{ b}\OtherTok{)}

\KeywordTok{open}\NormalTok{ Unit⇈ }\KeywordTok{public}

\KeywordTok{module}\NormalTok{ ◃⇈ }\OtherTok{\{}\NormalTok{ℓ0 ℓ1 ℓ2 κ0 κ1 κ2}\OtherTok{\}} \OtherTok{(}\NormalTok{p }\OtherTok{:}\NormalTok{ Poly ℓ0 κ0}\OtherTok{)} 
          \OtherTok{(}\NormalTok{q }\OtherTok{:}\NormalTok{ Poly ℓ1 κ1}\OtherTok{)} \OtherTok{(}\NormalTok{r }\OtherTok{:}\NormalTok{ Poly ℓ2 κ2}\OtherTok{)} \KeywordTok{where}

\NormalTok{    ⇈Curry }\OtherTok{:} \OtherTok{((}\NormalTok{p ◃ q}\OtherTok{)}\NormalTok{ ⇈ r}\OtherTok{)}\NormalTok{ ⇆ }\OtherTok{(}\NormalTok{p ⇈ }\OtherTok{(}\NormalTok{q ⇈ r}\OtherTok{))}
\NormalTok{    ⇈Curry }\OtherTok{=} \OtherTok{(} \OtherTok{(λ} \OtherTok{((}\NormalTok{a , h}\OtherTok{)}\NormalTok{ , k}\OtherTok{)} 
                  \OtherTok{→} \OtherTok{(}\NormalTok{a , }\OtherTok{(λ}\NormalTok{ b }\OtherTok{→} \OtherTok{(} \OtherTok{(}\NormalTok{h b}\OtherTok{)} 
\NormalTok{                                , }\OtherTok{(λ}\NormalTok{ d }\OtherTok{→}\NormalTok{ k }\OtherTok{(}\NormalTok{b , d}\OtherTok{))))))}
\NormalTok{             , }\OtherTok{(λ} \OtherTok{((}\NormalTok{a , h}\OtherTok{)}\NormalTok{ , k}\OtherTok{)}\NormalTok{ f }\OtherTok{(}\NormalTok{b , d}\OtherTok{)} \OtherTok{→}\NormalTok{ f b d}\OtherTok{)} \OtherTok{)}

\KeywordTok{open}\NormalTok{ ◃⇈ }\KeywordTok{public}
\end{Highlighting}
\end{Shaded}

The fact that \texttt{⇈Curry} is an equivalence corresponds to the usual
currying isomorphism relating dependent functions types to
dependent pair types: \[
\Pi (x , y) : \Sigma x : A . B[x] . C[x,y] \simeq \Pi x : A . \Pi y : B[x] . C[x,y]
\]

Similarly, \(\upuparrows\) is colax with respect to \(\triangleleft\) in
its second argument in that there are natural transformations
\[
p \upuparrows y \to y \quad \text{and} \quad
p \upuparrows (q \triangleleft r) \to (p \upuparrows q) \triangleleft (p \upuparrows r)
\]

\begin{Shaded}
\begin{Highlighting}[]
\KeywordTok{module}\NormalTok{ ⇈Unit }\OtherTok{\{}\NormalTok{ℓ κ}\OtherTok{\}} \OtherTok{(}\NormalTok{p }\OtherTok{:}\NormalTok{ Poly ℓ κ}\OtherTok{)} \KeywordTok{where}

\NormalTok{    ⇈𝕪 }\OtherTok{:} \OtherTok{(}\NormalTok{p ⇈ 𝕪}\OtherTok{)}\NormalTok{ ⇆ 𝕪}
\NormalTok{    ⇈𝕪 }\OtherTok{=} \OtherTok{(} \OtherTok{(λ} \OtherTok{(}\NormalTok{a , γ}\OtherTok{)} \OtherTok{→}\NormalTok{ tt}\OtherTok{)}\NormalTok{ , }\OtherTok{λ} \OtherTok{(}\NormalTok{a , γ}\OtherTok{)}\NormalTok{ tt b }\OtherTok{→}\NormalTok{ tt }\OtherTok{)}

\KeywordTok{open}\NormalTok{ ⇈Unit }\KeywordTok{public}

\KeywordTok{module}\NormalTok{ ⇈◃ }\OtherTok{\{}\NormalTok{ℓ0 ℓ1 ℓ2 κ0 κ1 κ2}\OtherTok{\}} \OtherTok{(}\NormalTok{p }\OtherTok{:}\NormalTok{ Poly ℓ0 κ0}\OtherTok{)} 
          \OtherTok{(}\NormalTok{q }\OtherTok{:}\NormalTok{ Poly ℓ1 κ1}\OtherTok{)} \OtherTok{(}\NormalTok{r }\OtherTok{:}\NormalTok{ Poly ℓ2 κ2}\OtherTok{)} \KeywordTok{where}

\NormalTok{    ⇈Distr }\OtherTok{:} \OtherTok{(}\NormalTok{p ⇈ }\OtherTok{(}\NormalTok{q ◃ r}\OtherTok{))}\NormalTok{ ⇆ }\OtherTok{((}\NormalTok{p ⇈ q}\OtherTok{)}\NormalTok{ ◃ }\OtherTok{(}\NormalTok{p ⇈ r}\OtherTok{))}
\NormalTok{    ⇈Distr }\OtherTok{=} \OtherTok{(} \OtherTok{(λ} \OtherTok{(}\NormalTok{a , h}\OtherTok{)} 
                  \OtherTok{→} \OtherTok{(} \OtherTok{(}\NormalTok{a , }\OtherTok{(λ}\NormalTok{ b }\OtherTok{→}\NormalTok{ fst }\OtherTok{(}\NormalTok{h b}\OtherTok{)))} 
\NormalTok{                    , }\OtherTok{λ}\NormalTok{ f }\OtherTok{→} \OtherTok{(}\NormalTok{a , }\OtherTok{(λ}\NormalTok{ b }\OtherTok{→}\NormalTok{ snd }\OtherTok{(}\NormalTok{h b}\OtherTok{)} \OtherTok{(}\NormalTok{f b}\OtherTok{)))} \OtherTok{))} 
\NormalTok{             , }\OtherTok{(λ} \OtherTok{(}\NormalTok{a , h}\OtherTok{)} \OtherTok{(}\NormalTok{f , g}\OtherTok{)}\NormalTok{ b }\OtherTok{→} \OtherTok{(}\NormalTok{f b , g b}\OtherTok{))} \OtherTok{)}

\KeywordTok{open}\NormalTok{ ⇈◃ }\KeywordTok{public}
\end{Highlighting}
\end{Shaded}

Moreover, this colax structure on ⇈ descends to $\mathbf{Poly^{Cart}}$ in that the natural transformations defined above are \emph{Cartesian}. In particular, the fact that \texttt{⇈Distr} is Cartesian corresponds to the
distributive law of \texttt{Π} types over \texttt{Σ} types, i.e.~\[
\Pi x : A . \Sigma y : B[x] . C[x,y] \simeq \Sigma f : \Pi x : A . B[x] . \Pi x : A . C[x, f(x)]
\] One may wonder whether this distributive law is
related to a distributive law of the monad structure on a
universe 𝔲 given by Σ types (as discussed in the previous subsection) over itself, i.e.~a morphism
\[ u \triangleleft u \to u \triangleleft u \] The answer to this question is affirmative, but in order to see why, we must first explain the machinery that allows us to straightforwardly derive the coherence data for such structures from the mere existence of cartesian morphisms as described above, namely \emph{univalence}.

\section{Polynomial Universes}\label{sec:polyuniv}

For any polynomial \texttt{𝔲}, we say that \texttt{𝔲} is
\emph{univalent} if it is a \emph{subterminal object} in
\(\mathbf{Poly^{Cart}}\), i.e.~for any other polynomial \texttt{p}, the
type of Cartesian lenses \texttt{p\ ⇆\ 𝔲} is a mere proposition, or in other words, any two
Cartesian lenses with codomain \texttt{𝔲} are equal.

\begin{Shaded}
\begin{Highlighting}[]
\NormalTok{isUnivalent }\OtherTok{:} \OtherTok{∀} \OtherTok{\{}\NormalTok{ℓ κ}\OtherTok{\}} \OtherTok{→}\NormalTok{ Poly ℓ κ }\OtherTok{→}\NormalTok{ Setω}
\NormalTok{isUnivalent u }\OtherTok{=} 
    \OtherTok{∀} \OtherTok{\{}\NormalTok{ℓ\textquotesingle{} κ\textquotesingle{}}\OtherTok{\}} \OtherTok{\{}\NormalTok{p }\OtherTok{:}\NormalTok{ Poly ℓ\textquotesingle{} κ\textquotesingle{}}\OtherTok{\}}
      \OtherTok{→} \OtherTok{\{}\NormalTok{f g }\OtherTok{:}\NormalTok{ p ⇆ u}\OtherTok{\}}
      \OtherTok{→}\NormalTok{ isCartesian u f}
      \OtherTok{→}\NormalTok{ isCartesian u g}
      \OtherTok{→}\NormalTok{ f ≡ g}
    
\end{Highlighting}
\end{Shaded} We call this property of polynomials univalence in analogy with the
usual univalence axiom of HoTT, since the univalence axiom for a universe $\mathcal{U}$ is equivalent to the statement that the polynomial $$
\sum_{A : \mathcal{U}} y^{A}
$$ is univalent in the above sense. Indeed, the statement that there are \emph{enough} univalent families in $\infty\mathbf{Grpd}^{\mathcal{C}^{\textnormal{op}}}$ is equivalent to the existence of a \emph{terminal family} in $\mathbf{Poly^{Cart}}$, i.e. a collection of subterminal objects $u_i \in \mathbf{Poly^{Cart}}$ such that every $p \in \mathbf{Poly^{Cart}}$ has a Cartesian morphism to \emph{some} $u_i$. In what follows, we therefore assume the existence of such a terminal family $u_i$, which, although not strictly necessary for the main theorems of this paper to do with semantics of dependent type theory,\footnote{which are in fact all valid in ordinary intensional Martin-Löf type theory without univalence} is useful for constructing illustrative examples.

We refer to univalent polynomial functors as \emph{polynomial
universes.} If we think of a polynomial \texttt{p} as representing a
family of types, then what this tells us is that if \texttt{𝔲} is a
polynomial universe, there is essentially at most one way for \texttt{𝔲}
to contain the types represented by \texttt{p}, where containment is
here understood as existence of a Cartesian lens \texttt{p\ ⇆\ 𝔲}. In
this case, we say that \texttt{𝔲} \emph{classifies} the types
represented by \texttt{p}.

As a direct consequence of this fact, it follows that every diagram
consisting of parallel Cartesian lenses into a polynomial universe
automatically commutes, and moreover, every higher diagram that can be
formed between the cells exhibiting such commutation must also commute,
etc.

Hence we have the following:

\begin{theorem}
If 𝔲 is a polynomial universe with Cartesian lenses \texttt{η\ :\ 𝕪\ ⇆\ 𝔲} and \texttt{μ\ :\ 𝔲\ ◃\ 𝔲\ ⇆\ 𝔲}, then the diagrams M1, M2 given in \S\ref{sec:composition} commute.
\end{theorem} The proof is essentially by observation that all paths through the diagrams M1 and M2 are Cartesian morphisms that terminate in 𝔲 and hence, by univalence of 𝔲, must be equal. Moreover, one can in principle reason similarly to derive all the higher coherence data of an $\infty$-monad, although, since it is not currently known how to express the totality of such data as a type in HoTT, this cannot be expressed directly as a theorem in HoTT at present.

For $\Pi$ types, the situation is more complex. In the remainder of this section, we will concern ourselves with showing that the closure of a univalent universe $u$ under $\Pi$ types gives rise to a \emph{distributive law} of the monad structure on $u$ over itself.

Recall that a (1-dimensional) distributive law of a monad $m$ over another monad $n$ is essentially an answer to the question ``when is the composite $m \triangleleft n$ also a monad?'' given by a morphism:  $$
\aleph : n \triangleleft m \to m \triangleleft n
$$ such that the following diagrams commute: \[
\footnotesize \def\arraystretch{3.5} \begin{array}{crccr} \begin{tikzcd}
	{n \triangleleft m \triangleleft m} & {m \triangleleft n \triangleleft m} & {m \triangleleft m \triangleleft n} \\
	{n \triangleleft m} && {m \triangleleft n}
	\arrow["{\aleph \triangleleft m}", from=1-1, to=1-2]
	\arrow["{n \triangleleft \mu_m}"', from=1-1, to=2-1]
	\arrow["{m \triangleleft \aleph}", from=1-2, to=1-3]
	\arrow["{\mu_m \triangleleft n}", from=1-3, to=2-3]
	\arrow["\aleph", from=2-1, to=2-3]
\end{tikzcd} & \text{(DL1)} &\quad&
\begin{tikzcd}
	{n \triangleleft n \triangleleft m} & {n \triangleleft m \triangleleft n} & {m \triangleleft n \triangleleft n} \\
	{n \triangleleft m} && {m \triangleleft n}
	\arrow["{n \triangleleft \aleph}", from=1-1, to=1-2]
	\arrow["{\mu_n \triangleleft m}"', from=1-1, to=2-1]
	\arrow["{\aleph \triangleleft n}", from=1-2, to=1-3]
	\arrow["{\mu_m \triangleleft n}", from=1-3, to=2-3]
	\arrow["\aleph", from=2-1, to=2-3]
\end{tikzcd} & \text{(DL2)}\\
\footnotesize \begin{tikzcd}
    {n \triangleleft \text{id}_1} & n & {\text{id}_1 \triangleleft n} \\
    {n \triangleleft m} && {m \triangleleft n}
    \arrow["\simeq"{description}, draw=none, from=1-1, to=1-2]
    \arrow["{n \triangleleft \eta_m}"{description}, from=1-1, to=2-1]
    \arrow["\simeq"{description}, draw=none, from=1-2, to=1-3]
    \arrow["{\eta_m \triangleleft n}"{description}, from=1-3, to=2-3]
    \arrow["\aleph", from=2-1, to=2-3]
\end{tikzcd} & \text{(DL3)} &\quad&
\footnotesize\begin{tikzcd}
    {\text{id}_1 \triangleleft m} & m & {m \triangleleft \text{id}_1} \\
    {n \triangleleft m} && {m \triangleleft n}
    \arrow["\simeq"{description}, draw=none, from=1-1, to=1-2]
    \arrow["{\eta_n \triangleleft m}"{description}, from=1-1, to=2-1]
    \arrow["\simeq"{description}, draw=none, from=1-2, to=1-3]
    \arrow["{m \triangleleft \eta_n}"{description}, from=1-3, to=2-3]
    \arrow["\aleph", from=2-1, to=2-3]
\end{tikzcd} & \text{(DL4)}
\end{array}
\] Note that, in the case where $m = n = u$, the morphisms identified by these diagrams are \emph{not} generally Cartesian morphisms into $u$, so we cannot immediately apply univalence as we did for the monad laws.

The solution to this problem proceeds in several steps: \begin{enumerate} 

  \item We first generalize from distributive laws $\aleph$, as above, to \emph{distributors},\footnote{Distributors as above arise in $\mathbf{EM}$-$\mathbb{C}\mathbf{at}^{\sharp}$ the Eilenberg-Moore completion \cite{lack2002formal} of the double category $\mathbb{C}\mathbf{at}^{\sharp}=\mathbb{C}\mathbf{omod}(\mathbf{Poly})$ of polynomial comonads and bicomodules \cite{shapiro2024polynomial}. Indeed, there we find morphisms of the form $m\tri p\to p\tri n$ for polynomial monads $m,n$, and distributive laws between monads are the formal monads in $\mathbf{EM}$-$\mathbb{C}\mathbf{at}^{\sharp}$. Distributors also arise in what Lynch et al.\ call \emph{effects handlers} \cite{lynch2023concepts}, where morphisms are of the form $s\tri c\to d\tri s$ for polynomial comonads $c,d\in\mathbb{C}\mathbf{at}^{\sharp}$.
  } which are morphisms of the form $$p \tri q \to r \tri s$$
  \item We likewise generalize the definition of the $\upuparrows$ functor given previously to an action on $\mathbf{Poly}$ of the \emph{twisted arrow category} $\mathsf{Tw}(\mathbf{Poly})$ of $\mathbf{Poly}$. In particular, since $\mathbf{Poly}^R$ as defined previously is equivalent to the full (monoidal) subcategory of $\mathsf{Tw}(\mathbf{Poly})$ spanned by identity morphisms on polynomial functors, our previous definition of $\upuparrows$ turns out to be the restriction of this generalized definition from an action of $\mathsf{Tw}(\mathbf{Poly})$ to one of $\mathbf{Poly}^R$ along the embedding $$\mathbf{Poly}^R \hookrightarrow \mathsf{Tw}(\mathbf{Poly})$$ In what follows, we therefore continue to write $p \upuparrows q$ as a shorthand for $\text{id}_p \upuparrows q$.
  
  \item So-generalized, $\upuparrows$ acquires a key property: given $\phi : p \to s$, every morphism $f : \phi \upuparrows q \to r$ gives rise to a distributor $\nabla_f : p \triangleleft q \to r \triangleleft s$.
  
  \item We show that the operations on distributors appearing in the equations for a distributive law correspond to compositions of morphisms of the form $\phi \upuparrows q \to r$ under the above translation between morphisms $\phi \upuparrows q \to r$ and distributors $p \triangleleft q \to r \triangleleft s$, making use of the structure of $\upuparrows$ as a colax monoidal action. This allows us to convert the diagrams for a distributive law of $u$ over itself into diagrams involving a (Cartesian) morphism $\text{id}_u \upuparrows u \to u$, whence we conclude as before in the case of the monad laws for $\Sigma$ types that, if $u$ is univalent, then these diagrams (and all higher-dimensional diagrams involving them) must commute.

\end{enumerate}

Hence we have the following:

\vspace{0.5em}

\begin{theorem}
If $u$ is a polynomial universe with Cartesian lenses $\eta : y \leftrightarrows u$, $\mu : u \triangleleft u \leftrightarrows u$, along with $\pi : u \upuparrows u \leftrightarrows u$, then letting $m = n = u$ and $\aleph = \nabla_\pi$, the diagrams DL1, DL2, DL3, and DL4 commute.
\end{theorem}

\vspace{0.5em}

A full proof, formalized in Agda, is given in the appendix.\footnote{As in the case of Theorem 4.2, we could in principle continue in this same manner to derive the higher coherences of an $\infty$-distributive law of $\infty$-monads, but since the \emph{type} of all such data is not currently definable, we cannot express this as a theorem in HoTT at present.} For the remainder of this section, we point out some high-level details of particular note in the proof.

\subsection{Distributors \& Jump Structures}

Not every distributor $\nabla : p \triangleleft q \to r \triangleleft s$ is of the form $\nabla_f$ for some lens $f : \phi \upuparrows q \to s$. We define a \emph{jump structure} on a distributor $\nabla$ to be a witness to the fact that $\nabla = \nabla_f$ for some $f$. Expressed this way, the type of jump structures on a distributor $\nabla$ may be written as $$
\sum_{\phi : p \to s} \sum_{f : \phi \upuparrows q \to r} \nabla = \nabla_f
$$ However, there is another description of jump structures purely in terms of the structure of polynomial functors. Intuitively, a jump structure witnesses that $\nabla$ essentially acts as $\phi$ on its components in $p,s$ while ``jumping over'' $q, r$. To express this directly in terms of polynomials, we need a notion of ``independence'' of the components of a distributor. For this, we make use of an additional monoidal product on $\mathbf{Poly}$, the \emph{tensor product} $\otimes$, defined as follows: given polynomials $p = \sum_{a : A} y^{B[a]}, q = \sum_{c : C} y^{D[c]}$, their tensor product $p \otimes q$ is defined as $$
p \otimes q = \sum_{(a,c) : A \times C} y^{B[a] \times D[c]}
$$ Unlike $\tri$, the monoidal product $\otimes$ is symmetric on $\mathbf{Poly}$. However, $\otimes$ and $\tri$ are compatible with one another, in that they together form the structure of a normal duoidal category \cite{SpivakNiu}. This in particular gives rise to a monoidal natural transformation $$\mathsf{indep}_{p,q} : p \otimes q \to p \tri q$$ between these two structures. Intuitively, thinking of $\otimes$ as independent and $\tri$ as dependent combination of polynomials, $\mathsf{indep}$ exhibits independence as a trivial kind of dependence.

We further note that, although lenses give a canonical notion of morphism between polynomial functors, they are not the only such. If we think of polynomial functors as corresponding to their representative display maps, then another candidate notion of morphism between polynomials is given by commutative squares. I.e. for $p : B \to A$ and $q : D \to C$, such a morphism $p \not \to q$ consists of a commuting square: $$
\begin{tikzcd}
	B & D \\
	A & C
	\arrow["{f_1}", from=1-1, to=1-2]
	\arrow["p"', from=1-1, to=2-1]
	\arrow["q", from=1-2, to=2-2]
	\arrow["{f^\flat}", from=2-1, to=2-2]
\end{tikzcd}
$$ Following \cite{catsys}, we refer to natural transformations of polynomial functors as \emph{lenses} and to commutative squares of their representative morphisms as \emph{charts}. Notably, the category of polynomial functors and charts has the following properties: \begin{enumerate} \item It is equivalent to the \emph{arrow category} $\mathbf{Type} \downarrow \mathbf{Type}$ of the category $\mathbf{Type}$ of types and functions. \item It is the fibrewise opposite of $\mathbf{Poly}$, viewed as a category fibred over $\mathbf{Type}$. \end{enumerate}

Straightforwardly, the category of polynomial functors and charts between them inherits $\otimes$ as a monoidal product, which is moreover the Cartesian product for that category. Although $\tri$ does not define a corresponding monoidal product on the category of polynomial functors and charts, it is the case that, for any $p,q$, there exists a chart $
\pi_1 : p \triangleleft q \not \to p
$ given by the first projection of both components of $p$.

Lenses and charts together form a \emph{double category}, i.e. a category with two distinct kinds of morphisms, along with a notion of \emph{commuting squares} formed by these morphisms (c.f. \cite{catsys} for a full definition of this double category). When depicting this double category, we draw lenses horizontally, and charts vertically. Then a square $$
\begin{tikzcd}
	p & q \\
	p' & q'
	\arrow["\phi", from=1-1, to=1-2]
	\arrow["f"', "\shortmid"{marking}, from=1-1, to=2-1]
	\arrow["g", "\shortmid"{marking}, from=1-2, to=2-2]
	\arrow["\phi'"', from=2-1, to=2-2]
\end{tikzcd}
$$ where $\phi = (\phi_1, \phi^\sharp), ~ \phi' = (\phi_1', {\phi'}^\sharp), ~ f = (f_1, f_\flat), ~ g = (g_1, g_\flat)$, is said to commute if the following diagrams commute $$
\footnotesize \begin{tikzcd}
	A & C \\
	{A'} & {C'}
	\arrow["{\phi_1}", from=1-1, to=1-2]
	\arrow["{f_1}"', from=1-1, to=2-1]
	\arrow["{g_1}", from=1-2, to=2-2]
	\arrow["{\phi'_1}"', from=2-1, to=2-2]
\end{tikzcd} \qquad
\begin{tikzcd}
	{D[\phi_1]} && B \\
	{D'[g_1 \circ \phi_1]} \\
	{D'[\phi_1' \circ f_1]} & {D'[\phi_1']} & {B'}
	\arrow["{\phi^\sharp}", from=1-1, to=1-3]
	\arrow["{g_\flat[\phi_1]}"', from=1-1, to=2-1]
	\arrow["{f_\flat}", from=1-3, to=3-3]
	\arrow["\simeq"{description}, draw=none, from=2-1, to=3-1]
	\arrow[from=3-1, to=3-2]
	\arrow["{{\phi'}^\sharp}"', from=3-2, to=3-3]
\end{tikzcd}
$$ Using these definitions, we can straightforwardly define a jump structure as follows:

\begin{definition}
A \emph{jump structure} of a morphism $\phi : p \to s$ on a distributor $\nabla : p \tri q \to r \tri s$ consists of homotopies witnessing that: \begin{enumerate} \item $\nabla$ factors through $\mathsf{indep}$, i.e. there exists $$\nabla' : p \triangleleft q \to r \otimes s$$ such that $\nabla = \mathsf{indep}_{r,s} \circ \nabla'$ \item The following square commutes $$
\begin{tikzcd}
	{p \triangleleft q} & {r \otimes s} \\
	p & s
	\arrow["{\nabla'}", from=1-1, to=1-2]
	\arrow["{\pi_1}"', "\shortmid"{marking}, from=1-1, to=2-1]
	\arrow["{\pi_2}", "\shortmid"{marking}, from=1-2, to=2-2]
	\arrow["\phi"', from=2-1, to=2-2]
\end{tikzcd}
$$ \end{enumerate}
\end{definition} We define a \emph{Cartesian} jump structure to be a jump structure whose corresponding morphism $\phi \upuparrows r \to s$ is Cartesian. It then follows, by Theorem 4.2 that, for $u$ a polynomial universe, the (mere) existence of a Cartesian jump structure on a distributive law $\aleph : u \triangleleft u \to u \triangleleft u$ is equivalent to $u$ being closed under $\Pi$ types, in that there exists a Cartesian morphism $u \upuparrows u \to u$.

\subsection{Induced Diagrams}

Under the translation between morphisms $\phi \upuparrows q \to r$ and distributors $p \triangleleft q \to r \triangleleft s$, for a (Cartesian) morphism $\pi : u \upuparrows u \to u$, the following diagrams correspond to those for a distributive law.

\noindent For DL1 and DL2, respectively, the corresponding diagrams are: $$
\begin{tikzcd}
	{u \upuparrows (u \triangleleft u)} & {(u \upuparrows u) \triangleleft (u \upuparrows u)} & {u \triangleleft u} \\
	{u \upuparrows u} && u
	\arrow["\delta", from=1-1, to=1-2]
	\arrow["{u \upuparrows \mu}"', from=1-1, to=2-1]
	\arrow["{\pi \triangleleft \pi}", from=1-2, to=1-3]
	\arrow["\mu", from=1-3, to=2-3]
	\arrow["\pi"', from=2-1, to=2-3]
\end{tikzcd} \quad \text{and} \quad
\begin{tikzcd}
	{\mu \upuparrows u} & {(u \triangleleft u) \upuparrows u} & {u\upuparrows (u \upuparrows u)} \\
	&& {u \upuparrows u} \\
	{u \upuparrows u} && u
	\arrow[from=1-1, to=1-2]
	\arrow[from=1-1, to=3-1]
	\arrow["\simeq"{description}, draw=none, from=1-2, to=1-3]
	\arrow["{u \upuparrows \pi}", from=1-3, to=2-3]
	\arrow["\pi", from=2-3, to=3-3]
	\arrow["\pi"', from=3-1, to=3-3]
\end{tikzcd}
$$ (where the unlabeled morphisms arise from the structure of $\upuparrows$ as an action of the twisted arrow category on $\mathbf{Poly}$). For DL3 and DL4, the corresponding diagrams are: $$
\begin{tikzcd}
	{u \upuparrows \text{id}_1} & {\text{id}_1} \\
	{u \upuparrows u} & u
	\arrow["\epsilon", from=1-1, to=1-2]
	\arrow["{u \upuparrows \eta}"', from=1-1, to=2-1]
	\arrow["\eta", from=1-2, to=2-2]
	\arrow["\pi"', from=2-1, to=2-2]
\end{tikzcd}
\quad \text{and} \quad 
\begin{tikzcd}
	{\eta \upuparrows u} & {u \upuparrows u} \\
	{\text{id}_1 \upuparrows u} & u
	\arrow[from=1-1, to=1-2]
	\arrow[from=1-1, to=2-1]
	\arrow["\pi"{description}, from=1-2, to=2-2]
	\arrow["\simeq"{marking, allow upside down}, draw=none, from=2-1, to=2-2]
\end{tikzcd}
$$ (where, as for DL2, the unlabeled morphisms arise from the structure of $\upuparrows$ as an action of the twisted arrow category on $\mathbf{Poly}$). We note that the morphisms identified by these diagrams terminate in $u$ and are Cartesian, assuming $\eta, \mu, \pi$ are Cartesian. Hence, if $u$ is univalent, these diagrams must commute.

\section{Examples of Polynomial Universes}

\subsection{Proof-Relevant Partiality Monads}

For a univalent universe $\mathcal{U}$, the corresponding polynomial functor looks like $$
X \mapsto \sum_{A : \mathcal{U}} X^A
$$ If we specialize this to the case where $\mathcal{U} = \mathbf{Prop}$, the type of propositions, this gives $$
X \mapsto \sum_{\phi : \mathbf{Prop}} X^\phi
$$ This monad is well-known in type theory by another name -- the \emph{partiality} monad. Specifically, this is the monad $M$ whose Kleisli morphisms $A \to M(B)$ correspond to partial functions $A \rightharpoonup B$, i.e. functions that associate to each element $a : A$ a proposition $\mathsf{def}_f(a)$ stating whether $f$ is defined at input $a$, such that if $\mathsf{def}_f(a)$ is true, then one can obtain a value $f(a) : B$.

It follows that one can more generally consider the polynomial monads derived from polynomial universes as \emph{proof-relevant partiality monads}.

\subsection{Rezk Completion, Lists \& Finite Sets}

Additionally, we can show that \emph{any} polynomial functor $p$ can be quotiented to a corresponding univalent polynomial, using a familiar construct
from the theory of categories in HoTT -- the \emph{Rezk Completion.} \cite{ahrens2015univalent}

We obtain the Rezk completion of $p$ as the image factorization in $\mathbf{Poly^{Cart}}$ of the classifying Cartesian morphism $p \to u_i$ from $p$ to a univalent polynomial $u_i$\footnote{where the existence of such a classifying map follows from the assumption previously mentioned in \S\ref{sec:polyuniv} that the ambient type theory has univalent families, which is equivalent to the existence of a \emph{terminal family} in $\mathbf{Poly^{Cart}}$, i.e. a family of polynomials $\{u_i\}_{i \in I}$ such that every other polynomial $p$ has a unique Cartesian morphism to some $u_i$.} $$
p \twoheadrightarrow \mathsf{Rezk}(p) \rightarrowtail u_i
$$ Then, since $\mathsf{Rezk}(p)$ is a subobject of a subterminal object in $\mathbf{Poly^{Cart}}$, it follows that it is itself also subterminal in $\mathbf{Poly^{Cart}}$.

\vspace{0.5em}

\begin{example}
The polynomial functor determined by the function $\left( (m,n) \mapsto n \right) : \{ m < n \in \mathbb{N} \} \to \mathbb{N}$ is $$
X \mapsto \sum_{n \in \mathbb{N}} X^{\{m \in \mathbb{N} \mid m < n \}} \cong \mathsf{List}(X)
$$ This polynomial isn't univalent, because $\mathbb{N}$ is a set (i.e. there is at most one path/identity between any two elements of $\mathbb{N}$), whereas the types $\{m \in \mathbb{N} \mid m < n\}$ form a groupoid (in general, there are $n!$ automorphisms of the path $\{m \in \mathbb{N} \mid m < n\}$, corresponding to permutations of finite sets). However, we can upgrade this to a univalent polynomial using the Rezk completion. If we write out an explicit description of $\mathsf{Rezk}(\mathsf{List})$, we see that it
is the subuniverse of types $X$ that are merely equivalent to
some finite type $\{m \in \mathbb{N} \mid m < n\}$. In constructive mathematics, these types (they are necessarily sets) are known as \emph{Bishop finite sets}. Hence the Rezk completion of the list monad is precisely the subuniverse of types spanned by (Bishop) finite sets.

Both $\mathsf{List}$ and $\mathsf{Rezk}(\mathsf{List})$ are Cartesian monads, hence closed under unit and $\Sigma$. However, although $\mathsf{List}$ does not have a self-distributive law corresponding to $\Pi$ types, $\mathsf{Rezk}(\mathsf{List})$ \emph{does} by Theorem 4.2, since finite sets are closed under finite products. Moreover, although $\mathsf{List}$ doesn't \emph{quite} have such a distributive law, it \emph{almost} does, where the pertinent morphism $\mathsf{List} \triangleleft \mathsf{List} \to \mathsf{List} \triangleleft \mathsf{List} \in \mathbf{Poly}$ is the ``Cartesian Product'' operation on lists that maps a list-of-lists $$
xss = [[x_{11}, \dots, x_{1i_1}], \dots, [x_{j1}, \dots, x_{ji_j}]]
$$ to the list of all $j$-element lists consisting of one element from each list in $xss$, ordered lexicographically by their occurrence in $xss$. This operation fails to satisfy some of the equational laws of a distributive law, namely DL1 \cite{zwart2022no}. However, passing to the Rezk completion of $\mathsf{List}$ essentially \emph{forces} this operation to satisfy these equations -- up to homotopy. It is therefore an interesting question, although beyond the scope of this paper, to consider what structure on a base (non-univalent) polynomial suffices to guarantee that its Rezk completion will be closed under $\Sigma$ types / $\Pi$ types.

Additionally, the fact that $\mathsf{Rezk}(\mathsf{List})$ posesses such a distributive law over itself has intriguing consequences for higher-categorical algebra. Since identities between elements of $\mathsf{Rezk}(\mathsf{List})$ correspond to permutations of finite sets, it follows that an \emph{algebra} for $\mathsf{Rezk}(\mathsf{List})$ should be a type equipped with an associative and unital operation for combining any finite number of elements of that type, that is invariant under all pemutations of finite tuples of elements of the type, i.e. (up to coherent homotopy) a commutative monoid. Hence $\mathsf{Rezk}(\mathsf{List})$ may be regarded as a higher-categorical analogue of the free commutative monoid monad. Moreover, this monad is both polynomial and Cartesian, unlike the ordinary free commutative monoid monad on the category of sets.

More strikingly still, because $\mathsf{Rezk}(\mathsf{List})$ possesses a distributive law over itself, $\mathsf{Rezk}(\mathsf{List}) \triangleleft \mathsf{Rezk}(\mathsf{List})$ is \emph{also} a monad, and we conjecture that this monad forms a higher-categorical analogue of the free commutative ring monad. Note that unlike $\mathsf{Rezk}(\mathsf{List})$, this monad is not Cartesian, although it is (by construction) still polynomial, again in contrast to the ordinary free commutative ring monad on the category of sets. We leave a further consideration of this monad and its algebras, hopefully along with a proof of this conjecture, to future work.
\end{example}
\newpage
\bibliographystyle{./entics}

\appendix
\section{Homotopy Type Theory in Agda}

The following Agda module is used to define key definitions from HoTT that will be used by subsequent modules in formalizing the results of this paper.

\begin{Shaded}
\begin{Highlighting}[]
\PreprocessorTok{\{{-}\# OPTIONS {-}{-}without{-}K {-}{-}rewriting \#{-}\}}
\KeywordTok{module}\NormalTok{ hott }\KeywordTok{where}

\KeywordTok{open} \KeywordTok{import}\NormalTok{ Agda}\OtherTok{.}\NormalTok{Primitive}
\KeywordTok{open} \KeywordTok{import}\NormalTok{ Agda}\OtherTok{.}\NormalTok{Builtin}\OtherTok{.}\NormalTok{Sigma}
\KeywordTok{open} \KeywordTok{import}\NormalTok{ Agda}\OtherTok{.}\NormalTok{Builtin}\OtherTok{.}\NormalTok{Unit}

\NormalTok{Type }\OtherTok{:} \OtherTok{(}\NormalTok{ℓ }\OtherTok{:}\NormalTok{ Level}\OtherTok{)} \OtherTok{→} \DataTypeTok{Set} \OtherTok{(}\NormalTok{lsuc ℓ}\OtherTok{)}
\NormalTok{Type ℓ }\OtherTok{=} \DataTypeTok{Set}\NormalTok{ ℓ}

\OtherTok{\_}\NormalTok{×}\OtherTok{\_} \OtherTok{:} \OtherTok{∀} \OtherTok{\{}\NormalTok{ℓ κ}\OtherTok{\}} \OtherTok{(}\NormalTok{A }\OtherTok{:}\NormalTok{ Type ℓ}\OtherTok{)} \OtherTok{(}\NormalTok{B }\OtherTok{:}\NormalTok{ Type κ}\OtherTok{)} \OtherTok{→}\NormalTok{ Type }\OtherTok{(}\NormalTok{ℓ ⊔ κ}\OtherTok{)}
\NormalTok{A × B }\OtherTok{=}\NormalTok{ Σ A }\OtherTok{(λ} \OtherTok{\_} \OtherTok{→}\NormalTok{ B}\OtherTok{)}
\end{Highlighting}
\end{Shaded}

Basic properties of the identity type, including reflexivity,
transitivity, symmetry, and congruence, and some convenient notation for
equality proofs.

\begin{Shaded}
\begin{Highlighting}[]
\KeywordTok{open} \KeywordTok{import}\NormalTok{ Agda}\OtherTok{.}\NormalTok{Builtin}\OtherTok{.}\NormalTok{Equality}
\KeywordTok{open} \KeywordTok{import}\NormalTok{ Agda}\OtherTok{.}\NormalTok{Builtin}\OtherTok{.}\NormalTok{Equality}\OtherTok{.}\NormalTok{Rewrite}
\end{Highlighting}
\end{Shaded}

\begin{Shaded}
\begin{Highlighting}[]
\OtherTok{\_}\NormalTok{□ }\OtherTok{:} \OtherTok{∀} \OtherTok{\{}\NormalTok{ℓ}\OtherTok{\}} \OtherTok{\{}\NormalTok{A }\OtherTok{:}\NormalTok{ Type ℓ}\OtherTok{\}} \OtherTok{(}\NormalTok{a }\OtherTok{:}\NormalTok{ A}\OtherTok{)} \OtherTok{→}\NormalTok{ a ≡ a}
\NormalTok{a □ }\OtherTok{=}\NormalTok{ refl}
\end{Highlighting}
\end{Shaded}

\begin{Shaded}
\begin{Highlighting}[]
\NormalTok{transp }\OtherTok{:} \OtherTok{∀} \OtherTok{\{}\NormalTok{ℓ κ}\OtherTok{\}} \OtherTok{\{}\NormalTok{A }\OtherTok{:}\NormalTok{ Type ℓ}\OtherTok{\}} \OtherTok{(}\NormalTok{B }\OtherTok{:}\NormalTok{ A }\OtherTok{→}\NormalTok{ Type κ}\OtherTok{)} \OtherTok{\{}\NormalTok{a a\textquotesingle{} }\OtherTok{:}\NormalTok{ A}\OtherTok{\}} 
         \OtherTok{→} \OtherTok{(}\NormalTok{e }\OtherTok{:}\NormalTok{ a ≡ a\textquotesingle{}}\OtherTok{)} \OtherTok{→}\NormalTok{ B a }\OtherTok{→}\NormalTok{ B a\textquotesingle{}}
\NormalTok{transp B refl b }\OtherTok{=}\NormalTok{ b}
\end{Highlighting}
\end{Shaded}

\begin{Shaded}
\begin{Highlighting}[]
\OtherTok{\_}\NormalTok{•}\OtherTok{\_} \OtherTok{:} \OtherTok{∀} \OtherTok{\{}\NormalTok{ℓ}\OtherTok{\}} \OtherTok{\{}\NormalTok{A }\OtherTok{:}\NormalTok{ Type ℓ}\OtherTok{\}} \OtherTok{\{}\NormalTok{a b c }\OtherTok{:}\NormalTok{ A}\OtherTok{\}}
      \OtherTok{→} \OtherTok{(}\NormalTok{a ≡ b}\OtherTok{)} \OtherTok{→} \OtherTok{(}\NormalTok{b ≡ c}\OtherTok{)} \OtherTok{→} \OtherTok{(}\NormalTok{a ≡ c}\OtherTok{)}
\NormalTok{e • refl }\OtherTok{=}\NormalTok{ e}
\end{Highlighting}
\end{Shaded}

\begin{Shaded}
\begin{Highlighting}[]
\OtherTok{\_}\NormalTok{≡〈}\OtherTok{\_}\NormalTok{〉}\OtherTok{\_} \OtherTok{:} \OtherTok{∀} \OtherTok{\{}\NormalTok{ℓ}\OtherTok{\}} \OtherTok{\{}\NormalTok{A }\OtherTok{:}\NormalTok{ Type ℓ}\OtherTok{\}} \OtherTok{(}\NormalTok{a }\OtherTok{:}\NormalTok{ A}\OtherTok{)} \OtherTok{\{}\NormalTok{b c }\OtherTok{:}\NormalTok{ A}\OtherTok{\}} 
          \OtherTok{→}\NormalTok{ a ≡ b }\OtherTok{→}\NormalTok{ b ≡ c }\OtherTok{→}\NormalTok{ a ≡ c}
\NormalTok{a ≡〈 e 〉 refl }\OtherTok{=}\NormalTok{ e}

\NormalTok{comprewrite }\OtherTok{:} \OtherTok{∀} \OtherTok{\{}\NormalTok{ℓ}\OtherTok{\}} \OtherTok{\{}\NormalTok{A }\OtherTok{:}\NormalTok{ Type ℓ}\OtherTok{\}} \OtherTok{\{}\NormalTok{a b c }\OtherTok{:}\NormalTok{ A}\OtherTok{\}}
              \OtherTok{→} \OtherTok{(}\NormalTok{e1 }\OtherTok{:}\NormalTok{ a ≡ b}\OtherTok{)} \OtherTok{(}\NormalTok{e2 }\OtherTok{:}\NormalTok{ b ≡ c}\OtherTok{)}
              \OtherTok{→} \OtherTok{(}\NormalTok{a ≡〈 e1 〉 e2}\OtherTok{)}\NormalTok{ ≡ }\OtherTok{(}\NormalTok{e1 • e2}\OtherTok{)}
\NormalTok{comprewrite refl refl }\OtherTok{=}\NormalTok{ refl}

\PreprocessorTok{\{{-}\# REWRITE comprewrite \#{-}\}}
\end{Highlighting}
\end{Shaded}

\begin{Shaded}
\begin{Highlighting}[]
\NormalTok{sym }\OtherTok{:} \OtherTok{∀} \OtherTok{\{}\NormalTok{ℓ}\OtherTok{\}} \OtherTok{\{}\NormalTok{A }\OtherTok{:}\NormalTok{ Type ℓ}\OtherTok{\}} \OtherTok{\{}\NormalTok{a a\textquotesingle{} }\OtherTok{:}\NormalTok{ A}\OtherTok{\}} \OtherTok{→}\NormalTok{ a ≡ a\textquotesingle{} }\OtherTok{→}\NormalTok{ a\textquotesingle{} ≡ a}
\NormalTok{sym refl }\OtherTok{=}\NormalTok{ refl}
\end{Highlighting}
\end{Shaded}

\begin{Shaded}
\begin{Highlighting}[]
\NormalTok{ap }\OtherTok{:} \OtherTok{∀} \OtherTok{\{}\NormalTok{ℓ κ}\OtherTok{\}} \OtherTok{\{}\NormalTok{A }\OtherTok{:}\NormalTok{ Type ℓ}\OtherTok{\}} \OtherTok{\{}\NormalTok{B }\OtherTok{:}\NormalTok{ Type κ}\OtherTok{\}} \OtherTok{\{}\NormalTok{a a\textquotesingle{} }\OtherTok{:}\NormalTok{ A}\OtherTok{\}}
     \OtherTok{→} \OtherTok{(}\NormalTok{f }\OtherTok{:}\NormalTok{ A }\OtherTok{→}\NormalTok{ B}\OtherTok{)} \OtherTok{→}\NormalTok{ a ≡ a\textquotesingle{} }\OtherTok{→} \OtherTok{(}\NormalTok{f a}\OtherTok{)}\NormalTok{ ≡ }\OtherTok{(}\NormalTok{f a\textquotesingle{}}\OtherTok{)}
\NormalTok{ap f refl }\OtherTok{=}\NormalTok{ refl}
\end{Highlighting}
\end{Shaded}

\begin{Shaded}
\begin{Highlighting}[]
\NormalTok{coAp }\OtherTok{:} \OtherTok{∀} \OtherTok{\{}\NormalTok{ℓ κ}\OtherTok{\}} \OtherTok{\{}\NormalTok{A }\OtherTok{:}\NormalTok{ Type ℓ}\OtherTok{\}} \OtherTok{\{}\NormalTok{B }\OtherTok{:}\NormalTok{ A }\OtherTok{→}\NormalTok{ Type κ}\OtherTok{\}} \OtherTok{\{}\NormalTok{f g }\OtherTok{:} \OtherTok{(}\NormalTok{x }\OtherTok{:}\NormalTok{ A}\OtherTok{)} \OtherTok{→}\NormalTok{ B x}\OtherTok{\}}
       \OtherTok{→}\NormalTok{ f ≡ g }\OtherTok{→} \OtherTok{(}\NormalTok{x }\OtherTok{:}\NormalTok{ A}\OtherTok{)} \OtherTok{→}\NormalTok{ f x ≡ g x}
\NormalTok{coAp refl x }\OtherTok{=}\NormalTok{ refl}
\end{Highlighting}
\end{Shaded}

\begin{Shaded}
\begin{Highlighting}[]
\NormalTok{apd }\OtherTok{:} \OtherTok{∀} \OtherTok{\{}\NormalTok{ℓ0 ℓ1 κ}\OtherTok{\}} \OtherTok{\{}\NormalTok{A }\OtherTok{:}\NormalTok{ Type ℓ0}\OtherTok{\}} \OtherTok{\{}\NormalTok{B }\OtherTok{:}\NormalTok{ Type ℓ1}\OtherTok{\}} \OtherTok{\{}\NormalTok{f }\OtherTok{:}\NormalTok{ A }\OtherTok{→}\NormalTok{ B}\OtherTok{\}}
      \OtherTok{→} \OtherTok{(}\NormalTok{C }\OtherTok{:}\NormalTok{ B }\OtherTok{→}\NormalTok{ Type κ}\OtherTok{)} \OtherTok{\{}\NormalTok{a a\textquotesingle{} }\OtherTok{:}\NormalTok{ A}\OtherTok{\}}
      \OtherTok{→} \OtherTok{(}\NormalTok{g }\OtherTok{:} \OtherTok{(}\NormalTok{x }\OtherTok{:}\NormalTok{ A}\OtherTok{)} \OtherTok{→}\NormalTok{ C }\OtherTok{(}\NormalTok{f x}\OtherTok{))} \OtherTok{→} \OtherTok{(}\NormalTok{e }\OtherTok{:}\NormalTok{ a ≡ a\textquotesingle{}}\OtherTok{)} \OtherTok{→}\NormalTok{ transp C }\OtherTok{(}\NormalTok{ap f e}\OtherTok{)} \OtherTok{(}\NormalTok{g a}\OtherTok{)}\NormalTok{ ≡ g a\textquotesingle{}}
\NormalTok{apd B f refl }\OtherTok{=}\NormalTok{ refl}
\end{Highlighting}
\end{Shaded}

Equality for pairs:

\begin{Shaded}
\begin{Highlighting}[]
\KeywordTok{module}\NormalTok{ PairEq }\OtherTok{\{}\NormalTok{ℓ κ}\OtherTok{\}} \OtherTok{\{}\NormalTok{A }\OtherTok{:}\NormalTok{ Type ℓ}\OtherTok{\}} \OtherTok{\{}\NormalTok{B }\OtherTok{:}\NormalTok{ A }\OtherTok{→}\NormalTok{ Type κ}\OtherTok{\}} 
              \OtherTok{\{}\NormalTok{a a\textquotesingle{} }\OtherTok{:}\NormalTok{ A}\OtherTok{\}} \OtherTok{\{}\NormalTok{b }\OtherTok{:}\NormalTok{ B a}\OtherTok{\}} \OtherTok{\{}\NormalTok{b\textquotesingle{} }\OtherTok{:}\NormalTok{ B a\textquotesingle{}}\OtherTok{\}} \KeywordTok{where}

\NormalTok{    pairEq }\OtherTok{:} \OtherTok{(}\NormalTok{e }\OtherTok{:}\NormalTok{ a ≡ a\textquotesingle{}}\OtherTok{)} \OtherTok{(}\NormalTok{e\textquotesingle{} }\OtherTok{:}\NormalTok{ transp B e b ≡ b\textquotesingle{}}\OtherTok{)} \OtherTok{→} \OtherTok{(}\NormalTok{a , b}\OtherTok{)}\NormalTok{ ≡ }\OtherTok{(}\NormalTok{a\textquotesingle{} , b\textquotesingle{}}\OtherTok{)}
\NormalTok{    pairEq refl refl }\OtherTok{=}\NormalTok{ refl}
\end{Highlighting}
\end{Shaded}

\begin{Shaded}
\begin{Highlighting}[]
\NormalTok{    pairEqβ1 }\OtherTok{:} \OtherTok{(}\NormalTok{e }\OtherTok{:}\NormalTok{ a ≡ a\textquotesingle{}}\OtherTok{)} \OtherTok{(}\NormalTok{e\textquotesingle{} }\OtherTok{:}\NormalTok{ transp B e b ≡ b\textquotesingle{}}\OtherTok{)} \OtherTok{→}\NormalTok{ ap fst }\OtherTok{(}\NormalTok{pairEq e e\textquotesingle{}}\OtherTok{)}\NormalTok{ ≡ e}
\NormalTok{    pairEqβ1 refl refl }\OtherTok{=}\NormalTok{ refl}

\NormalTok{    pairEqη }\OtherTok{:} \OtherTok{(}\NormalTok{e }\OtherTok{:} \OtherTok{(}\NormalTok{a , b}\OtherTok{)}\NormalTok{ ≡ }\OtherTok{(}\NormalTok{a\textquotesingle{} , b\textquotesingle{}}\OtherTok{))} \OtherTok{→}\NormalTok{ pairEq }\OtherTok{(}\NormalTok{ap fst e}\OtherTok{)} \OtherTok{(}\NormalTok{apd B snd e}\OtherTok{)}\NormalTok{ ≡ e}
\NormalTok{    pairEqη refl }\OtherTok{=}\NormalTok{ refl}

\KeywordTok{open}\NormalTok{ PairEq }\KeywordTok{public}
\end{Highlighting}
\end{Shaded}

The bi-invertible maps definition of equivalence, and closure of
equivalences under identity and composition:

\begin{Shaded}
\begin{Highlighting}[]
\NormalTok{isEquiv }\OtherTok{:} \OtherTok{∀} \OtherTok{\{}\NormalTok{ℓ κ}\OtherTok{\}} \OtherTok{\{}\NormalTok{A }\OtherTok{:}\NormalTok{ Type ℓ}\OtherTok{\}} \OtherTok{\{}\NormalTok{B }\OtherTok{:}\NormalTok{ Type κ}\OtherTok{\}} \OtherTok{→} \OtherTok{(}\NormalTok{A }\OtherTok{→}\NormalTok{ B}\OtherTok{)} \OtherTok{→}\NormalTok{ Type }\OtherTok{(}\NormalTok{ℓ ⊔ κ}\OtherTok{)}
\NormalTok{isEquiv }\OtherTok{\{}\NormalTok{A }\OtherTok{=}\NormalTok{ A}\OtherTok{\}} \OtherTok{\{}\NormalTok{B }\OtherTok{=}\NormalTok{ B}\OtherTok{\}}\NormalTok{ f }\OtherTok{=} 
      \OtherTok{(}\NormalTok{Σ }\OtherTok{(}\NormalTok{B }\OtherTok{→}\NormalTok{ A}\OtherTok{)} \OtherTok{(λ}\NormalTok{ g }\OtherTok{→} \OtherTok{(}\NormalTok{a }\OtherTok{:}\NormalTok{ A}\OtherTok{)} \OtherTok{→}\NormalTok{ g }\OtherTok{(}\NormalTok{f a}\OtherTok{)}\NormalTok{ ≡ a}\OtherTok{))} 
\NormalTok{    × }\OtherTok{(}\NormalTok{Σ }\OtherTok{(}\NormalTok{B }\OtherTok{→}\NormalTok{ A}\OtherTok{)} \OtherTok{(λ}\NormalTok{ h }\OtherTok{→} \OtherTok{(}\NormalTok{b }\OtherTok{:}\NormalTok{ B}\OtherTok{)} \OtherTok{→}\NormalTok{ f }\OtherTok{(}\NormalTok{h b}\OtherTok{)}\NormalTok{ ≡ b}\OtherTok{))}
\end{Highlighting}
\end{Shaded}

\begin{Shaded}
\begin{Highlighting}[]
\NormalTok{idIsEquiv }\OtherTok{:} \OtherTok{∀} \OtherTok{\{}\NormalTok{ℓ}\OtherTok{\}} \OtherTok{\{}\NormalTok{A }\OtherTok{:}\NormalTok{ Type ℓ}\OtherTok{\}} \OtherTok{→}\NormalTok{ isEquiv }\OtherTok{\{}\NormalTok{A }\OtherTok{=}\NormalTok{ A}\OtherTok{\}} \OtherTok{(λ}\NormalTok{ x }\OtherTok{→}\NormalTok{ x}\OtherTok{)}
\NormalTok{idIsEquiv }\OtherTok{=} \OtherTok{((λ}\NormalTok{ x }\OtherTok{→}\NormalTok{ x}\OtherTok{)}\NormalTok{ , }\OtherTok{(λ}\NormalTok{ x }\OtherTok{→}\NormalTok{ refl}\OtherTok{))}\NormalTok{ , }\OtherTok{((λ}\NormalTok{ x }\OtherTok{→}\NormalTok{ x}\OtherTok{)}\NormalTok{ , }\OtherTok{(λ}\NormalTok{ x }\OtherTok{→}\NormalTok{ refl}\OtherTok{))}

\NormalTok{compIsEquiv }\OtherTok{:} \OtherTok{∀} \OtherTok{\{}\NormalTok{ℓ0 ℓ1 ℓ2}\OtherTok{\}} \OtherTok{\{}\NormalTok{A }\OtherTok{:}\NormalTok{ Type ℓ0}\OtherTok{\}} \OtherTok{\{}\NormalTok{B }\OtherTok{:}\NormalTok{ Type ℓ1}\OtherTok{\}} \OtherTok{\{}\NormalTok{C }\OtherTok{:}\NormalTok{ Type ℓ2}\OtherTok{\}}
              \OtherTok{→} \OtherTok{\{}\NormalTok{g }\OtherTok{:}\NormalTok{ B }\OtherTok{→}\NormalTok{ C}\OtherTok{\}} \OtherTok{\{}\NormalTok{f }\OtherTok{:}\NormalTok{ A }\OtherTok{→}\NormalTok{ B}\OtherTok{\}} \OtherTok{→}\NormalTok{ isEquiv g }\OtherTok{→}\NormalTok{ isEquiv f}
              \OtherTok{→}\NormalTok{ isEquiv }\OtherTok{(λ}\NormalTok{ a }\OtherTok{→}\NormalTok{ g }\OtherTok{(}\NormalTok{f a}\OtherTok{))}
\NormalTok{compIsEquiv }\OtherTok{\{}\NormalTok{g }\OtherTok{=}\NormalTok{ g}\OtherTok{\}} \OtherTok{\{}\NormalTok{f }\OtherTok{=}\NormalTok{ f}\OtherTok{\}} 
            \OtherTok{((}\NormalTok{g\textquotesingle{} , lg}\OtherTok{)}\NormalTok{ , }\OtherTok{(}\NormalTok{g\textquotesingle{}\textquotesingle{} , rg}\OtherTok{))} 
            \OtherTok{((}\NormalTok{f\textquotesingle{} , lf}\OtherTok{)}\NormalTok{ , }\OtherTok{(}\NormalTok{f\textquotesingle{}\textquotesingle{} , rf}\OtherTok{))} \OtherTok{=}
      \OtherTok{(} \OtherTok{(λ}\NormalTok{ c }\OtherTok{→}\NormalTok{ f\textquotesingle{} }\OtherTok{(}\NormalTok{g\textquotesingle{} c}\OtherTok{))}   
\NormalTok{      , }\OtherTok{λ}\NormalTok{ a }\OtherTok{→} \OtherTok{(}\NormalTok{f\textquotesingle{} }\OtherTok{(}\NormalTok{g\textquotesingle{} }\OtherTok{(}\NormalTok{g }\OtherTok{(}\NormalTok{f a}\OtherTok{))))}\NormalTok{   ≡〈 ap f\textquotesingle{} }\OtherTok{(}\NormalTok{lg }\OtherTok{(}\NormalTok{f a}\OtherTok{))}\NormalTok{ 〉 }
              \OtherTok{(}\NormalTok{f\textquotesingle{} }\OtherTok{(}\NormalTok{f a}\OtherTok{)}\NormalTok{             ≡〈 lf a 〉 }
              \OtherTok{(}\NormalTok{a                    □}\OtherTok{)))} 
\NormalTok{    , }\OtherTok{((λ}\NormalTok{ c }\OtherTok{→}\NormalTok{ f\textquotesingle{}\textquotesingle{} }\OtherTok{(}\NormalTok{g\textquotesingle{}\textquotesingle{} c}\OtherTok{))} 
\NormalTok{      , }\OtherTok{λ}\NormalTok{ c }\OtherTok{→} \OtherTok{(}\NormalTok{g }\OtherTok{(}\NormalTok{f }\OtherTok{(}\NormalTok{f\textquotesingle{}\textquotesingle{} }\OtherTok{(}\NormalTok{g\textquotesingle{}\textquotesingle{} c}\OtherTok{))))}\NormalTok{ ≡〈 ap g  }\OtherTok{(}\NormalTok{rf }\OtherTok{(}\NormalTok{g\textquotesingle{}\textquotesingle{} c}\OtherTok{))}\NormalTok{ 〉 }
              \OtherTok{(}\NormalTok{g }\OtherTok{(}\NormalTok{g\textquotesingle{}\textquotesingle{} c}\OtherTok{)}\NormalTok{            ≡〈 rg c 〉}
              \OtherTok{(}\NormalTok{c                    □}\OtherTok{)))}
\end{Highlighting}
\end{Shaded}

Isomorphisms and equi-inhabitation of isomorphism and equivalence:

\begin{Shaded}
\begin{Highlighting}[]
\NormalTok{Iso }\OtherTok{:} \OtherTok{∀} \OtherTok{\{}\NormalTok{ℓ κ}\OtherTok{\}} \OtherTok{\{}\NormalTok{A }\OtherTok{:}\NormalTok{ Type ℓ}\OtherTok{\}} \OtherTok{\{}\NormalTok{B }\OtherTok{:}\NormalTok{ Type κ}\OtherTok{\}} \OtherTok{→} \OtherTok{(}\NormalTok{A }\OtherTok{→}\NormalTok{ B}\OtherTok{)} \OtherTok{→}\NormalTok{ Type }\OtherTok{(}\NormalTok{ℓ ⊔ κ}\OtherTok{)}
\NormalTok{Iso }\OtherTok{\{}\NormalTok{A }\OtherTok{=}\NormalTok{ A}\OtherTok{\}} \OtherTok{\{}\NormalTok{B }\OtherTok{=}\NormalTok{ B}\OtherTok{\}}\NormalTok{ f }\OtherTok{=} 
    \OtherTok{(}\NormalTok{Σ }\OtherTok{(}\NormalTok{B }\OtherTok{→}\NormalTok{ A}\OtherTok{)} \OtherTok{(λ}\NormalTok{ g }\OtherTok{→} \OtherTok{((}\NormalTok{a }\OtherTok{:}\NormalTok{ A}\OtherTok{)} \OtherTok{→}\NormalTok{ g }\OtherTok{(}\NormalTok{f a}\OtherTok{)}\NormalTok{ ≡ a}\OtherTok{)} 
\NormalTok{                    × }\OtherTok{((}\NormalTok{b }\OtherTok{:}\NormalTok{ B}\OtherTok{)} \OtherTok{→}\NormalTok{ f }\OtherTok{(}\NormalTok{g b}\OtherTok{)}\NormalTok{ ≡ b}\OtherTok{)))}
\end{Highlighting}
\end{Shaded}

\begin{Shaded}
\begin{Highlighting}[]
\KeywordTok{module}\NormalTok{ Iso↔Equiv }\OtherTok{\{}\NormalTok{ℓ κ}\OtherTok{\}} \OtherTok{\{}\NormalTok{A }\OtherTok{:}\NormalTok{ Type ℓ}\OtherTok{\}} \OtherTok{\{}\NormalTok{B }\OtherTok{:}\NormalTok{ Type κ}\OtherTok{\}} \OtherTok{\{}\NormalTok{f }\OtherTok{:}\NormalTok{ A }\OtherTok{→}\NormalTok{ B}\OtherTok{\}} \KeywordTok{where}

\NormalTok{    Iso→isEquiv }\OtherTok{:}\NormalTok{ Iso f }\OtherTok{→}\NormalTok{ isEquiv f}
\NormalTok{    Iso→isEquiv }\OtherTok{(}\NormalTok{g , l , r}\OtherTok{)} \OtherTok{=} \OtherTok{((}\NormalTok{g , l}\OtherTok{)}\NormalTok{ , }\OtherTok{(}\NormalTok{g , r}\OtherTok{))}

\NormalTok{    isEquiv→Iso }\OtherTok{:}\NormalTok{ isEquiv f }\OtherTok{→}\NormalTok{ Iso f}
\NormalTok{    isEquiv→Iso }\OtherTok{((}\NormalTok{g , l}\OtherTok{)}\NormalTok{ , }\OtherTok{(}\NormalTok{h , r}\OtherTok{))} \OtherTok{=} 
\NormalTok{        h , }\OtherTok{(λ}\NormalTok{ x }\OtherTok{→} \OtherTok{(}\NormalTok{h }\OtherTok{(}\NormalTok{f x}\OtherTok{))}\NormalTok{        ≡〈 sym }\OtherTok{(}\NormalTok{l }\OtherTok{(}\NormalTok{h }\OtherTok{(}\NormalTok{f x}\OtherTok{)))}\NormalTok{ 〉 }
                   \OtherTok{(}\NormalTok{g }\OtherTok{(}\NormalTok{f }\OtherTok{(}\NormalTok{h }\OtherTok{(}\NormalTok{f x}\OtherTok{)))}\NormalTok{ ≡〈 ap g }\OtherTok{(}\NormalTok{r }\OtherTok{(}\NormalTok{f x}\OtherTok{))}\NormalTok{ 〉}
                   \OtherTok{((}\NormalTok{g }\OtherTok{(}\NormalTok{f x}\OtherTok{))}\NormalTok{       ≡〈 l x 〉 }
                   \OtherTok{(}\NormalTok{x □}\OtherTok{))))}\NormalTok{ , r}

\KeywordTok{open}\NormalTok{ Iso↔Equiv }\KeywordTok{public}
\end{Highlighting}
\end{Shaded}

Proof that the inverse of an equivalence is an equivalence:

\begin{Shaded}
\begin{Highlighting}[]
\KeywordTok{module}\NormalTok{ InvEquiv }\OtherTok{\{}\NormalTok{ℓ κ}\OtherTok{\}} \OtherTok{\{}\NormalTok{A }\OtherTok{:}\NormalTok{ Type ℓ}\OtherTok{\}} \OtherTok{\{}\NormalTok{B }\OtherTok{:}\NormalTok{ Type κ}\OtherTok{\}} \OtherTok{\{}\NormalTok{f }\OtherTok{:}\NormalTok{ A }\OtherTok{→}\NormalTok{ B}\OtherTok{\}} \KeywordTok{where}

\NormalTok{    inv }\OtherTok{:}\NormalTok{ isEquiv f }\OtherTok{→}\NormalTok{ B }\OtherTok{→}\NormalTok{ A}
\NormalTok{    inv }\OtherTok{(\_}\NormalTok{ , }\OtherTok{(}\NormalTok{h , }\OtherTok{\_))} \OtherTok{=}\NormalTok{ h}

\NormalTok{    isoInv }\OtherTok{:} \OtherTok{(}\NormalTok{isof }\OtherTok{:}\NormalTok{ Iso f}\OtherTok{)} \OtherTok{→}\NormalTok{ Iso }\OtherTok{(}\NormalTok{fst isof}\OtherTok{)}
\NormalTok{    isoInv }\OtherTok{(}\NormalTok{g , l , r}\OtherTok{)} \OtherTok{=} \OtherTok{(}\NormalTok{f , r , l}\OtherTok{)}

\NormalTok{    invIsEquiv }\OtherTok{:} \OtherTok{(}\NormalTok{ef }\OtherTok{:}\NormalTok{ isEquiv f}\OtherTok{)} \OtherTok{→}\NormalTok{ isEquiv }\OtherTok{(}\NormalTok{inv ef}\OtherTok{)}
\NormalTok{    invIsEquiv ef }\OtherTok{=}\NormalTok{ Iso→isEquiv }\OtherTok{(}\NormalTok{isoInv }\OtherTok{(}\NormalTok{isEquiv→Iso ef}\OtherTok{))}
    
\KeywordTok{open}\NormalTok{ InvEquiv }\KeywordTok{public}
\end{Highlighting}
\end{Shaded}

Proof that transport along a path is an equivalence:

\begin{Shaded}
\begin{Highlighting}[]
\KeywordTok{module}\NormalTok{ TranspEquiv }\OtherTok{\{}\NormalTok{ℓ κ}\OtherTok{\}} \OtherTok{\{}\NormalTok{A }\OtherTok{:}\NormalTok{ Type ℓ}\OtherTok{\}} \OtherTok{\{}\NormalTok{B }\OtherTok{:}\NormalTok{ A }\OtherTok{→}\NormalTok{ Type κ}\OtherTok{\}} \OtherTok{\{}\NormalTok{a b }\OtherTok{:}\NormalTok{ A}\OtherTok{\}} \OtherTok{(}\NormalTok{e }\OtherTok{:}\NormalTok{ a ≡ b}\OtherTok{)} \KeywordTok{where}

\NormalTok{    syml }\OtherTok{:} \OtherTok{(}\NormalTok{x }\OtherTok{:}\NormalTok{ B a}\OtherTok{)} \OtherTok{→}\NormalTok{ transp B }\OtherTok{(}\NormalTok{sym e}\OtherTok{)} \OtherTok{(}\NormalTok{transp B e x}\OtherTok{)}\NormalTok{ ≡ x}
\NormalTok{    syml x }\KeywordTok{rewrite}\NormalTok{ e }\OtherTok{=}\NormalTok{ refl}

\NormalTok{    symr }\OtherTok{:} \OtherTok{(}\NormalTok{y }\OtherTok{:}\NormalTok{ B b}\OtherTok{)} \OtherTok{→}\NormalTok{ transp B e }\OtherTok{(}\NormalTok{transp B }\OtherTok{(}\NormalTok{sym e}\OtherTok{)}\NormalTok{ y}\OtherTok{)}\NormalTok{ ≡ y}
\NormalTok{    symr y }\KeywordTok{rewrite}\NormalTok{ e }\OtherTok{=}\NormalTok{ refl}

\NormalTok{    transpIsEquiv }\OtherTok{:}\NormalTok{ isEquiv }\OtherTok{\{}\NormalTok{A }\OtherTok{=}\NormalTok{ B a}\OtherTok{\}} \OtherTok{\{}\NormalTok{B }\OtherTok{=}\NormalTok{ B b}\OtherTok{\}} \OtherTok{(λ}\NormalTok{ x }\OtherTok{→}\NormalTok{ transp B e x}\OtherTok{)}
\NormalTok{    transpIsEquiv }\OtherTok{=}\NormalTok{ Iso→isEquiv }\OtherTok{((λ}\NormalTok{ x }\OtherTok{→}\NormalTok{ transp B }\OtherTok{(}\NormalTok{sym e}\OtherTok{)}\NormalTok{ x}\OtherTok{)}\NormalTok{ , }\OtherTok{(}\NormalTok{syml , symr}\OtherTok{))}

\KeywordTok{open}\NormalTok{ TranspEquiv }\KeywordTok{public}
\end{Highlighting}
\end{Shaded}

Some additional facts about the identity type that will be used
throughout this formalization, are as follows:

\begin{Shaded}
\begin{Highlighting}[]
\NormalTok{transpAp }\OtherTok{:} \OtherTok{∀} \OtherTok{\{}\NormalTok{ℓ ℓ\textquotesingle{} κ}\OtherTok{\}} \OtherTok{\{}\NormalTok{A }\OtherTok{:}\NormalTok{ Type ℓ}\OtherTok{\}} \OtherTok{\{}\NormalTok{A\textquotesingle{} }\OtherTok{:}\NormalTok{ Type ℓ\textquotesingle{}}\OtherTok{\}} \OtherTok{\{}\NormalTok{a b }\OtherTok{:}\NormalTok{ A}\OtherTok{\}}
           \OtherTok{→} \OtherTok{(}\NormalTok{B }\OtherTok{:}\NormalTok{ A\textquotesingle{} }\OtherTok{→}\NormalTok{ Type κ}\OtherTok{)} \OtherTok{(}\NormalTok{f }\OtherTok{:}\NormalTok{ A }\OtherTok{→}\NormalTok{ A\textquotesingle{}}\OtherTok{)} \OtherTok{(}\NormalTok{e }\OtherTok{:}\NormalTok{ a ≡ b}\OtherTok{)} \OtherTok{(}\NormalTok{x }\OtherTok{:}\NormalTok{ B }\OtherTok{(}\NormalTok{f a}\OtherTok{))}
           \OtherTok{→}\NormalTok{ transp }\OtherTok{(λ}\NormalTok{ x }\OtherTok{→}\NormalTok{ B }\OtherTok{(}\NormalTok{f x}\OtherTok{))}\NormalTok{ e x ≡ transp B }\OtherTok{(}\NormalTok{ap f e}\OtherTok{)}\NormalTok{ x}
\NormalTok{transpAp B f refl x }\OtherTok{=}\NormalTok{ refl}

\NormalTok{•invr }\OtherTok{:} \OtherTok{∀} \OtherTok{\{}\NormalTok{ℓ}\OtherTok{\}} \OtherTok{\{}\NormalTok{A }\OtherTok{:}\NormalTok{ Type ℓ}\OtherTok{\}} \OtherTok{\{}\NormalTok{a b }\OtherTok{:}\NormalTok{ A}\OtherTok{\}}
        \OtherTok{→} \OtherTok{(}\NormalTok{e }\OtherTok{:}\NormalTok{ a ≡ b}\OtherTok{)} \OtherTok{→} \OtherTok{(}\NormalTok{sym e}\OtherTok{)}\NormalTok{ • e ≡ refl}
\NormalTok{•invr refl }\OtherTok{=}\NormalTok{ refl}

\NormalTok{≡siml }\OtherTok{:} \OtherTok{∀} \OtherTok{\{}\NormalTok{ℓ}\OtherTok{\}} \OtherTok{\{}\NormalTok{A }\OtherTok{:}\NormalTok{ Type ℓ}\OtherTok{\}} \OtherTok{\{}\NormalTok{a b }\OtherTok{:}\NormalTok{ A}\OtherTok{\}}
        \OtherTok{→} \OtherTok{(}\NormalTok{e }\OtherTok{:}\NormalTok{ a ≡ b}\OtherTok{)} \OtherTok{→}\NormalTok{ refl ≡ }\OtherTok{(}\NormalTok{b ≡〈 sym e 〉 e}\OtherTok{)}
\NormalTok{≡siml refl }\OtherTok{=}\NormalTok{ refl}

\NormalTok{≡idr }\OtherTok{:} \OtherTok{∀} \OtherTok{\{}\NormalTok{ℓ}\OtherTok{\}} \OtherTok{\{}\NormalTok{A }\OtherTok{:}\NormalTok{ Type ℓ}\OtherTok{\}} \OtherTok{\{}\NormalTok{a b }\OtherTok{:}\NormalTok{ A}\OtherTok{\}}
       \OtherTok{→} \OtherTok{(}\NormalTok{e }\OtherTok{:}\NormalTok{ a ≡ b}\OtherTok{)} \OtherTok{→}\NormalTok{ e ≡ }\OtherTok{(}\NormalTok{a ≡〈 refl 〉 e}\OtherTok{)}
\NormalTok{≡idr refl }\OtherTok{=}\NormalTok{ refl}

\NormalTok{conj }\OtherTok{:} \OtherTok{∀} \OtherTok{\{}\NormalTok{ℓ}\OtherTok{\}} \OtherTok{\{}\NormalTok{A }\OtherTok{:}\NormalTok{ Type ℓ}\OtherTok{\}} \OtherTok{\{}\NormalTok{a b c d }\OtherTok{:}\NormalTok{ A}\OtherTok{\}}
       \OtherTok{→} \OtherTok{(}\NormalTok{e1 }\OtherTok{:}\NormalTok{ a ≡ b}\OtherTok{)} \OtherTok{(}\NormalTok{e2 }\OtherTok{:}\NormalTok{ a ≡ c}\OtherTok{)} \OtherTok{(}\NormalTok{e3 }\OtherTok{:}\NormalTok{ b ≡ d}\OtherTok{)} \OtherTok{(}\NormalTok{e4 }\OtherTok{:}\NormalTok{ c ≡ d}\OtherTok{)}
       \OtherTok{→} \OtherTok{(}\NormalTok{a ≡〈 e1 〉 e3}\OtherTok{)}\NormalTok{ ≡ }\OtherTok{(}\NormalTok{a ≡〈 e2 〉 e4}\OtherTok{)}
       \OtherTok{→}\NormalTok{ e3 ≡ }\OtherTok{(}\NormalTok{b ≡〈 sym e1 〉}\OtherTok{(}\NormalTok{a ≡〈 e2 〉 e4}\OtherTok{))}
\NormalTok{conj e1 e2 refl refl refl }\OtherTok{=}\NormalTok{ ≡siml e1}

\NormalTok{nat }\OtherTok{:} \OtherTok{∀} \OtherTok{\{}\NormalTok{ℓ κ}\OtherTok{\}} \OtherTok{\{}\NormalTok{A }\OtherTok{:}\NormalTok{ Type ℓ}\OtherTok{\}} \OtherTok{\{}\NormalTok{B }\OtherTok{:}\NormalTok{ Type κ}\OtherTok{\}} \OtherTok{\{}\NormalTok{f g }\OtherTok{:}\NormalTok{ A }\OtherTok{→}\NormalTok{ B}\OtherTok{\}} \OtherTok{\{}\NormalTok{a b }\OtherTok{:}\NormalTok{ A}\OtherTok{\}}
      \OtherTok{→} \OtherTok{(}\NormalTok{α }\OtherTok{:} \OtherTok{(}\NormalTok{x }\OtherTok{:}\NormalTok{ A}\OtherTok{)} \OtherTok{→}\NormalTok{ f x ≡ g x}\OtherTok{)} \OtherTok{(}\NormalTok{e }\OtherTok{:}\NormalTok{ a ≡ b}\OtherTok{)}
      \OtherTok{→} \OtherTok{((}\NormalTok{f a}\OtherTok{)}\NormalTok{ ≡〈 α a 〉 }\OtherTok{(}\NormalTok{ap g e}\OtherTok{))}\NormalTok{ ≡ }\OtherTok{((}\NormalTok{f a}\OtherTok{)}\NormalTok{ ≡〈 ap f e 〉 }\OtherTok{(}\NormalTok{α b}\OtherTok{))}
\NormalTok{nat }\OtherTok{\{}\NormalTok{a }\OtherTok{=}\NormalTok{ a}\OtherTok{\}}\NormalTok{ α refl }\OtherTok{=}\NormalTok{ ≡idr }\OtherTok{(}\NormalTok{α a}\OtherTok{)}

\NormalTok{cancel }\OtherTok{:} \OtherTok{∀} \OtherTok{\{}\NormalTok{ℓ}\OtherTok{\}} \OtherTok{\{}\NormalTok{A }\OtherTok{:}\NormalTok{ Type ℓ}\OtherTok{\}} \OtherTok{\{}\NormalTok{a b c }\OtherTok{:}\NormalTok{ A}\OtherTok{\}}
         \OtherTok{→} \OtherTok{(}\NormalTok{e1 e2 }\OtherTok{:}\NormalTok{ a ≡ b}\OtherTok{)} \OtherTok{(}\NormalTok{e3 }\OtherTok{:}\NormalTok{ b ≡ c}\OtherTok{)}
         \OtherTok{→} \OtherTok{(}\NormalTok{a ≡〈 e1 〉 e3}\OtherTok{)}\NormalTok{ ≡ }\OtherTok{(}\NormalTok{a ≡〈 e2 〉 e3}\OtherTok{)}
         \OtherTok{→}\NormalTok{ e1 ≡ e2}
\NormalTok{cancel e1 e2 refl refl }\OtherTok{=}\NormalTok{ refl}

\NormalTok{apId }\OtherTok{:} \OtherTok{∀} \OtherTok{\{}\NormalTok{ℓ}\OtherTok{\}} \OtherTok{\{}\NormalTok{A }\OtherTok{:}\NormalTok{ Type ℓ}\OtherTok{\}} \OtherTok{\{}\NormalTok{a b }\OtherTok{:}\NormalTok{ A}\OtherTok{\}}
       \OtherTok{→} \OtherTok{(}\NormalTok{e }\OtherTok{:}\NormalTok{ a ≡ b}\OtherTok{)} \OtherTok{→}\NormalTok{ ap }\OtherTok{(λ}\NormalTok{ x }\OtherTok{→}\NormalTok{ x}\OtherTok{)}\NormalTok{ e ≡ e}
\NormalTok{apId refl }\OtherTok{=}\NormalTok{ refl}

\NormalTok{apComp }\OtherTok{:} \OtherTok{∀} \OtherTok{\{}\NormalTok{ℓ ℓ\textquotesingle{} ℓ\textquotesingle{}\textquotesingle{}}\OtherTok{\}} \OtherTok{\{}\NormalTok{A }\OtherTok{:}\NormalTok{ Type ℓ}\OtherTok{\}} \OtherTok{\{}\NormalTok{B }\OtherTok{:}\NormalTok{ Type ℓ\textquotesingle{}}\OtherTok{\}} \OtherTok{\{}\NormalTok{C }\OtherTok{:}\NormalTok{ Type ℓ\textquotesingle{}\textquotesingle{}}\OtherTok{\}} \OtherTok{\{}\NormalTok{a b }\OtherTok{:}\NormalTok{ A}\OtherTok{\}}
         \OtherTok{→} \OtherTok{(}\NormalTok{f }\OtherTok{:}\NormalTok{ A }\OtherTok{→}\NormalTok{ B}\OtherTok{)} \OtherTok{(}\NormalTok{g }\OtherTok{:}\NormalTok{ B }\OtherTok{→}\NormalTok{ C}\OtherTok{)} \OtherTok{(}\NormalTok{e }\OtherTok{:}\NormalTok{ a ≡ b}\OtherTok{)}
         \OtherTok{→}\NormalTok{ ap }\OtherTok{(λ}\NormalTok{ x }\OtherTok{→}\NormalTok{ g }\OtherTok{(}\NormalTok{f x}\OtherTok{))}\NormalTok{ e ≡ ap g }\OtherTok{(}\NormalTok{ap f e}\OtherTok{)}
\NormalTok{apComp f g refl }\OtherTok{=}\NormalTok{ refl}

\NormalTok{apHtpy }\OtherTok{:} \OtherTok{∀} \OtherTok{\{}\NormalTok{ℓ}\OtherTok{\}} \OtherTok{\{}\NormalTok{A }\OtherTok{:}\NormalTok{ Type ℓ}\OtherTok{\}} \OtherTok{\{}\NormalTok{a }\OtherTok{:}\NormalTok{ A}\OtherTok{\}}
         \OtherTok{→} \OtherTok{(}\NormalTok{i }\OtherTok{:}\NormalTok{ A }\OtherTok{→}\NormalTok{ A}\OtherTok{)} \OtherTok{(}\NormalTok{α }\OtherTok{:} \OtherTok{(}\NormalTok{x }\OtherTok{:}\NormalTok{ A}\OtherTok{)} \OtherTok{→}\NormalTok{ i x ≡ x}\OtherTok{)}
         \OtherTok{→}\NormalTok{ ap i }\OtherTok{(}\NormalTok{α a}\OtherTok{)}\NormalTok{ ≡ α }\OtherTok{(}\NormalTok{i a}\OtherTok{)}
\NormalTok{apHtpy }\OtherTok{\{}\NormalTok{a }\OtherTok{=}\NormalTok{ a}\OtherTok{\}}\NormalTok{ i α }\OtherTok{=} 
\NormalTok{    cancel }\OtherTok{(}\NormalTok{ap i }\OtherTok{(}\NormalTok{α a}\OtherTok{))} \OtherTok{(}\NormalTok{α }\OtherTok{(}\NormalTok{i a}\OtherTok{))} \OtherTok{(}\NormalTok{α a}\OtherTok{)} 
           \OtherTok{((}\NormalTok{i }\OtherTok{(}\NormalTok{i a}\OtherTok{)}\NormalTok{ ≡〈 ap i }\OtherTok{(}\NormalTok{α a}\OtherTok{)}\NormalTok{ 〉 α a}\OtherTok{)} 
\NormalTok{           ≡〈 sym }\OtherTok{(}\NormalTok{nat α }\OtherTok{(}\NormalTok{α a}\OtherTok{))}\NormalTok{ 〉 }
           \OtherTok{((}\NormalTok{i }\OtherTok{(}\NormalTok{i a}\OtherTok{)}\NormalTok{ ≡〈 α }\OtherTok{(}\NormalTok{i a}\OtherTok{)}\NormalTok{ 〉 ap }\OtherTok{(λ}\NormalTok{ z }\OtherTok{→}\NormalTok{ z}\OtherTok{)} \OtherTok{(}\NormalTok{α a}\OtherTok{))} 
\NormalTok{           ≡〈 ap }\OtherTok{(λ}\NormalTok{ e }\OtherTok{→}\NormalTok{ i }\OtherTok{(}\NormalTok{i a}\OtherTok{)}\NormalTok{ ≡〈 α }\OtherTok{(}\NormalTok{i a}\OtherTok{)}\NormalTok{ 〉 e}\OtherTok{)} \OtherTok{(}\NormalTok{apId }\OtherTok{(}\NormalTok{α a}\OtherTok{))}\NormalTok{ 〉 }
           \OtherTok{((}\NormalTok{i }\OtherTok{(}\NormalTok{i a}\OtherTok{)}\NormalTok{ ≡〈 α }\OtherTok{(}\NormalTok{i a}\OtherTok{)}\NormalTok{ 〉 α a}\OtherTok{)}\NormalTok{ □}\OtherTok{)))}
\end{Highlighting}
\end{Shaded}

Half-adjoint equivalences and converting an isomorphism to a
half-adjoint equivalence:

\begin{Shaded}
\begin{Highlighting}[]
\NormalTok{HAdj }\OtherTok{:} \OtherTok{∀} \OtherTok{\{}\NormalTok{ℓ κ}\OtherTok{\}} \OtherTok{\{}\NormalTok{A }\OtherTok{:}\NormalTok{ Type ℓ}\OtherTok{\}} \OtherTok{\{}\NormalTok{B }\OtherTok{:}\NormalTok{ Type κ}\OtherTok{\}}
       \OtherTok{→} \OtherTok{(}\NormalTok{A }\OtherTok{→}\NormalTok{ B}\OtherTok{)} \OtherTok{→} \DataTypeTok{Set} \OtherTok{(}\NormalTok{ℓ ⊔ κ}\OtherTok{)}
\NormalTok{HAdj }\OtherTok{\{}\NormalTok{A }\OtherTok{=}\NormalTok{ A}\OtherTok{\}} \OtherTok{\{}\NormalTok{B }\OtherTok{=}\NormalTok{ B}\OtherTok{\}}\NormalTok{ f }\OtherTok{=}
\NormalTok{    Σ }\OtherTok{(}\NormalTok{B }\OtherTok{→}\NormalTok{ A}\OtherTok{)} \OtherTok{(λ}\NormalTok{ g }\OtherTok{→} 
\NormalTok{      Σ }\OtherTok{((}\NormalTok{x }\OtherTok{:}\NormalTok{ A}\OtherTok{)} \OtherTok{→}\NormalTok{ g }\OtherTok{(}\NormalTok{f x}\OtherTok{)}\NormalTok{ ≡ x}\OtherTok{)} \OtherTok{(λ}\NormalTok{ η }\OtherTok{→} 
\NormalTok{        Σ }\OtherTok{((}\NormalTok{y }\OtherTok{:}\NormalTok{ B}\OtherTok{)} \OtherTok{→}\NormalTok{ f }\OtherTok{(}\NormalTok{g y}\OtherTok{)}\NormalTok{ ≡ y}\OtherTok{)} \OtherTok{(λ}\NormalTok{ ε }\OtherTok{→} 
          \OtherTok{(}\NormalTok{x }\OtherTok{:}\NormalTok{ A}\OtherTok{)} \OtherTok{→}\NormalTok{ ap f }\OtherTok{(}\NormalTok{η x}\OtherTok{)}\NormalTok{ ≡ ε }\OtherTok{(}\NormalTok{f x}\OtherTok{))))}

\NormalTok{Iso→HAdj }\OtherTok{:} \OtherTok{∀} \OtherTok{\{}\NormalTok{ℓ κ}\OtherTok{\}} \OtherTok{\{}\NormalTok{A }\OtherTok{:}\NormalTok{ Type ℓ}\OtherTok{\}} \OtherTok{\{}\NormalTok{B }\OtherTok{:}\NormalTok{ Type κ}\OtherTok{\}} \OtherTok{\{}\NormalTok{f }\OtherTok{:}\NormalTok{ A }\OtherTok{→}\NormalTok{ B}\OtherTok{\}}
           \OtherTok{→}\NormalTok{ Iso f }\OtherTok{→}\NormalTok{ HAdj f}
\NormalTok{Iso→HAdj }\OtherTok{\{}\NormalTok{f }\OtherTok{=}\NormalTok{ f}\OtherTok{\}} \OtherTok{(}\NormalTok{g , η , ε}\OtherTok{)} \OtherTok{=}
\NormalTok{    g , }\OtherTok{(}\NormalTok{η }
\NormalTok{    , }\OtherTok{(} \OtherTok{(λ}\NormalTok{ y }\OtherTok{→} 
\NormalTok{           f }\OtherTok{(}\NormalTok{g y}\OtherTok{)}\NormalTok{         ≡〈 sym }\OtherTok{(}\NormalTok{ε }\OtherTok{(}\NormalTok{f }\OtherTok{(}\NormalTok{g y}\OtherTok{)))}\NormalTok{ 〉 }
          \OtherTok{(}\NormalTok{f }\OtherTok{(}\NormalTok{g }\OtherTok{(}\NormalTok{f }\OtherTok{(}\NormalTok{g y}\OtherTok{)))}\NormalTok{ ≡〈 ap f }\OtherTok{(}\NormalTok{η }\OtherTok{(}\NormalTok{g y}\OtherTok{))}\NormalTok{ 〉 }
          \OtherTok{(}\NormalTok{f }\OtherTok{(}\NormalTok{g y}\OtherTok{)}\NormalTok{         ≡〈 ε y 〉 }
          \OtherTok{(}\NormalTok{y               □}\OtherTok{))))} 
\NormalTok{      , }\OtherTok{λ}\NormalTok{ x }\OtherTok{→}\NormalTok{ conj }\OtherTok{(}\NormalTok{ε }\OtherTok{(}\NormalTok{f }\OtherTok{(}\NormalTok{g }\OtherTok{(}\NormalTok{f x}\OtherTok{))))} \OtherTok{(}\NormalTok{ap f }\OtherTok{(}\NormalTok{η }\OtherTok{(}\NormalTok{g }\OtherTok{(}\NormalTok{f x}\OtherTok{))))} 
                   \OtherTok{(}\NormalTok{ap f }\OtherTok{(}\NormalTok{η x}\OtherTok{))} \OtherTok{(}\NormalTok{ε }\OtherTok{(}\NormalTok{f x}\OtherTok{))} 
                   \OtherTok{(((}\NormalTok{f }\OtherTok{(}\NormalTok{g }\OtherTok{(}\NormalTok{f }\OtherTok{(}\NormalTok{g }\OtherTok{(}\NormalTok{f x}\OtherTok{))))}\NormalTok{ ≡〈 ε }\OtherTok{(}\NormalTok{f }\OtherTok{(}\NormalTok{g }\OtherTok{(}\NormalTok{f x}\OtherTok{)))}\NormalTok{ 〉 ap f }\OtherTok{(}\NormalTok{η x}\OtherTok{)))} 
\NormalTok{                    ≡〈 nat }\OtherTok{(λ}\NormalTok{ z }\OtherTok{→}\NormalTok{ ε }\OtherTok{(}\NormalTok{f z}\OtherTok{))} \OtherTok{(}\NormalTok{η x}\OtherTok{)}\NormalTok{ 〉 }
                    \OtherTok{(((}\NormalTok{f }\OtherTok{(}\NormalTok{g }\OtherTok{(}\NormalTok{f }\OtherTok{(}\NormalTok{g }\OtherTok{(}\NormalTok{f x}\OtherTok{))))}\NormalTok{ ≡〈 ap }\OtherTok{(λ}\NormalTok{ z }\OtherTok{→}\NormalTok{ f }\OtherTok{(}\NormalTok{g }\OtherTok{(}\NormalTok{f z}\OtherTok{)))} \OtherTok{(}\NormalTok{η x}\OtherTok{)}\NormalTok{ 〉 ε }\OtherTok{(}\NormalTok{f x}\OtherTok{)))} 
\NormalTok{                    ≡〈 ap }\OtherTok{(λ}\NormalTok{ e }\OtherTok{→} \OtherTok{(}\NormalTok{f }\OtherTok{(}\NormalTok{g }\OtherTok{(}\NormalTok{f }\OtherTok{(}\NormalTok{g }\OtherTok{(}\NormalTok{f x}\OtherTok{))))}\NormalTok{ ≡〈 e 〉 ε }\OtherTok{(}\NormalTok{f x}\OtherTok{)))} 
                          \OtherTok{((}\NormalTok{ap }\OtherTok{(λ}\NormalTok{ z }\OtherTok{→}\NormalTok{ f }\OtherTok{(}\NormalTok{g }\OtherTok{(}\NormalTok{f z}\OtherTok{)))} \OtherTok{(}\NormalTok{η x}\OtherTok{))} 
\NormalTok{                           ≡〈 apComp }\OtherTok{(λ}\NormalTok{ z }\OtherTok{→}\NormalTok{ g }\OtherTok{(}\NormalTok{f z}\OtherTok{))}\NormalTok{ f }\OtherTok{(}\NormalTok{η x}\OtherTok{)}\NormalTok{ 〉 }
                           \OtherTok{((}\NormalTok{ap f }\OtherTok{(}\NormalTok{ap }\OtherTok{(λ}\NormalTok{ z }\OtherTok{→}\NormalTok{ g }\OtherTok{(}\NormalTok{f z}\OtherTok{))} \OtherTok{(}\NormalTok{η x}\OtherTok{)))} 
\NormalTok{                           ≡〈 ap }\OtherTok{(}\NormalTok{ap f}\OtherTok{)} \OtherTok{(}\NormalTok{apHtpy }\OtherTok{(λ}\NormalTok{ z }\OtherTok{→}\NormalTok{ g }\OtherTok{(}\NormalTok{f z}\OtherTok{))}\NormalTok{ η}\OtherTok{)}\NormalTok{ 〉 }
                           \OtherTok{(}\NormalTok{ap f }\OtherTok{(}\NormalTok{η }\OtherTok{(}\NormalTok{g }\OtherTok{(}\NormalTok{f x}\OtherTok{)))}\NormalTok{ □}\OtherTok{)))}\NormalTok{ 〉 }
                    \OtherTok{(((}\NormalTok{f }\OtherTok{(}\NormalTok{g }\OtherTok{(}\NormalTok{f }\OtherTok{(}\NormalTok{g }\OtherTok{(}\NormalTok{f x}\OtherTok{))))}\NormalTok{ ≡〈 ap f }\OtherTok{(}\NormalTok{η }\OtherTok{(}\NormalTok{g }\OtherTok{(}\NormalTok{f x}\OtherTok{)))}\NormalTok{ 〉 ε }\OtherTok{(}\NormalTok{f x}\OtherTok{)))}\NormalTok{ □}\OtherTok{)))))}
\end{Highlighting}
\end{Shaded}

Equivalences of dependent pair types:

\begin{Shaded}
\begin{Highlighting}[]
\NormalTok{pairEquiv1 }\OtherTok{:} \OtherTok{∀} \OtherTok{\{}\NormalTok{ℓ ℓ\textquotesingle{} κ}\OtherTok{\}} \OtherTok{\{}\NormalTok{A }\OtherTok{:}\NormalTok{ Type ℓ}\OtherTok{\}} \OtherTok{\{}\NormalTok{A\textquotesingle{} }\OtherTok{:}\NormalTok{ Type ℓ\textquotesingle{}}\OtherTok{\}} \OtherTok{\{}\NormalTok{B }\OtherTok{:}\NormalTok{ A\textquotesingle{} }\OtherTok{→}\NormalTok{ Type κ}\OtherTok{\}}
             \OtherTok{→} \OtherTok{(}\NormalTok{f }\OtherTok{:}\NormalTok{ A }\OtherTok{→}\NormalTok{ A\textquotesingle{}}\OtherTok{)} \OtherTok{→}\NormalTok{ isEquiv f}
             \OtherTok{→}\NormalTok{ isEquiv }\OtherTok{\{}\NormalTok{A }\OtherTok{=}\NormalTok{ Σ A }\OtherTok{(λ}\NormalTok{ x }\OtherTok{→}\NormalTok{ B }\OtherTok{(}\NormalTok{f x}\OtherTok{))\}} \OtherTok{\{}\NormalTok{B }\OtherTok{=}\NormalTok{ Σ A\textquotesingle{} B}\OtherTok{\}} 
                       \OtherTok{(λ} \OtherTok{(}\NormalTok{x , y}\OtherTok{)} \OtherTok{→} \OtherTok{(}\NormalTok{f x , y}\OtherTok{))}
\NormalTok{pairEquiv1 }\OtherTok{\{}\NormalTok{A }\OtherTok{=}\NormalTok{ A}\OtherTok{\}} \OtherTok{\{}\NormalTok{A\textquotesingle{} }\OtherTok{=}\NormalTok{ A\textquotesingle{}}\OtherTok{\}} \OtherTok{\{}\NormalTok{B }\OtherTok{=}\NormalTok{ B}\OtherTok{\}}\NormalTok{ f ef }\OtherTok{=}
\NormalTok{  Iso→isEquiv}
    \OtherTok{(} \OtherTok{(λ} \OtherTok{(}\NormalTok{x , y}\OtherTok{)} \OtherTok{→} \OtherTok{(}\NormalTok{g x , transp B }\OtherTok{(}\NormalTok{sym }\OtherTok{(}\NormalTok{ε x}\OtherTok{))}\NormalTok{ y}\OtherTok{))}
\NormalTok{    , }\OtherTok{(} \OtherTok{(λ} \OtherTok{(}\NormalTok{x , y}\OtherTok{)} \OtherTok{→}\NormalTok{ pairEq }\OtherTok{(}\NormalTok{η x}\OtherTok{)} \OtherTok{(}\NormalTok{lemma x y}\OtherTok{))} 
\NormalTok{      , }\OtherTok{λ} \OtherTok{(}\NormalTok{x , y}\OtherTok{)} \OtherTok{→}\NormalTok{ pairEq }\OtherTok{(}\NormalTok{ε x}\OtherTok{)} \OtherTok{(}\NormalTok{symr }\OtherTok{(}\NormalTok{ε x}\OtherTok{)}\NormalTok{ y}\OtherTok{)} \OtherTok{)} \OtherTok{)}
  \KeywordTok{where}
\NormalTok{    g }\OtherTok{:}\NormalTok{ A\textquotesingle{} }\OtherTok{→}\NormalTok{ A}
\NormalTok{    g }\OtherTok{=}\NormalTok{ fst }\OtherTok{(}\NormalTok{Iso→HAdj }\OtherTok{(}\NormalTok{isEquiv→Iso ef}\OtherTok{))}
\NormalTok{    η }\OtherTok{:} \OtherTok{(}\NormalTok{x }\OtherTok{:}\NormalTok{ A}\OtherTok{)} \OtherTok{→}\NormalTok{ g }\OtherTok{(}\NormalTok{f x}\OtherTok{)}\NormalTok{ ≡ x}
\NormalTok{    η }\OtherTok{=}\NormalTok{ fst }\OtherTok{(}\NormalTok{snd }\OtherTok{(}\NormalTok{Iso→HAdj }\OtherTok{(}\NormalTok{isEquiv→Iso ef}\OtherTok{)))}
\NormalTok{    ε }\OtherTok{:} \OtherTok{(}\NormalTok{y }\OtherTok{:}\NormalTok{ A\textquotesingle{}}\OtherTok{)} \OtherTok{→}\NormalTok{ f }\OtherTok{(}\NormalTok{g y}\OtherTok{)}\NormalTok{ ≡ y}
\NormalTok{    ε }\OtherTok{=}\NormalTok{ fst }\OtherTok{(}\NormalTok{snd }\OtherTok{(}\NormalTok{snd }\OtherTok{(}\NormalTok{Iso→HAdj }\OtherTok{(}\NormalTok{isEquiv→Iso ef}\OtherTok{))))}
\NormalTok{    ρ }\OtherTok{:} \OtherTok{(}\NormalTok{x }\OtherTok{:}\NormalTok{ A}\OtherTok{)} \OtherTok{→}\NormalTok{ ap f }\OtherTok{(}\NormalTok{η x}\OtherTok{)}\NormalTok{ ≡ ε }\OtherTok{(}\NormalTok{f x}\OtherTok{)}
\NormalTok{    ρ }\OtherTok{=}\NormalTok{ snd }\OtherTok{(}\NormalTok{snd }\OtherTok{(}\NormalTok{snd }\OtherTok{(}\NormalTok{Iso→HAdj }\OtherTok{(}\NormalTok{isEquiv→Iso ef}\OtherTok{))))}
\NormalTok{    lemma }\OtherTok{:} \OtherTok{(}\NormalTok{x }\OtherTok{:}\NormalTok{ A}\OtherTok{)} \OtherTok{(}\NormalTok{y }\OtherTok{:}\NormalTok{ B }\OtherTok{(}\NormalTok{f x}\OtherTok{))}
            \OtherTok{→}\NormalTok{ transp }\OtherTok{(λ}\NormalTok{ z }\OtherTok{→}\NormalTok{ B }\OtherTok{(}\NormalTok{f z}\OtherTok{))} \OtherTok{(}\NormalTok{η x}\OtherTok{)}
                     \OtherTok{(}\NormalTok{transp B }\OtherTok{(}\NormalTok{sym }\OtherTok{(}\NormalTok{ε }\OtherTok{(}\NormalTok{f x}\OtherTok{)))}\NormalTok{ y}\OtherTok{)}
\NormalTok{              ≡ y}
\NormalTok{    lemma x y }\OtherTok{=} \OtherTok{(}\NormalTok{transp }\OtherTok{(λ}\NormalTok{ z }\OtherTok{→}\NormalTok{ B }\OtherTok{(}\NormalTok{f z}\OtherTok{))} \OtherTok{(}\NormalTok{η x}\OtherTok{)} 
                        \OtherTok{(}\NormalTok{transp B }\OtherTok{(}\NormalTok{sym }\OtherTok{(}\NormalTok{ε }\OtherTok{(}\NormalTok{f x}\OtherTok{)))}\NormalTok{ y}\OtherTok{))} 
\NormalTok{                ≡〈 transpAp B f }\OtherTok{(}\NormalTok{η x}\OtherTok{)} 
                            \OtherTok{(}\NormalTok{transp B }\OtherTok{(}\NormalTok{sym }\OtherTok{(}\NormalTok{ε }\OtherTok{(}\NormalTok{f x}\OtherTok{)))}\NormalTok{ y}\OtherTok{)}\NormalTok{ 〉 }
                \OtherTok{(}\NormalTok{ transp B }\OtherTok{(}\NormalTok{ap f }\OtherTok{(}\NormalTok{η x}\OtherTok{))} 
                           \OtherTok{(}\NormalTok{transp B }\OtherTok{(}\NormalTok{sym }\OtherTok{(}\NormalTok{ε }\OtherTok{(}\NormalTok{f x}\OtherTok{)))}\NormalTok{ y}\OtherTok{)} 
\NormalTok{                ≡〈 ap }\OtherTok{(λ}\NormalTok{ e }\OtherTok{→}\NormalTok{ transp B e }
                                \OtherTok{(}\NormalTok{transp B }\OtherTok{(}\NormalTok{sym }\OtherTok{(}\NormalTok{ε }\OtherTok{(}\NormalTok{f x}\OtherTok{)))}\NormalTok{ y}\OtherTok{))} 
                      \OtherTok{(}\NormalTok{ρ x}\OtherTok{)}\NormalTok{ 〉 }
                \OtherTok{(} \OtherTok{(}\NormalTok{transp B }\OtherTok{(}\NormalTok{ε }\OtherTok{(}\NormalTok{f x}\OtherTok{))} 
                          \OtherTok{(}\NormalTok{transp B }\OtherTok{(}\NormalTok{sym }\OtherTok{(}\NormalTok{ε }\OtherTok{(}\NormalTok{f x}\OtherTok{)))}\NormalTok{ y}\OtherTok{))} 
\NormalTok{                ≡〈 }\OtherTok{(}\NormalTok{symr }\OtherTok{(}\NormalTok{ε }\OtherTok{(}\NormalTok{f x}\OtherTok{))}\NormalTok{ y}\OtherTok{)}\NormalTok{ 〉 }
                \OtherTok{(}\NormalTok{y □}\OtherTok{)))}

\NormalTok{pairEquiv2 }\OtherTok{:} \OtherTok{∀} \OtherTok{\{}\NormalTok{ℓ κ κ\textquotesingle{}}\OtherTok{\}} \OtherTok{\{}\NormalTok{A }\OtherTok{:}\NormalTok{ Type ℓ}\OtherTok{\}} \OtherTok{\{}\NormalTok{B }\OtherTok{:}\NormalTok{ A }\OtherTok{→}\NormalTok{ Type κ}\OtherTok{\}} \OtherTok{\{}\NormalTok{B\textquotesingle{} }\OtherTok{:}\NormalTok{ A }\OtherTok{→}\NormalTok{ Type κ\textquotesingle{}}\OtherTok{\}}
             \OtherTok{→} \OtherTok{(}\NormalTok{g }\OtherTok{:} \OtherTok{(}\NormalTok{x }\OtherTok{:}\NormalTok{ A}\OtherTok{)} \OtherTok{→}\NormalTok{ B x }\OtherTok{→}\NormalTok{ B\textquotesingle{} x}\OtherTok{)} \OtherTok{→} \OtherTok{((}\NormalTok{x }\OtherTok{:}\NormalTok{ A}\OtherTok{)} \OtherTok{→}\NormalTok{ isEquiv }\OtherTok{(}\NormalTok{g x}\OtherTok{))}
             \OtherTok{→}\NormalTok{ isEquiv }\OtherTok{\{}\NormalTok{A }\OtherTok{=}\NormalTok{ Σ A B}\OtherTok{\}} \OtherTok{\{}\NormalTok{B }\OtherTok{=}\NormalTok{ Σ A B\textquotesingle{}}\OtherTok{\}}
                       \OtherTok{(λ} \OtherTok{(}\NormalTok{x , y}\OtherTok{)} \OtherTok{→} \OtherTok{(}\NormalTok{x , g x y}\OtherTok{))}
\NormalTok{pairEquiv2 g eg }\OtherTok{=}
    \KeywordTok{let}\NormalTok{ isog }\OtherTok{=} \OtherTok{(λ}\NormalTok{ x }\OtherTok{→}\NormalTok{ isEquiv→Iso }\OtherTok{(}\NormalTok{eg x}\OtherTok{))} 
    \KeywordTok{in}\NormalTok{ Iso→isEquiv }\OtherTok{(} \OtherTok{(λ} \OtherTok{(}\NormalTok{x , y}\OtherTok{)} \OtherTok{→} \OtherTok{(}\NormalTok{x , fst }\OtherTok{(}\NormalTok{isog x}\OtherTok{)}\NormalTok{ y}\OtherTok{))} 
\NormalTok{                   , }\OtherTok{(} \OtherTok{(λ} \OtherTok{(}\NormalTok{x , y}\OtherTok{)} \OtherTok{→} 
\NormalTok{                          pairEq refl }\OtherTok{(}\NormalTok{fst }\OtherTok{(}\NormalTok{snd }\OtherTok{(}\NormalTok{isog x}\OtherTok{))}\NormalTok{ y}\OtherTok{))} 
\NormalTok{                     , }\OtherTok{λ} \OtherTok{(}\NormalTok{x , y}\OtherTok{)} \OtherTok{→} 
\NormalTok{                         pairEq refl }\OtherTok{(}\NormalTok{snd }\OtherTok{(}\NormalTok{snd }\OtherTok{(}\NormalTok{isog x}\OtherTok{))}\NormalTok{ y}\OtherTok{)))}

\NormalTok{pairEquiv }\OtherTok{:} \OtherTok{∀} \OtherTok{\{}\NormalTok{ℓ ℓ\textquotesingle{} κ κ\textquotesingle{}}\OtherTok{\}} \OtherTok{\{}\NormalTok{A }\OtherTok{:}\NormalTok{ Type ℓ}\OtherTok{\}} \OtherTok{\{}\NormalTok{A\textquotesingle{} }\OtherTok{:}\NormalTok{ Type ℓ\textquotesingle{}}\OtherTok{\}}
            \OtherTok{→} \OtherTok{\{}\NormalTok{B }\OtherTok{:}\NormalTok{ A }\OtherTok{→}\NormalTok{ Type κ}\OtherTok{\}} \OtherTok{\{}\NormalTok{B\textquotesingle{} }\OtherTok{:}\NormalTok{ A\textquotesingle{} }\OtherTok{→}\NormalTok{ Type κ\textquotesingle{}}\OtherTok{\}}
            \OtherTok{→} \OtherTok{(}\NormalTok{f }\OtherTok{:}\NormalTok{ A }\OtherTok{→}\NormalTok{ A\textquotesingle{}}\OtherTok{)} \OtherTok{(}\NormalTok{g }\OtherTok{:} \OtherTok{(}\NormalTok{x }\OtherTok{:}\NormalTok{ A}\OtherTok{)} \OtherTok{→}\NormalTok{ B x }\OtherTok{→}\NormalTok{ B\textquotesingle{} }\OtherTok{(}\NormalTok{f x}\OtherTok{))}
            \OtherTok{→}\NormalTok{ isEquiv f }\OtherTok{→} \OtherTok{((}\NormalTok{x }\OtherTok{:}\NormalTok{ A}\OtherTok{)} \OtherTok{→}\NormalTok{ isEquiv }\OtherTok{(}\NormalTok{g x}\OtherTok{))}
            \OtherTok{→}\NormalTok{ isEquiv }\OtherTok{\{}\NormalTok{A }\OtherTok{=}\NormalTok{ Σ A B}\OtherTok{\}} \OtherTok{\{}\NormalTok{B }\OtherTok{=}\NormalTok{ Σ A\textquotesingle{} B\textquotesingle{}}\OtherTok{\}}
                      \OtherTok{(λ} \OtherTok{(}\NormalTok{x , y}\OtherTok{)} \OtherTok{→} \OtherTok{(}\NormalTok{f x , g x y}\OtherTok{))}
\NormalTok{pairEquiv f g ef eg }\OtherTok{=} 
\NormalTok{    compIsEquiv }\OtherTok{(}\NormalTok{pairEquiv1 f ef}\OtherTok{)} 
                \OtherTok{(}\NormalTok{pairEquiv2 g eg}\OtherTok{)}
\end{Highlighting}
\end{Shaded}

The J rule, i.e.~the induction principle for the identity type:

\begin{Shaded}
\begin{Highlighting}[]
\NormalTok{J }\OtherTok{:} \OtherTok{∀} \OtherTok{\{}\NormalTok{ℓ κ}\OtherTok{\}} \OtherTok{\{}\NormalTok{A }\OtherTok{:}\NormalTok{ Type ℓ}\OtherTok{\}} \OtherTok{\{}\NormalTok{a }\OtherTok{:}\NormalTok{ A}\OtherTok{\}} \OtherTok{(}\NormalTok{B }\OtherTok{:} \OtherTok{(}\NormalTok{x }\OtherTok{:}\NormalTok{ A}\OtherTok{)} \OtherTok{→}\NormalTok{ a ≡ x }\OtherTok{→}\NormalTok{ Type κ}\OtherTok{)}
    \OtherTok{→} \OtherTok{\{}\NormalTok{a\textquotesingle{} }\OtherTok{:}\NormalTok{ A}\OtherTok{\}} \OtherTok{(}\NormalTok{e }\OtherTok{:}\NormalTok{ a ≡ a\textquotesingle{}}\OtherTok{)} \OtherTok{→}\NormalTok{ B a refl }\OtherTok{→}\NormalTok{ B a\textquotesingle{} e}
\NormalTok{J B refl b }\OtherTok{=}\NormalTok{ b}
\end{Highlighting}
\end{Shaded}

Function extensionality and derived results:

\begin{Shaded}
\begin{Highlighting}[]
\KeywordTok{postulate}
\NormalTok{    funext }\OtherTok{:} \OtherTok{∀} \OtherTok{\{}\NormalTok{ℓ κ}\OtherTok{\}} \OtherTok{\{}\NormalTok{A }\OtherTok{:}\NormalTok{ Type ℓ}\OtherTok{\}} \OtherTok{\{}\NormalTok{B }\OtherTok{:}\NormalTok{ A }\OtherTok{→}\NormalTok{ Type κ}\OtherTok{\}} \OtherTok{\{}\NormalTok{f g }\OtherTok{:} \OtherTok{(}\NormalTok{x }\OtherTok{:}\NormalTok{ A}\OtherTok{)} \OtherTok{→}\NormalTok{ B x}\OtherTok{\}}
             \OtherTok{→} \OtherTok{((}\NormalTok{x }\OtherTok{:}\NormalTok{ A}\OtherTok{)} \OtherTok{→}\NormalTok{ f x ≡ g x}\OtherTok{)} \OtherTok{→}\NormalTok{ f ≡ g}
\NormalTok{    funextr }\OtherTok{:} \OtherTok{∀} \OtherTok{\{}\NormalTok{ℓ κ}\OtherTok{\}} \OtherTok{\{}\NormalTok{A }\OtherTok{:}\NormalTok{ Type ℓ}\OtherTok{\}} \OtherTok{\{}\NormalTok{B }\OtherTok{:}\NormalTok{ A }\OtherTok{→}\NormalTok{ Type κ}\OtherTok{\}} \OtherTok{\{}\NormalTok{f g }\OtherTok{:} \OtherTok{(}\NormalTok{x }\OtherTok{:}\NormalTok{ A}\OtherTok{)} \OtherTok{→}\NormalTok{ B x}\OtherTok{\}}
              \OtherTok{→} \OtherTok{(}\NormalTok{e }\OtherTok{:} \OtherTok{(}\NormalTok{x }\OtherTok{:}\NormalTok{ A}\OtherTok{)} \OtherTok{→}\NormalTok{ f x ≡ g x}\OtherTok{)} \OtherTok{→}\NormalTok{ coAp }\OtherTok{(}\NormalTok{funext e}\OtherTok{)}\NormalTok{ ≡ e}
\NormalTok{    funextl }\OtherTok{:} \OtherTok{∀} \OtherTok{\{}\NormalTok{ℓ κ}\OtherTok{\}} \OtherTok{\{}\NormalTok{A }\OtherTok{:}\NormalTok{ Type ℓ}\OtherTok{\}} \OtherTok{\{}\NormalTok{B }\OtherTok{:}\NormalTok{ A }\OtherTok{→}\NormalTok{ Type κ}\OtherTok{\}} \OtherTok{\{}\NormalTok{f g }\OtherTok{:} \OtherTok{(}\NormalTok{x }\OtherTok{:}\NormalTok{ A}\OtherTok{)} \OtherTok{→}\NormalTok{ B x}\OtherTok{\}}
              \OtherTok{→} \OtherTok{(}\NormalTok{e }\OtherTok{:}\NormalTok{ f ≡ g}\OtherTok{)} \OtherTok{→}\NormalTok{ funext }\OtherTok{(}\NormalTok{coAp e}\OtherTok{)}\NormalTok{ ≡ e}

\NormalTok{transpD }\OtherTok{:} \OtherTok{∀} \OtherTok{\{}\NormalTok{ℓ κ}\OtherTok{\}} \OtherTok{\{}\NormalTok{A }\OtherTok{:}\NormalTok{ Type ℓ}\OtherTok{\}} \OtherTok{\{}\NormalTok{B }\OtherTok{:}\NormalTok{ A }\OtherTok{→}\NormalTok{ Type κ}\OtherTok{\}} \OtherTok{\{}\NormalTok{a a\textquotesingle{} }\OtherTok{:}\NormalTok{ A}\OtherTok{\}}
          \OtherTok{→} \OtherTok{(}\NormalTok{f }\OtherTok{:} \OtherTok{(}\NormalTok{x }\OtherTok{:}\NormalTok{ A}\OtherTok{)} \OtherTok{→}\NormalTok{ B x}\OtherTok{)} \OtherTok{(}\NormalTok{e }\OtherTok{:}\NormalTok{ a ≡ a\textquotesingle{}}\OtherTok{)}
          \OtherTok{→}\NormalTok{ transp B e }\OtherTok{(}\NormalTok{f a}\OtherTok{)}\NormalTok{ ≡ f a\textquotesingle{}}
\NormalTok{transpD f refl }\OtherTok{=}\NormalTok{ refl}

\NormalTok{transpHAdj }\OtherTok{:} \OtherTok{∀} \OtherTok{\{}\NormalTok{ℓ ℓ\textquotesingle{} κ}\OtherTok{\}} \OtherTok{\{}\NormalTok{A }\OtherTok{:}\NormalTok{ Type ℓ}\OtherTok{\}} \OtherTok{\{}\NormalTok{B }\OtherTok{:}\NormalTok{ Type ℓ\textquotesingle{}}\OtherTok{\}} 
            \OtherTok{→} \OtherTok{\{}\NormalTok{C }\OtherTok{:}\NormalTok{ B }\OtherTok{→}\NormalTok{ Type κ}\OtherTok{\}} \OtherTok{\{}\NormalTok{a }\OtherTok{:}\NormalTok{ A}\OtherTok{\}}
            \OtherTok{→} \OtherTok{\{}\NormalTok{g }\OtherTok{:}\NormalTok{ A }\OtherTok{→}\NormalTok{ B}\OtherTok{\}} \OtherTok{\{}\NormalTok{h }\OtherTok{:}\NormalTok{ B }\OtherTok{→}\NormalTok{ A}\OtherTok{\}} 
            \OtherTok{→} \OtherTok{(}\NormalTok{f }\OtherTok{:} \OtherTok{(}\NormalTok{x }\OtherTok{:}\NormalTok{ A}\OtherTok{)} \OtherTok{→}\NormalTok{ C }\OtherTok{(}\NormalTok{g x}\OtherTok{))} 
            \OtherTok{→} \OtherTok{(}\NormalTok{e }\OtherTok{:} \OtherTok{(}\NormalTok{y }\OtherTok{:}\NormalTok{ B}\OtherTok{)} \OtherTok{→}\NormalTok{ g }\OtherTok{(}\NormalTok{h y}\OtherTok{)}\NormalTok{ ≡ y}\OtherTok{)}
            \OtherTok{→} \OtherTok{(}\NormalTok{e\textquotesingle{} }\OtherTok{:} \OtherTok{(}\NormalTok{x }\OtherTok{:}\NormalTok{ A}\OtherTok{)} \OtherTok{→}\NormalTok{ h }\OtherTok{(}\NormalTok{g x}\OtherTok{)}\NormalTok{ ≡ x}\OtherTok{)}
            \OtherTok{→} \OtherTok{(}\NormalTok{e\textquotesingle{}\textquotesingle{} }\OtherTok{:} \OtherTok{(}\NormalTok{x }\OtherTok{:}\NormalTok{ A}\OtherTok{)} \OtherTok{→}\NormalTok{ e }\OtherTok{(}\NormalTok{g x}\OtherTok{)}\NormalTok{ ≡ ap g }\OtherTok{(}\NormalTok{e\textquotesingle{} x}\OtherTok{))}
            \OtherTok{→}\NormalTok{ transp C }\OtherTok{(}\NormalTok{e }\OtherTok{(}\NormalTok{g a}\OtherTok{))} \OtherTok{(}\NormalTok{f }\OtherTok{(}\NormalTok{h }\OtherTok{(}\NormalTok{g a}\OtherTok{)))}\NormalTok{ ≡ f a}
\NormalTok{transpHAdj }\OtherTok{\{}\NormalTok{C }\OtherTok{=}\NormalTok{ C}\OtherTok{\}} \OtherTok{\{}\NormalTok{a }\OtherTok{=}\NormalTok{ a}\OtherTok{\}} \OtherTok{\{}\NormalTok{g }\OtherTok{=}\NormalTok{ g}\OtherTok{\}} \OtherTok{\{}\NormalTok{h }\OtherTok{=}\NormalTok{ h}\OtherTok{\}}\NormalTok{ f e e\textquotesingle{} e\textquotesingle{}\textquotesingle{} }\OtherTok{=} 
\NormalTok{    transp C }\OtherTok{(}\NormalTok{e }\OtherTok{(}\NormalTok{g a}\OtherTok{))} \OtherTok{(}\NormalTok{f }\OtherTok{(}\NormalTok{h }\OtherTok{(}\NormalTok{g a}\OtherTok{)))}               
\NormalTok{        ≡〈 ap }\OtherTok{(λ}\NormalTok{ ee }\OtherTok{→}\NormalTok{ transp C ee }\OtherTok{(}\NormalTok{f }\OtherTok{(}\NormalTok{h }\OtherTok{(}\NormalTok{g a}\OtherTok{))))} \OtherTok{(}\NormalTok{e\textquotesingle{}\textquotesingle{} a}\OtherTok{)}\NormalTok{ 〉 }
    \OtherTok{(}\NormalTok{transp C }\OtherTok{(}\NormalTok{ap g }\OtherTok{(}\NormalTok{e\textquotesingle{} a}\OtherTok{))} \OtherTok{(}\NormalTok{f }\OtherTok{(}\NormalTok{h }\OtherTok{(}\NormalTok{g a}\OtherTok{)))} 
\NormalTok{        ≡〈 sym }\OtherTok{(}\NormalTok{transpAp C g }\OtherTok{(}\NormalTok{e\textquotesingle{} a}\OtherTok{)} \OtherTok{(}\NormalTok{f }\OtherTok{(}\NormalTok{h }\OtherTok{(}\NormalTok{g a}\OtherTok{))))}\NormalTok{ 〉 }
    \OtherTok{((}\NormalTok{transp }\OtherTok{(λ}\NormalTok{ x }\OtherTok{→}\NormalTok{ C }\OtherTok{(}\NormalTok{g x}\OtherTok{))} \OtherTok{(}\NormalTok{e\textquotesingle{} a}\OtherTok{)} \OtherTok{(}\NormalTok{f }\OtherTok{(}\NormalTok{h }\OtherTok{(}\NormalTok{g a}\OtherTok{))))} 
\NormalTok{        ≡〈 transpD f }\OtherTok{(}\NormalTok{e\textquotesingle{} a}\OtherTok{)}\NormalTok{ 〉}
    \OtherTok{((}\NormalTok{f a}\OtherTok{)}\NormalTok{ □}\OtherTok{)))}

\NormalTok{PreCompEquiv }\OtherTok{:} \OtherTok{∀} \OtherTok{\{}\NormalTok{ℓ ℓ\textquotesingle{} κ}\OtherTok{\}} \OtherTok{\{}\NormalTok{A }\OtherTok{:}\NormalTok{ Type ℓ}\OtherTok{\}} \OtherTok{\{}\NormalTok{B }\OtherTok{:}\NormalTok{ Type ℓ\textquotesingle{}}\OtherTok{\}} \OtherTok{\{}\NormalTok{C }\OtherTok{:}\NormalTok{ B }\OtherTok{→}\NormalTok{ Type κ}\OtherTok{\}}
               \OtherTok{→} \OtherTok{(}\NormalTok{f }\OtherTok{:}\NormalTok{ A }\OtherTok{→}\NormalTok{ B}\OtherTok{)} \OtherTok{→}\NormalTok{ isEquiv f }
               \OtherTok{→}\NormalTok{ isEquiv }\OtherTok{\{}\NormalTok{A }\OtherTok{=} \OtherTok{(}\NormalTok{b }\OtherTok{:}\NormalTok{ B}\OtherTok{)} \OtherTok{→}\NormalTok{ C b}\OtherTok{\}} 
                         \OtherTok{\{}\NormalTok{B }\OtherTok{=} \OtherTok{(}\NormalTok{a }\OtherTok{:}\NormalTok{ A}\OtherTok{)} \OtherTok{→}\NormalTok{ C }\OtherTok{(}\NormalTok{f a}\OtherTok{)\}} 
                         \OtherTok{(λ}\NormalTok{ g }\OtherTok{→} \OtherTok{λ}\NormalTok{ a }\OtherTok{→}\NormalTok{ g }\OtherTok{(}\NormalTok{f a}\OtherTok{))}
\NormalTok{PreCompEquiv }\OtherTok{\{}\NormalTok{C }\OtherTok{=}\NormalTok{ C}\OtherTok{\}}\NormalTok{ f ef }\OtherTok{=}
    \KeywordTok{let} \OtherTok{(}\NormalTok{f⁻¹ , l , r , e}\OtherTok{)} \OtherTok{=}\NormalTok{ Iso→HAdj }\OtherTok{(}\NormalTok{isEquiv→Iso ef}\OtherTok{)} 
    \KeywordTok{in}\NormalTok{ Iso→isEquiv }\OtherTok{(} \OtherTok{(λ}\NormalTok{ g b }\OtherTok{→}\NormalTok{ transp C }\OtherTok{(}\NormalTok{r b}\OtherTok{)} \OtherTok{(}\NormalTok{g }\OtherTok{(}\NormalTok{f⁻¹ b}\OtherTok{)))} 
\NormalTok{                   , }\OtherTok{(} \OtherTok{(λ}\NormalTok{ g }\OtherTok{→}\NormalTok{ funext }\OtherTok{(λ}\NormalTok{ b }\OtherTok{→}\NormalTok{ transpD g }\OtherTok{(}\NormalTok{r b}\OtherTok{)))} 
\NormalTok{                     , }\OtherTok{λ}\NormalTok{ g }\OtherTok{→}\NormalTok{ funext }\OtherTok{(λ}\NormalTok{ a }\OtherTok{→}\NormalTok{ transpHAdj g r l }\OtherTok{(λ}\NormalTok{ x }\OtherTok{→}\NormalTok{ sym }\OtherTok{(}\NormalTok{e x}\OtherTok{)))))}

\NormalTok{PostCompEquiv }\OtherTok{:} \OtherTok{∀} \OtherTok{\{}\NormalTok{ℓ κ κ\textquotesingle{}}\OtherTok{\}} \OtherTok{\{}\NormalTok{A }\OtherTok{:}\NormalTok{ Type ℓ}\OtherTok{\}} \OtherTok{\{}\NormalTok{B }\OtherTok{:}\NormalTok{ A }\OtherTok{→}\NormalTok{ Type κ}\OtherTok{\}} \OtherTok{\{}\NormalTok{C }\OtherTok{:}\NormalTok{ A }\OtherTok{→}\NormalTok{ Type κ\textquotesingle{}}\OtherTok{\}}
                \OtherTok{→} \OtherTok{(}\NormalTok{f }\OtherTok{:} \OtherTok{(}\NormalTok{x }\OtherTok{:}\NormalTok{ A}\OtherTok{)} \OtherTok{→}\NormalTok{ B x }\OtherTok{→}\NormalTok{ C x}\OtherTok{)} \OtherTok{→} \OtherTok{((}\NormalTok{x }\OtherTok{:}\NormalTok{ A}\OtherTok{)} \OtherTok{→}\NormalTok{ isEquiv }\OtherTok{(}\NormalTok{f x}\OtherTok{))}
                \OtherTok{→}\NormalTok{ isEquiv }\OtherTok{\{}\NormalTok{A }\OtherTok{=} \OtherTok{(}\NormalTok{x }\OtherTok{:}\NormalTok{ A}\OtherTok{)} \OtherTok{→}\NormalTok{ B x}\OtherTok{\}} 
                          \OtherTok{\{}\NormalTok{B }\OtherTok{=} \OtherTok{(}\NormalTok{x }\OtherTok{:}\NormalTok{ A}\OtherTok{)} \OtherTok{→}\NormalTok{ C x}\OtherTok{\}}
                          \OtherTok{(λ}\NormalTok{ g x }\OtherTok{→}\NormalTok{ f x }\OtherTok{(}\NormalTok{g x}\OtherTok{))}
\NormalTok{PostCompEquiv f ef }\OtherTok{=} 
    \OtherTok{(} \OtherTok{(} \OtherTok{(λ}\NormalTok{ g x }\OtherTok{→}\NormalTok{ fst }\OtherTok{(}\NormalTok{fst }\OtherTok{(}\NormalTok{ef x}\OtherTok{))} \OtherTok{(}\NormalTok{g x}\OtherTok{))}
\NormalTok{      , }\OtherTok{λ}\NormalTok{ g }\OtherTok{→}\NormalTok{ funext }\OtherTok{(λ}\NormalTok{ x }\OtherTok{→}\NormalTok{ snd }\OtherTok{(}\NormalTok{fst }\OtherTok{(}\NormalTok{ef x}\OtherTok{))} \OtherTok{(}\NormalTok{g x}\OtherTok{))))}
\NormalTok{    , }\OtherTok{(} \OtherTok{(λ}\NormalTok{ g x }\OtherTok{→}\NormalTok{ fst }\OtherTok{(}\NormalTok{snd }\OtherTok{(}\NormalTok{ef x}\OtherTok{))} \OtherTok{(}\NormalTok{g x}\OtherTok{))} 
\NormalTok{      , }\OtherTok{λ}\NormalTok{ g }\OtherTok{→}\NormalTok{ funext }\OtherTok{(λ}\NormalTok{ x }\OtherTok{→}\NormalTok{ snd }\OtherTok{(}\NormalTok{snd }\OtherTok{(}\NormalTok{ef x}\OtherTok{))} \OtherTok{(}\NormalTok{g x}\OtherTok{)))}
\end{Highlighting}
\end{Shaded}

\section{Polynomial Functors in Agda}

This module gives basic definitions involving polynomial functors, lenses, Cartesian lenses, and univalence.

\begin{Shaded}
\begin{Highlighting}[]
\PreprocessorTok{\{{-}\# OPTIONS {-}{-}without{-}K {-}{-}rewriting {-}{-}lossy{-}unification \#{-}\}}
\KeywordTok{module}\NormalTok{ poly }\KeywordTok{where}

\KeywordTok{open} \KeywordTok{import}\NormalTok{ Agda}\OtherTok{.}\NormalTok{Primitive}
\KeywordTok{open} \KeywordTok{import}\NormalTok{ Agda}\OtherTok{.}\NormalTok{Builtin}\OtherTok{.}\NormalTok{Sigma}
\KeywordTok{open} \KeywordTok{import}\NormalTok{ Agda}\OtherTok{.}\NormalTok{Builtin}\OtherTok{.}\NormalTok{Unit}
\KeywordTok{open} \KeywordTok{import}\NormalTok{ Agda}\OtherTok{.}\NormalTok{Builtin}\OtherTok{.}\NormalTok{Equality}
\KeywordTok{open} \KeywordTok{import}\NormalTok{ Agda}\OtherTok{.}\NormalTok{Builtin}\OtherTok{.}\NormalTok{Equality}\OtherTok{.}\NormalTok{Rewrite}
\KeywordTok{open} \KeywordTok{import}\NormalTok{ hott}
\end{Highlighting}
\end{Shaded}

Definition of polynomial functors:

\begin{Shaded}
\begin{Highlighting}[]
\NormalTok{Poly }\OtherTok{:} \OtherTok{(}\NormalTok{ℓ κ }\OtherTok{:}\NormalTok{ Level}\OtherTok{)} \OtherTok{→}\NormalTok{ Type }\OtherTok{((}\NormalTok{lsuc ℓ}\OtherTok{)}\NormalTok{ ⊔ }\OtherTok{(}\NormalTok{lsuc κ}\OtherTok{))}
\NormalTok{Poly ℓ κ }\OtherTok{=}\NormalTok{ Σ }\OtherTok{(}\NormalTok{Type ℓ}\OtherTok{)} \OtherTok{(λ}\NormalTok{ A }\OtherTok{→}\NormalTok{ A }\OtherTok{→}\NormalTok{ Type κ}\OtherTok{)}
\end{Highlighting}
\end{Shaded}

Lenses:

\begin{Shaded}
\begin{Highlighting}[]
\OtherTok{\_}\NormalTok{⇆}\OtherTok{\_} \OtherTok{:} \OtherTok{∀} \OtherTok{\{}\NormalTok{ℓ0 ℓ1 κ0 κ1}\OtherTok{\}} \OtherTok{→}\NormalTok{ Poly ℓ0 κ0 }\OtherTok{→}\NormalTok{ Poly ℓ1 κ1 }\OtherTok{→}\NormalTok{ Type }\OtherTok{(}\NormalTok{ℓ0 ⊔ ℓ1 ⊔ κ0 ⊔ κ1}\OtherTok{)}
\OtherTok{(}\NormalTok{A , B}\OtherTok{)}\NormalTok{ ⇆ }\OtherTok{(}\NormalTok{C , D}\OtherTok{)} \OtherTok{=}\NormalTok{ Σ }\OtherTok{(}\NormalTok{A }\OtherTok{→}\NormalTok{ C}\OtherTok{)} \OtherTok{(λ}\NormalTok{ f }\OtherTok{→} \OtherTok{(}\NormalTok{a }\OtherTok{:}\NormalTok{ A}\OtherTok{)} \OtherTok{→}\NormalTok{ D }\OtherTok{(}\NormalTok{f a}\OtherTok{)} \OtherTok{→}\NormalTok{ B a}\OtherTok{)}
\end{Highlighting}
\end{Shaded}

Type of equality proofs for lenses:

\begin{Shaded}
\begin{Highlighting}[]
\NormalTok{EqLens }\OtherTok{:} \OtherTok{∀} \OtherTok{\{}\NormalTok{ℓ0 ℓ1 κ0 κ1}\OtherTok{\}}
         \OtherTok{→} \OtherTok{\{}\NormalTok{p }\OtherTok{:}\NormalTok{ Poly ℓ0 κ0}\OtherTok{\}} \OtherTok{(}\NormalTok{q }\OtherTok{:}\NormalTok{ Poly ℓ1 κ1}\OtherTok{)}
         \OtherTok{→} \OtherTok{(}\NormalTok{f g }\OtherTok{:}\NormalTok{ p ⇆ q}\OtherTok{)} \OtherTok{→}\NormalTok{ Type }\OtherTok{(}\NormalTok{ℓ0 ⊔ ℓ1 ⊔ κ0 ⊔ κ1}\OtherTok{)}
\NormalTok{EqLens }\OtherTok{\{}\NormalTok{p }\OtherTok{=} \OtherTok{(}\NormalTok{A , B}\OtherTok{)\}} \OtherTok{(}\NormalTok{C , D}\OtherTok{)} \OtherTok{(}\NormalTok{f , f♯}\OtherTok{)} \OtherTok{(}\NormalTok{g , g♯}\OtherTok{)} \OtherTok{=}
\NormalTok{  Σ }\OtherTok{((}\NormalTok{a }\OtherTok{:}\NormalTok{ A}\OtherTok{)} \OtherTok{→}\NormalTok{ f a ≡ g a}\OtherTok{)}
    \OtherTok{(λ}\NormalTok{ e }\OtherTok{→} \OtherTok{(}\NormalTok{a }\OtherTok{:}\NormalTok{ A}\OtherTok{)} \OtherTok{(}\NormalTok{d }\OtherTok{:}\NormalTok{ D }\OtherTok{(}\NormalTok{f a}\OtherTok{))} 
           \OtherTok{→}\NormalTok{ f♯ a d ≡ g♯ a }\OtherTok{(}\NormalTok{transp D }\OtherTok{(}\NormalTok{e a}\OtherTok{)}\NormalTok{ d}\OtherTok{))}
\end{Highlighting}
\end{Shaded}

Identity and composition of lenses:

\begin{Shaded}
\begin{Highlighting}[]
\NormalTok{id }\OtherTok{:} \OtherTok{∀} \OtherTok{\{}\NormalTok{ℓ κ}\OtherTok{\}} \OtherTok{(}\NormalTok{p }\OtherTok{:}\NormalTok{ Poly ℓ κ}\OtherTok{)} \OtherTok{→}\NormalTok{ p ⇆ p}
\NormalTok{id p }\OtherTok{=} \OtherTok{(} \OtherTok{(λ}\NormalTok{ a }\OtherTok{→}\NormalTok{ a}\OtherTok{)}\NormalTok{ , }\OtherTok{λ}\NormalTok{ a b }\OtherTok{→}\NormalTok{ b }\OtherTok{)}
\end{Highlighting}
\end{Shaded}

\begin{Shaded}
\begin{Highlighting}[]
\NormalTok{comp }\OtherTok{:} \OtherTok{∀} \OtherTok{\{}\NormalTok{ℓ0 ℓ1 ℓ2 κ0 κ1 κ2}\OtherTok{\}}
       \OtherTok{→} \OtherTok{\{}\NormalTok{p }\OtherTok{:}\NormalTok{ Poly ℓ0 κ0}\OtherTok{\}} \OtherTok{\{}\NormalTok{q }\OtherTok{:}\NormalTok{ Poly ℓ1 κ1}\OtherTok{\}} \OtherTok{(}\NormalTok{r }\OtherTok{:}\NormalTok{ Poly ℓ2 κ2}\OtherTok{)}
       \OtherTok{→}\NormalTok{ p ⇆ q }\OtherTok{→}\NormalTok{ q ⇆ r }\OtherTok{→}\NormalTok{ p ⇆ r}
\NormalTok{comp r }\OtherTok{(}\NormalTok{f , f♯}\OtherTok{)} \OtherTok{(}\NormalTok{g , g♯}\OtherTok{)} \OtherTok{=} 
     \OtherTok{(} \OtherTok{(λ}\NormalTok{ a }\OtherTok{→}\NormalTok{ g }\OtherTok{(}\NormalTok{f a}\OtherTok{))}\NormalTok{ , }\OtherTok{λ}\NormalTok{ a z }\OtherTok{→}\NormalTok{ f♯ a }\OtherTok{(}\NormalTok{g♯ }\OtherTok{(}\NormalTok{f a}\OtherTok{)}\NormalTok{ z}\OtherTok{)} \OtherTok{)}
\end{Highlighting}
\end{Shaded}

Cartesian lenses:

\begin{Shaded}
\begin{Highlighting}[]
\KeywordTok{module}\NormalTok{ Cart }\OtherTok{\{}\NormalTok{ℓ0 ℓ1 κ0 κ1}\OtherTok{\}} \OtherTok{\{}\NormalTok{p }\OtherTok{:}\NormalTok{ Poly ℓ0 κ0}\OtherTok{\}} 
            \OtherTok{(}\NormalTok{q }\OtherTok{:}\NormalTok{ Poly ℓ1 κ1}\OtherTok{)} \OtherTok{(}\NormalTok{f }\OtherTok{:}\NormalTok{ p ⇆ q}\OtherTok{)} \KeywordTok{where}

\NormalTok{    isCartesian }\OtherTok{:} \DataTypeTok{Set} \OtherTok{(}\NormalTok{ℓ0 ⊔ κ0 ⊔ κ1}\OtherTok{)}
\NormalTok{    isCartesian }\OtherTok{=} \OtherTok{(}\NormalTok{a }\OtherTok{:}\NormalTok{ fst p}\OtherTok{)} \OtherTok{→}\NormalTok{ isEquiv }\OtherTok{(}\NormalTok{snd f a}\OtherTok{)}

\KeywordTok{open}\NormalTok{ Cart }\KeywordTok{public}
\end{Highlighting}
\end{Shaded}

Identity and composition of Cartesian lenses:

\begin{Shaded}
\begin{Highlighting}[]
\NormalTok{idCart }\OtherTok{:} \OtherTok{∀} \OtherTok{\{}\NormalTok{ℓ κ}\OtherTok{\}} \OtherTok{(}\NormalTok{p }\OtherTok{:}\NormalTok{ Poly ℓ κ}\OtherTok{)}
         \OtherTok{→}\NormalTok{ isCartesian p }\OtherTok{(}\NormalTok{id p}\OtherTok{)}
\NormalTok{idCart p a }\OtherTok{=}\NormalTok{ idIsEquiv}

\NormalTok{compCartesian }\OtherTok{:} \OtherTok{∀} \OtherTok{\{}\NormalTok{ℓ0 ℓ1 ℓ2 κ0 κ1 κ2}\OtherTok{\}}
                \OtherTok{→} \OtherTok{\{}\NormalTok{p }\OtherTok{:}\NormalTok{ Poly ℓ0 κ0}\OtherTok{\}} \OtherTok{\{}\NormalTok{q }\OtherTok{:}\NormalTok{ Poly ℓ1 κ1}\OtherTok{\}} \OtherTok{(}\NormalTok{r }\OtherTok{:}\NormalTok{ Poly ℓ2 κ2}\OtherTok{)}
                \OtherTok{→} \OtherTok{\{}\NormalTok{f }\OtherTok{:}\NormalTok{ p ⇆ q}\OtherTok{\}} \OtherTok{\{}\NormalTok{g }\OtherTok{:}\NormalTok{ q ⇆ r}\OtherTok{\}}
                \OtherTok{→}\NormalTok{ isCartesian q f }\OtherTok{→}\NormalTok{ isCartesian r g }
                \OtherTok{→}\NormalTok{ isCartesian r }\OtherTok{(}\NormalTok{comp r f g}\OtherTok{)}
\NormalTok{compCartesian r }\OtherTok{\{}\NormalTok{f }\OtherTok{=} \OtherTok{(}\NormalTok{f , f♯}\OtherTok{)\}} \OtherTok{\{}\NormalTok{g }\OtherTok{=} \OtherTok{(}\NormalTok{g , g♯}\OtherTok{)\}}\NormalTok{ cf cg a }\OtherTok{=} 
\NormalTok{    compIsEquiv }\OtherTok{(}\NormalTok{cf a}\OtherTok{)} \OtherTok{(}\NormalTok{cg }\OtherTok{(}\NormalTok{f a}\OtherTok{))}
\end{Highlighting}
\end{Shaded}

Univalent polynomials:

\begin{Shaded}
\begin{Highlighting}[]
\NormalTok{isUnivalent }\OtherTok{:} \OtherTok{∀} \OtherTok{\{}\NormalTok{ℓ κ}\OtherTok{\}} \OtherTok{→}\NormalTok{ Poly ℓ κ }\OtherTok{→}\NormalTok{ Setω}
\NormalTok{isUnivalent u }\OtherTok{=} 
    \OtherTok{∀} \OtherTok{\{}\NormalTok{ℓ\textquotesingle{} κ\textquotesingle{}}\OtherTok{\}} \OtherTok{\{}\NormalTok{p }\OtherTok{:}\NormalTok{ Poly ℓ\textquotesingle{} κ\textquotesingle{}}\OtherTok{\}}
      \OtherTok{→} \OtherTok{\{}\NormalTok{f g }\OtherTok{:}\NormalTok{ p ⇆ u}\OtherTok{\}}
      \OtherTok{→}\NormalTok{ isCartesian u f}
      \OtherTok{→}\NormalTok{ isCartesian u g}
      \OtherTok{→}\NormalTok{ EqLens u f g}
    
\end{Highlighting}
\end{Shaded}

\section{Composition of Polynomials and Monads in Agda}

\begin{Shaded}
\begin{Highlighting}[]
\PreprocessorTok{\{{-}\# OPTIONS {-}{-}without{-}K {-}{-}rewriting \#{-}\}}
\KeywordTok{module}\NormalTok{ sum }\KeywordTok{where}

\KeywordTok{open} \KeywordTok{import}\NormalTok{ Agda}\OtherTok{.}\NormalTok{Primitive}
\KeywordTok{open} \KeywordTok{import}\NormalTok{ Agda}\OtherTok{.}\NormalTok{Builtin}\OtherTok{.}\NormalTok{Sigma}
\KeywordTok{open} \KeywordTok{import}\NormalTok{ Agda}\OtherTok{.}\NormalTok{Builtin}\OtherTok{.}\NormalTok{Unit}
\KeywordTok{open} \KeywordTok{import}\NormalTok{ Agda}\OtherTok{.}\NormalTok{Builtin}\OtherTok{.}\NormalTok{Equality}
\KeywordTok{open} \KeywordTok{import}\NormalTok{ Agda}\OtherTok{.}\NormalTok{Builtin}\OtherTok{.}\NormalTok{Equality}\OtherTok{.}\NormalTok{Rewrite}
\KeywordTok{open} \KeywordTok{import}\NormalTok{ hott}
\KeywordTok{open} \KeywordTok{import}\NormalTok{ poly}
\end{Highlighting}
\end{Shaded}

Composition and identity of polynomial endofunctors, and functoriality
of composition:

\begin{Shaded}
\begin{Highlighting}[]
\NormalTok{𝕪 }\OtherTok{:}\NormalTok{ Poly lzero lzero}
\NormalTok{𝕪 }\OtherTok{=} \OtherTok{(}\NormalTok{⊤ , }\OtherTok{λ} \OtherTok{\_} \OtherTok{→}\NormalTok{ ⊤}\OtherTok{)}

\OtherTok{\_}\NormalTok{◃}\OtherTok{\_} \OtherTok{:} \OtherTok{∀} \OtherTok{\{}\NormalTok{ℓ0 ℓ1 κ0 κ1}\OtherTok{\}} \OtherTok{→}\NormalTok{ Poly ℓ0 κ0 }\OtherTok{→}\NormalTok{ Poly ℓ1 κ1 }\OtherTok{→}\NormalTok{ Poly }\OtherTok{(}\NormalTok{ℓ0 ⊔ κ0 ⊔ ℓ1}\OtherTok{)} \OtherTok{(}\NormalTok{κ0 ⊔ κ1}\OtherTok{)}
\OtherTok{(}\NormalTok{A , B}\OtherTok{)}\NormalTok{ ◃ }\OtherTok{(}\NormalTok{C , D}\OtherTok{)} \OtherTok{=} \OtherTok{(}\NormalTok{Σ A }\OtherTok{(λ}\NormalTok{ a }\OtherTok{→}\NormalTok{ B a }\OtherTok{→}\NormalTok{ C}\OtherTok{)}\NormalTok{ , }\OtherTok{λ} \OtherTok{(}\NormalTok{a , f}\OtherTok{)} \OtherTok{→}\NormalTok{ Σ }\OtherTok{(}\NormalTok{B a}\OtherTok{)} \OtherTok{(λ}\NormalTok{ b }\OtherTok{→}\NormalTok{ D }\OtherTok{(}\NormalTok{f b}\OtherTok{)))}

\OtherTok{\_}\NormalTok{◃◃[}\OtherTok{\_}\NormalTok{]}\OtherTok{\_} \OtherTok{:} \OtherTok{∀} \OtherTok{\{}\NormalTok{ℓ0 ℓ1 ℓ2 ℓ3 κ0 κ1 κ2 κ3}\OtherTok{\}}
        \OtherTok{→} \OtherTok{\{}\NormalTok{p }\OtherTok{:}\NormalTok{ Poly ℓ0 κ0}\OtherTok{\}} \OtherTok{\{}\NormalTok{q }\OtherTok{:}\NormalTok{ Poly ℓ2 κ2}\OtherTok{\}} \OtherTok{→}\NormalTok{ p ⇆ q}
        \OtherTok{→} \OtherTok{\{}\NormalTok{r }\OtherTok{:}\NormalTok{ Poly ℓ1 κ1}\OtherTok{\}} \OtherTok{(}\NormalTok{s }\OtherTok{:}\NormalTok{ Poly ℓ3 κ3}\OtherTok{)} \OtherTok{→}\NormalTok{ r ⇆ s }
        \OtherTok{→} \OtherTok{(}\NormalTok{p ◃ r}\OtherTok{)}\NormalTok{ ⇆ }\OtherTok{(}\NormalTok{q ◃ s}\OtherTok{)}
\OtherTok{(}\NormalTok{f , f♯}\OtherTok{)}\NormalTok{ ◃◃[ s ] }\OtherTok{(}\NormalTok{g , g♯}\OtherTok{)} \OtherTok{=}
    \OtherTok{((λ} \OtherTok{(}\NormalTok{a , γ}\OtherTok{)} \OtherTok{→} \OtherTok{(}\NormalTok{f a , }\OtherTok{λ}\NormalTok{ b\textquotesingle{} }\OtherTok{→}\NormalTok{ g }\OtherTok{(}\NormalTok{γ }\OtherTok{(}\NormalTok{f♯ a b\textquotesingle{}}\OtherTok{))))}
\NormalTok{    , }\OtherTok{λ} \OtherTok{(}\NormalTok{a , γ}\OtherTok{)} \OtherTok{(}\NormalTok{b\textquotesingle{} , d\textquotesingle{}}\OtherTok{)} \OtherTok{→} \OtherTok{((}\NormalTok{f♯ a b\textquotesingle{}}\OtherTok{)}\NormalTok{ , g♯ }\OtherTok{(}\NormalTok{γ }\OtherTok{(}\NormalTok{f♯ a b\textquotesingle{}}\OtherTok{))}\NormalTok{ d\textquotesingle{}}\OtherTok{))}
\end{Highlighting}
\end{Shaded}

Associativity of \texttt{◃}:

\begin{Shaded}
\begin{Highlighting}[]
\KeywordTok{module}\NormalTok{ ◃Assoc }\OtherTok{\{}\NormalTok{ℓ0 ℓ1 ℓ2 κ0 κ1 κ2}\OtherTok{\}} \OtherTok{(}\NormalTok{p }\OtherTok{:}\NormalTok{ Poly ℓ0 κ0}\OtherTok{)} 
              \OtherTok{(}\NormalTok{q }\OtherTok{:}\NormalTok{ Poly ℓ1 κ1}\OtherTok{)} \OtherTok{(}\NormalTok{r }\OtherTok{:}\NormalTok{ Poly ℓ2 κ2}\OtherTok{)} \KeywordTok{where}

\NormalTok{    ◃assoc }\OtherTok{:} \OtherTok{((}\NormalTok{p ◃ q}\OtherTok{)}\NormalTok{ ◃ r}\OtherTok{)}\NormalTok{ ⇆ }\OtherTok{(}\NormalTok{p ◃ }\OtherTok{(}\NormalTok{q ◃ r}\OtherTok{))}
\NormalTok{    ◃assoc }\OtherTok{=} \OtherTok{(} \OtherTok{(λ} \OtherTok{((}\NormalTok{a , γ}\OtherTok{)}\NormalTok{ , δ}\OtherTok{)} 
                  \OtherTok{→} \OtherTok{(}\NormalTok{a , }\OtherTok{(λ}\NormalTok{ b }\OtherTok{→} \OtherTok{(}\NormalTok{γ b , }\OtherTok{λ}\NormalTok{ d }\OtherTok{→}\NormalTok{ δ }\OtherTok{(}\NormalTok{b , d}\OtherTok{)))))} 
\NormalTok{             , }\OtherTok{(λ} \OtherTok{\_} \OtherTok{(}\NormalTok{b , }\OtherTok{(}\NormalTok{d , x}\OtherTok{))} \OtherTok{→} \OtherTok{((}\NormalTok{b , d}\OtherTok{)}\NormalTok{ , x}\OtherTok{))} \OtherTok{)}
    
\NormalTok{    ◃assoc⁻¹ }\OtherTok{:} \OtherTok{(}\NormalTok{p ◃ }\OtherTok{(}\NormalTok{q ◃ r}\OtherTok{))}\NormalTok{ ⇆ }\OtherTok{((}\NormalTok{p ◃ q}\OtherTok{)}\NormalTok{ ◃ r}\OtherTok{)}
\NormalTok{    ◃assoc⁻¹ }\OtherTok{=} \OtherTok{(} \OtherTok{(λ} \OtherTok{(}\NormalTok{a , γ}\OtherTok{)} \OtherTok{→} \OtherTok{(} \OtherTok{(}\NormalTok{a , }\OtherTok{(λ}\NormalTok{ x }\OtherTok{→}\NormalTok{ fst }\OtherTok{(}\NormalTok{γ x}\OtherTok{)))} 
\NormalTok{                              , }\OtherTok{(λ} \OtherTok{(}\NormalTok{x , y}\OtherTok{)} \OtherTok{→}\NormalTok{ snd }\OtherTok{(}\NormalTok{γ x}\OtherTok{)}\NormalTok{ y}\OtherTok{)} \OtherTok{))}
\NormalTok{               , }\OtherTok{λ} \OtherTok{\_} \OtherTok{((}\NormalTok{x , y}\OtherTok{)}\NormalTok{ , z}\OtherTok{)} \OtherTok{→} \OtherTok{(}\NormalTok{x , }\OtherTok{(}\NormalTok{y , z}\OtherTok{))} \OtherTok{)}

\KeywordTok{open}\NormalTok{ ◃Assoc }\KeywordTok{public}
\end{Highlighting}
\end{Shaded}

Left and right unit laws for \texttt{◃}:

\begin{Shaded}
\begin{Highlighting}[]
\KeywordTok{module}\NormalTok{ ◃LRUnit }\OtherTok{\{}\NormalTok{ℓ κ}\OtherTok{\}} \OtherTok{(}\NormalTok{p }\OtherTok{:}\NormalTok{ Poly ℓ κ}\OtherTok{)} \KeywordTok{where}

\NormalTok{    ◃unitl }\OtherTok{:} \OtherTok{(}\NormalTok{𝕪 ◃ p}\OtherTok{)}\NormalTok{ ⇆ p}
\NormalTok{    ◃unitl }\OtherTok{=} \OtherTok{(} \OtherTok{(λ} \OtherTok{(\_}\NormalTok{ , a}\OtherTok{)} \OtherTok{→}\NormalTok{ a tt}\OtherTok{)}\NormalTok{ , }\OtherTok{λ} \OtherTok{(\_}\NormalTok{ , a}\OtherTok{)}\NormalTok{ x }\OtherTok{→} \OtherTok{(}\NormalTok{tt , x}\OtherTok{)} \OtherTok{)}

\NormalTok{    ◃unitl⁻¹ }\OtherTok{:}\NormalTok{ p ⇆ }\OtherTok{(}\NormalTok{𝕪 ◃ p}\OtherTok{)}
\NormalTok{    ◃unitl⁻¹ }\OtherTok{=} \OtherTok{(} \OtherTok{(λ}\NormalTok{ a }\OtherTok{→} \OtherTok{(}\NormalTok{tt , }\OtherTok{λ} \OtherTok{\_} \OtherTok{→}\NormalTok{ a}\OtherTok{))}\NormalTok{ , }\OtherTok{(λ}\NormalTok{ a }\OtherTok{(\_}\NormalTok{ , b}\OtherTok{)} \OtherTok{→}\NormalTok{ b }\OtherTok{)} \OtherTok{)}

\NormalTok{    ◃unitr }\OtherTok{:} \OtherTok{(}\NormalTok{p ◃ 𝕪}\OtherTok{)}\NormalTok{ ⇆ p}
\NormalTok{    ◃unitr }\OtherTok{=} \OtherTok{(} \OtherTok{(λ} \OtherTok{(}\NormalTok{a , γ}\OtherTok{)} \OtherTok{→}\NormalTok{ a}\OtherTok{)}\NormalTok{ , }\OtherTok{(λ} \OtherTok{(}\NormalTok{a , γ}\OtherTok{)}\NormalTok{ b }\OtherTok{→} \OtherTok{(}\NormalTok{b , tt}\OtherTok{))} \OtherTok{)}

\NormalTok{    ◃unitr⁻¹ }\OtherTok{:}\NormalTok{ p ⇆ }\OtherTok{(}\NormalTok{p ◃ 𝕪}\OtherTok{)}
\NormalTok{    ◃unitr⁻¹ }\OtherTok{=} \OtherTok{(} \OtherTok{(λ}\NormalTok{ a }\OtherTok{→}\NormalTok{ a , }\OtherTok{(λ} \OtherTok{\_} \OtherTok{→}\NormalTok{ tt}\OtherTok{))}\NormalTok{ , }\OtherTok{(λ}\NormalTok{ a }\OtherTok{(}\NormalTok{b , }\OtherTok{\_)} \OtherTok{→}\NormalTok{ b}\OtherTok{)} \OtherTok{)}

\KeywordTok{open}\NormalTok{ ◃LRUnit }\KeywordTok{public}
\end{Highlighting}
\end{Shaded}

Restriction of \texttt{◃} to a monoidal product on
\(\mathbf{Poly^{Cart}}\):

\begin{Shaded}
\begin{Highlighting}[]
\NormalTok{◃◃Cart }\OtherTok{:} \OtherTok{∀} \OtherTok{\{}\NormalTok{ℓ0 ℓ1 ℓ2 ℓ3 κ0 κ1 κ2 κ3}\OtherTok{\}}
         \OtherTok{→} \OtherTok{\{}\NormalTok{p }\OtherTok{:}\NormalTok{ Poly ℓ0 κ0}\OtherTok{\}} \OtherTok{(}\NormalTok{q }\OtherTok{:}\NormalTok{ Poly ℓ2 κ2}\OtherTok{)} \OtherTok{\{}\NormalTok{f }\OtherTok{:}\NormalTok{ p ⇆ q}\OtherTok{\}}
         \OtherTok{→} \OtherTok{\{}\NormalTok{r }\OtherTok{:}\NormalTok{ Poly ℓ1 κ1}\OtherTok{\}} \OtherTok{(}\NormalTok{s }\OtherTok{:}\NormalTok{ Poly ℓ3 κ3}\OtherTok{)} \OtherTok{\{}\NormalTok{g }\OtherTok{:}\NormalTok{ r ⇆ s}\OtherTok{\}}
         \OtherTok{→}\NormalTok{ isCartesian q f }\OtherTok{→}\NormalTok{ isCartesian s g}
         \OtherTok{→}\NormalTok{ isCartesian }\OtherTok{(}\NormalTok{q ◃ s}\OtherTok{)} \OtherTok{(}\NormalTok{f ◃◃[ s ] g}\OtherTok{)}
\NormalTok{◃◃Cart q }\OtherTok{\{}\NormalTok{f }\OtherTok{=} \OtherTok{(}\NormalTok{f , f♯}\OtherTok{)\}}\NormalTok{ s }\OtherTok{\{}\NormalTok{g }\OtherTok{=} \OtherTok{(}\NormalTok{g , g♯}\OtherTok{)\}}\NormalTok{ cf cg }\OtherTok{(}\NormalTok{a , γ}\OtherTok{)} \OtherTok{=} 
\NormalTok{    pairEquiv }\OtherTok{(}\NormalTok{f♯ a}\OtherTok{)} \OtherTok{(λ}\NormalTok{ x }\OtherTok{→}\NormalTok{ g♯ }\OtherTok{(}\NormalTok{γ }\OtherTok{(}\NormalTok{f♯ a x}\OtherTok{)))} 
              \OtherTok{(}\NormalTok{cf a}\OtherTok{)} \OtherTok{(λ}\NormalTok{ x }\OtherTok{→}\NormalTok{ cg }\OtherTok{(}\NormalTok{γ }\OtherTok{(}\NormalTok{f♯ a x}\OtherTok{)))}
\end{Highlighting}
\end{Shaded}

\begin{Shaded}
\begin{Highlighting}[]
\KeywordTok{module}\NormalTok{ ◃AssocCart }\OtherTok{\{}\NormalTok{ℓ0 ℓ1 ℓ2 κ0 κ1 κ2}\OtherTok{\}} \OtherTok{(}\NormalTok{p }\OtherTok{:}\NormalTok{ Poly ℓ0 κ0}\OtherTok{)} 
                  \OtherTok{(}\NormalTok{q }\OtherTok{:}\NormalTok{ Poly ℓ1 κ1}\OtherTok{)} \OtherTok{(}\NormalTok{r }\OtherTok{:}\NormalTok{ Poly ℓ2 κ2}\OtherTok{)} \KeywordTok{where}

\NormalTok{    ◃assocCart }\OtherTok{:}\NormalTok{ isCartesian }\OtherTok{(}\NormalTok{p ◃ }\OtherTok{(}\NormalTok{q ◃ r}\OtherTok{))} \OtherTok{(}\NormalTok{◃assoc p q r}\OtherTok{)}
\NormalTok{    ◃assocCart }\OtherTok{\_} \OtherTok{=} 
\NormalTok{        Iso→isEquiv }\OtherTok{(}\NormalTok{snd }\OtherTok{(}\NormalTok{◃assoc⁻¹ p q r}\OtherTok{)} \OtherTok{\_}\NormalTok{ , }\OtherTok{((λ} \OtherTok{\_} \OtherTok{→}\NormalTok{ refl}\OtherTok{)}\NormalTok{ , }\OtherTok{(λ} \OtherTok{\_} \OtherTok{→}\NormalTok{ refl}\OtherTok{)))}
    
\NormalTok{    ◃assoc⁻¹Cart }\OtherTok{:}\NormalTok{ isCartesian }\OtherTok{((}\NormalTok{p ◃ q}\OtherTok{)}\NormalTok{ ◃ r}\OtherTok{)} \OtherTok{(}\NormalTok{◃assoc⁻¹ p q r}\OtherTok{)}
\NormalTok{    ◃assoc⁻¹Cart }\OtherTok{\_} \OtherTok{=} 
\NormalTok{        Iso→isEquiv }\OtherTok{(}\NormalTok{snd }\OtherTok{(}\NormalTok{◃assoc p q r}\OtherTok{)} \OtherTok{\_}\NormalTok{ , }\OtherTok{((λ} \OtherTok{\_} \OtherTok{→}\NormalTok{ refl}\OtherTok{)}\NormalTok{ , }\OtherTok{(λ} \OtherTok{\_} \OtherTok{→}\NormalTok{ refl}\OtherTok{)))}

\KeywordTok{open}\NormalTok{ ◃AssocCart }\KeywordTok{public}

\KeywordTok{module}\NormalTok{ ◃LRUnitCart }\OtherTok{\{}\NormalTok{ℓ κ}\OtherTok{\}} \OtherTok{(}\NormalTok{p }\OtherTok{:}\NormalTok{ Poly ℓ κ}\OtherTok{)} \KeywordTok{where}

\NormalTok{    ◃unitlCart }\OtherTok{:}\NormalTok{ isCartesian p }\OtherTok{(}\NormalTok{◃unitl p}\OtherTok{)}
\NormalTok{    ◃unitlCart }\OtherTok{\_} \OtherTok{=}\NormalTok{ Iso→isEquiv }\OtherTok{(}\NormalTok{snd }\OtherTok{(}\NormalTok{◃unitl⁻¹ p}\OtherTok{)} \OtherTok{\_}\NormalTok{ , }\OtherTok{((λ} \OtherTok{\_} \OtherTok{→}\NormalTok{ refl}\OtherTok{)}\NormalTok{ , }\OtherTok{(λ} \OtherTok{\_} \OtherTok{→}\NormalTok{ refl}\OtherTok{)))}

\NormalTok{    ◃unitl⁻¹Cart }\OtherTok{:}\NormalTok{ isCartesian }\OtherTok{(}\NormalTok{𝕪 ◃ p}\OtherTok{)} \OtherTok{(}\NormalTok{◃unitl⁻¹ p}\OtherTok{)}
\NormalTok{    ◃unitl⁻¹Cart }\OtherTok{\_} \OtherTok{=}\NormalTok{ Iso→isEquiv }\OtherTok{(}\NormalTok{snd }\OtherTok{(}\NormalTok{◃unitl p}\OtherTok{)} \OtherTok{\_}\NormalTok{ , }\OtherTok{((λ} \OtherTok{\_} \OtherTok{→}\NormalTok{ refl}\OtherTok{)}\NormalTok{ , }\OtherTok{(λ} \OtherTok{\_} \OtherTok{→}\NormalTok{ refl}\OtherTok{)))}

\NormalTok{    ◃unitrCart }\OtherTok{:}\NormalTok{ isCartesian p }\OtherTok{(}\NormalTok{◃unitr p}\OtherTok{)}
\NormalTok{    ◃unitrCart }\OtherTok{\_} \OtherTok{=}\NormalTok{ Iso→isEquiv }\OtherTok{(}\NormalTok{snd }\OtherTok{(}\NormalTok{◃unitr⁻¹ p}\OtherTok{)} \OtherTok{\_}\NormalTok{ , }\OtherTok{((λ} \OtherTok{\_} \OtherTok{→}\NormalTok{ refl}\OtherTok{)}\NormalTok{ , }\OtherTok{(λ} \OtherTok{\_} \OtherTok{→}\NormalTok{ refl}\OtherTok{)))}

\NormalTok{    ◃unitr⁻¹Cart }\OtherTok{:}\NormalTok{ isCartesian }\OtherTok{(}\NormalTok{p ◃ 𝕪}\OtherTok{)} \OtherTok{(}\NormalTok{◃unitr⁻¹ p}\OtherTok{)}
\NormalTok{    ◃unitr⁻¹Cart }\OtherTok{\_} \OtherTok{=}\NormalTok{ Iso→isEquiv }\OtherTok{(}\NormalTok{snd }\OtherTok{(}\NormalTok{◃unitr p}\OtherTok{)} \OtherTok{\_}\NormalTok{ , }\OtherTok{((λ} \OtherTok{\_} \OtherTok{→}\NormalTok{ refl}\OtherTok{)}\NormalTok{ , }\OtherTok{(λ} \OtherTok{\_} \OtherTok{→}\NormalTok{ refl}\OtherTok{)))}

\KeywordTok{open}\NormalTok{ ◃LRUnitCart }\KeywordTok{public}
\end{Highlighting}
\end{Shaded}

Proof of Theorem 4.1:

\begin{Shaded}
\begin{Highlighting}[]
\KeywordTok{module}\NormalTok{ PolyMonad }\OtherTok{\{}\NormalTok{ℓ κ}\OtherTok{\}} \OtherTok{(}\NormalTok{u }\OtherTok{:}\NormalTok{ Poly ℓ κ}\OtherTok{)} \OtherTok{(}\NormalTok{univ }\OtherTok{:}\NormalTok{ isUnivalent u}\OtherTok{)}
                 \OtherTok{(}\NormalTok{η }\OtherTok{:}\NormalTok{ 𝕪 ⇆ u}\OtherTok{)} \OtherTok{(}\NormalTok{ηcart }\OtherTok{:}\NormalTok{ isCartesian u η}\OtherTok{)}
                 \OtherTok{(}\NormalTok{μ }\OtherTok{:} \OtherTok{(}\NormalTok{u ◃ u}\OtherTok{)}\NormalTok{ ⇆ u}\OtherTok{)} \OtherTok{(}\NormalTok{μcart }\OtherTok{:}\NormalTok{ isCartesian u μ}\OtherTok{)} \KeywordTok{where}

\NormalTok{    idl }\OtherTok{:}\NormalTok{ EqLens u }\OtherTok{(}\NormalTok{◃unitl u}\OtherTok{)} \OtherTok{(}\NormalTok{comp u }\OtherTok{(}\NormalTok{η ◃◃[ u ] }\OtherTok{(}\NormalTok{id u}\OtherTok{))}\NormalTok{ μ}\OtherTok{)}
\NormalTok{    idl }\OtherTok{=}\NormalTok{ univ }\OtherTok{(}\NormalTok{◃unitlCart u}\OtherTok{)} \OtherTok{(}\NormalTok{compCartesian u }\OtherTok{(}\NormalTok{◃◃Cart u u ηcart }\OtherTok{(}\NormalTok{idCart u}\OtherTok{))}\NormalTok{ μcart}\OtherTok{)}

\NormalTok{    idr }\OtherTok{:}\NormalTok{ EqLens u }\OtherTok{(}\NormalTok{◃unitr u}\OtherTok{)} \OtherTok{(}\NormalTok{comp u }\OtherTok{(}\NormalTok{id u ◃◃[ u ] η}\OtherTok{)}\NormalTok{ μ}\OtherTok{)}
\NormalTok{    idr }\OtherTok{=}\NormalTok{ univ }\OtherTok{(}\NormalTok{◃unitrCart u}\OtherTok{)} \OtherTok{(}\NormalTok{compCartesian u }\OtherTok{(}\NormalTok{◃◃Cart u u }\OtherTok{(}\NormalTok{idCart u}\OtherTok{)}\NormalTok{ ηcart}\OtherTok{)}\NormalTok{ μcart}\OtherTok{)}

\NormalTok{    assoc }\OtherTok{:}\NormalTok{ EqLens u }\OtherTok{(}\NormalTok{comp u }\OtherTok{(}\NormalTok{◃assoc u u u}\OtherTok{)} \OtherTok{(}\NormalTok{comp u }\OtherTok{((}\NormalTok{id u}\OtherTok{)}\NormalTok{ ◃◃[ u ] μ}\OtherTok{)}\NormalTok{ μ}\OtherTok{))} 
                     \OtherTok{(}\NormalTok{comp u }\OtherTok{(}\NormalTok{μ ◃◃[ u ] }\OtherTok{(}\NormalTok{id u}\OtherTok{))}\NormalTok{ μ}\OtherTok{)}
\NormalTok{    assoc }\OtherTok{=}\NormalTok{ univ }\OtherTok{(}\NormalTok{compCartesian u }\OtherTok{(}\NormalTok{◃assocCart u u u}\OtherTok{)} 
                                  \OtherTok{(}\NormalTok{compCartesian u }\OtherTok{(}\NormalTok{◃◃Cart u u }\OtherTok{(}\NormalTok{idCart u}\OtherTok{)}\NormalTok{ μcart}\OtherTok{)} 
\NormalTok{                                                   μcart}\OtherTok{))} 
                 \OtherTok{(}\NormalTok{compCartesian u }\OtherTok{(}\NormalTok{◃◃Cart u u μcart }\OtherTok{(}\NormalTok{idCart u}\OtherTok{))}\NormalTok{ μcart}\OtherTok{)} 

\KeywordTok{open}\NormalTok{ PolyMonad }\KeywordTok{public}
\end{Highlighting}
\end{Shaded}

\section{The $\upuparrows$ Functor and Distributive Laws in Agda}

This module sets up the necessary definitions and lemmas for the proof of Theorem 4.2.

\begin{Shaded}
\begin{Highlighting}[]
\PreprocessorTok{\{{-}\# OPTIONS {-}{-}without{-}K {-}{-}rewriting \#{-}\}}
\KeywordTok{module}\NormalTok{ prod }\KeywordTok{where}

\KeywordTok{open} \KeywordTok{import}\NormalTok{ Agda}\OtherTok{.}\NormalTok{Primitive}
\KeywordTok{open} \KeywordTok{import}\NormalTok{ Agda}\OtherTok{.}\NormalTok{Builtin}\OtherTok{.}\NormalTok{Sigma}
\KeywordTok{open} \KeywordTok{import}\NormalTok{ Agda}\OtherTok{.}\NormalTok{Builtin}\OtherTok{.}\NormalTok{Unit}
\KeywordTok{open} \KeywordTok{import}\NormalTok{ Agda}\OtherTok{.}\NormalTok{Builtin}\OtherTok{.}\NormalTok{Equality}
\KeywordTok{open} \KeywordTok{import}\NormalTok{ Agda}\OtherTok{.}\NormalTok{Builtin}\OtherTok{.}\NormalTok{Equality}\OtherTok{.}\NormalTok{Rewrite}
\KeywordTok{open} \KeywordTok{import}\NormalTok{ hott}
\KeywordTok{open} \KeywordTok{import}\NormalTok{ poly}
\KeywordTok{open} \KeywordTok{import}\NormalTok{ sum}
\end{Highlighting}
\end{Shaded}

Definition of the \(\upuparrows\) functor:

\begin{Shaded}
\begin{Highlighting}[]
\OtherTok{\_}\NormalTok{⇈}\OtherTok{\_} \OtherTok{:} \OtherTok{∀} \OtherTok{\{}\NormalTok{ℓ0 ℓ1 κ0 κ1}\OtherTok{\}} \OtherTok{→}\NormalTok{ Poly ℓ0 κ0 }\OtherTok{→}\NormalTok{ Poly ℓ1 κ1 }
      \OtherTok{→}\NormalTok{ Poly }\OtherTok{(}\NormalTok{ℓ0 ⊔ κ0 ⊔ ℓ1}\OtherTok{)} \OtherTok{(}\NormalTok{κ0 ⊔ κ1}\OtherTok{)}
\OtherTok{(}\NormalTok{A , B}\OtherTok{)}\NormalTok{ ⇈ }\OtherTok{(}\NormalTok{C , D}\OtherTok{)} \OtherTok{=} 
    \OtherTok{(}\NormalTok{ Σ A }\OtherTok{(λ}\NormalTok{ a }\OtherTok{→}\NormalTok{ B a }\OtherTok{→}\NormalTok{ C}\OtherTok{)} 
\NormalTok{    , }\OtherTok{(λ} \OtherTok{(}\NormalTok{a , f}\OtherTok{)} \OtherTok{→} \OtherTok{(}\NormalTok{b }\OtherTok{:}\NormalTok{ B a}\OtherTok{)} \OtherTok{→}\NormalTok{ D }\OtherTok{(}\NormalTok{f b}\OtherTok{)))}
\end{Highlighting}
\end{Shaded}

Functoriality of \(\upuparrows\):

\begin{Shaded}
\begin{Highlighting}[]
\NormalTok{⇈Lens }\OtherTok{:} \OtherTok{∀} \OtherTok{\{}\NormalTok{ℓ0 ℓ1 ℓ2 ℓ3 κ0 κ1 κ2 κ3}\OtherTok{\}}
        \OtherTok{→} \OtherTok{\{}\NormalTok{p }\OtherTok{:}\NormalTok{ Poly ℓ0 κ0}\OtherTok{\}} \OtherTok{(}\NormalTok{r }\OtherTok{:}\NormalTok{ Poly ℓ2 κ2}\OtherTok{)}
        \OtherTok{→} \OtherTok{\{}\NormalTok{q }\OtherTok{:}\NormalTok{ Poly ℓ1 κ1}\OtherTok{\}} \OtherTok{(}\NormalTok{s }\OtherTok{:}\NormalTok{ Poly ℓ3 κ3}\OtherTok{)}
        \OtherTok{→} \OtherTok{(}\NormalTok{f }\OtherTok{:}\NormalTok{ p ⇆ r}\OtherTok{)} \OtherTok{(}\NormalTok{f\textquotesingle{} }\OtherTok{:}\NormalTok{ r ⇆ p}\OtherTok{)} 
        \OtherTok{→}\NormalTok{ EqLens p }\OtherTok{(}\NormalTok{id p}\OtherTok{)} \OtherTok{(}\NormalTok{comp p f f\textquotesingle{}}\OtherTok{)}
        \OtherTok{→} \OtherTok{(}\NormalTok{g }\OtherTok{:}\NormalTok{ q ⇆ s}\OtherTok{)} \OtherTok{→} \OtherTok{(}\NormalTok{p ⇈ q}\OtherTok{)}\NormalTok{ ⇆ }\OtherTok{(}\NormalTok{r ⇈ s}\OtherTok{)}
\NormalTok{⇈Lens }\OtherTok{\{}\NormalTok{p }\OtherTok{=}\NormalTok{ p}\OtherTok{\}}\NormalTok{ r s }\OtherTok{(}\NormalTok{f , f♯}\OtherTok{)} \OtherTok{(}\NormalTok{f\textquotesingle{} , f\textquotesingle{}♯}\OtherTok{)} \OtherTok{(}\NormalTok{e , e♯}\OtherTok{)} \OtherTok{(}\NormalTok{g , g♯}\OtherTok{)} \OtherTok{=} 
    \OtherTok{(} \OtherTok{(λ} \OtherTok{(}\NormalTok{a , γ}\OtherTok{)} \OtherTok{→} \OtherTok{(}\NormalTok{f a , }\OtherTok{(λ}\NormalTok{ x }\OtherTok{→}\NormalTok{ g }\OtherTok{(}\NormalTok{γ }\OtherTok{(}\NormalTok{f♯ a x}\OtherTok{)))))}
\NormalTok{    , }\OtherTok{(λ} \OtherTok{(}\NormalTok{a , γ}\OtherTok{)}\NormalTok{ Ϝ x }\OtherTok{→} 
\NormalTok{         g♯ }\OtherTok{(}\NormalTok{γ x}\OtherTok{)} 
            \OtherTok{(}\NormalTok{transp }\OtherTok{(λ}\NormalTok{ y }\OtherTok{→}\NormalTok{ snd s }\OtherTok{(}\NormalTok{g }\OtherTok{(}\NormalTok{γ y}\OtherTok{)))} 
                    \OtherTok{(}\NormalTok{sym }\OtherTok{(}\NormalTok{e♯ a x}\OtherTok{))} 
                    \OtherTok{(}\NormalTok{Ϝ }\OtherTok{(}\NormalTok{f\textquotesingle{}♯ }\OtherTok{(}\NormalTok{f a}\OtherTok{)} \OtherTok{(}\NormalTok{transp }\OtherTok{(}\NormalTok{snd p}\OtherTok{)} \OtherTok{(}\NormalTok{e a}\OtherTok{)}\NormalTok{ x}\OtherTok{)))))} \OtherTok{)}
\end{Highlighting}
\end{Shaded}

Interaction of \(\upuparrows\) with \texttt{◃} in its first argument,
exhibiting \(\upuparrows\) as a monoidal action, which moreover descends
to a monoidal action on \(\mathbf{Poly}^\mathbf{Cart}\):

\begin{Shaded}
\begin{Highlighting}[]
\KeywordTok{module}\NormalTok{ Unit⇈ }\OtherTok{\{}\NormalTok{ℓ κ}\OtherTok{\}} \OtherTok{(}\NormalTok{p }\OtherTok{:}\NormalTok{ Poly ℓ κ}\OtherTok{)} \KeywordTok{where}

\NormalTok{    𝕪⇈ }\OtherTok{:} \OtherTok{(}\NormalTok{𝕪 ⇈ p}\OtherTok{)}\NormalTok{ ⇆ p}
\NormalTok{    𝕪⇈ }\OtherTok{=} \OtherTok{(} \OtherTok{(λ} \OtherTok{(\_}\NormalTok{ , a}\OtherTok{)} \OtherTok{→}\NormalTok{ a tt}\OtherTok{)}\NormalTok{ , }\OtherTok{λ} \OtherTok{(\_}\NormalTok{ , a}\OtherTok{)}\NormalTok{ b tt }\OtherTok{→}\NormalTok{ b}\OtherTok{)}

\NormalTok{    𝕪⇈Cart }\OtherTok{:}\NormalTok{ isCartesian p 𝕪⇈}
\NormalTok{    𝕪⇈Cart }\OtherTok{(\_}\NormalTok{ , x}\OtherTok{)} \OtherTok{=} 
\NormalTok{        Iso→isEquiv }\OtherTok{(} \OtherTok{(λ}\NormalTok{ Ϝ }\OtherTok{→}\NormalTok{ Ϝ tt}\OtherTok{)} 
\NormalTok{                    , }\OtherTok{(} \OtherTok{(λ}\NormalTok{ a }\OtherTok{→}\NormalTok{ refl}\OtherTok{)} 
\NormalTok{                      , }\OtherTok{λ}\NormalTok{ b }\OtherTok{→}\NormalTok{ refl}\OtherTok{))}

\KeywordTok{open}\NormalTok{ Unit⇈ }\KeywordTok{public}

\KeywordTok{module}\NormalTok{ ◃⇈ }\OtherTok{\{}\NormalTok{ℓ0 ℓ1 ℓ2 κ0 κ1 κ2}\OtherTok{\}} \OtherTok{(}\NormalTok{p }\OtherTok{:}\NormalTok{ Poly ℓ0 κ0}\OtherTok{)} 
          \OtherTok{(}\NormalTok{q }\OtherTok{:}\NormalTok{ Poly ℓ1 κ1}\OtherTok{)} \OtherTok{(}\NormalTok{r }\OtherTok{:}\NormalTok{ Poly ℓ2 κ2}\OtherTok{)} \KeywordTok{where}

\NormalTok{    ⇈Curry }\OtherTok{:} \OtherTok{((}\NormalTok{p ◃ q}\OtherTok{)}\NormalTok{ ⇈ r}\OtherTok{)}\NormalTok{ ⇆ }\OtherTok{(}\NormalTok{p ⇈ }\OtherTok{(}\NormalTok{q ⇈ r}\OtherTok{))}
\NormalTok{    ⇈Curry }\OtherTok{=} \OtherTok{(} \OtherTok{(λ} \OtherTok{((}\NormalTok{a , h}\OtherTok{)}\NormalTok{ , k}\OtherTok{)} 
                  \OtherTok{→} \OtherTok{(}\NormalTok{a , }\OtherTok{(λ}\NormalTok{ b }\OtherTok{→} \OtherTok{(} \OtherTok{(}\NormalTok{h b}\OtherTok{)} 
\NormalTok{                                , }\OtherTok{(λ}\NormalTok{ d }\OtherTok{→}\NormalTok{ k }\OtherTok{(}\NormalTok{b , d}\OtherTok{))))))}
\NormalTok{             , }\OtherTok{(λ} \OtherTok{((}\NormalTok{a , h}\OtherTok{)}\NormalTok{ , k}\OtherTok{)}\NormalTok{ f }\OtherTok{(}\NormalTok{b , d}\OtherTok{)} \OtherTok{→}\NormalTok{ f b d}\OtherTok{)} \OtherTok{)}
    
\NormalTok{    ⇈CurryCart }\OtherTok{:}\NormalTok{ isCartesian }\OtherTok{(}\NormalTok{p ⇈ }\OtherTok{(}\NormalTok{q ⇈ r}\OtherTok{))}\NormalTok{ ⇈Curry}
\NormalTok{    ⇈CurryCart }\OtherTok{((}\NormalTok{a , h}\OtherTok{)}\NormalTok{ , k}\OtherTok{)} \OtherTok{=} 
\NormalTok{        Iso→isEquiv }\OtherTok{(} \OtherTok{(λ}\NormalTok{ f b d }\OtherTok{→}\NormalTok{ f }\OtherTok{(}\NormalTok{b , d}\OtherTok{))} 
\NormalTok{                    , }\OtherTok{(} \OtherTok{(λ}\NormalTok{ f }\OtherTok{→}\NormalTok{ refl}\OtherTok{)}
\NormalTok{                      , }\OtherTok{(λ}\NormalTok{ f }\OtherTok{→}\NormalTok{ refl}\OtherTok{)} \OtherTok{)} \OtherTok{)}

\KeywordTok{open}\NormalTok{ ◃⇈ }\KeywordTok{public}
\end{Highlighting}
\end{Shaded}

Interaction of \(\upuparrows\) with \texttt{◃} in its second argument,
exhibiting \(\upuparrows\) as a colax monoidal functor, which moreover
descends to a colax monoidal functor on \(\mathbf{Poly}^\mathbf{Cart}\):

\begin{Shaded}
\begin{Highlighting}[]
\KeywordTok{module}\NormalTok{ ⇈Unit }\OtherTok{\{}\NormalTok{ℓ κ}\OtherTok{\}} \OtherTok{(}\NormalTok{p }\OtherTok{:}\NormalTok{ Poly ℓ κ}\OtherTok{)} \KeywordTok{where}

\NormalTok{    ⇈𝕪 }\OtherTok{:} \OtherTok{(}\NormalTok{p ⇈ 𝕪}\OtherTok{)}\NormalTok{ ⇆ 𝕪}
\NormalTok{    ⇈𝕪 }\OtherTok{=} \OtherTok{(} \OtherTok{(λ} \OtherTok{(}\NormalTok{a , γ}\OtherTok{)} \OtherTok{→}\NormalTok{ tt}\OtherTok{)}\NormalTok{ , }\OtherTok{λ} \OtherTok{(}\NormalTok{a , γ}\OtherTok{)}\NormalTok{ tt b }\OtherTok{→}\NormalTok{ tt }\OtherTok{)}

\NormalTok{    ⇈𝕪Cart }\OtherTok{:}\NormalTok{ isCartesian 𝕪 ⇈𝕪}
\NormalTok{    ⇈𝕪Cart }\OtherTok{(}\NormalTok{x , γ}\OtherTok{)} \OtherTok{=} 
\NormalTok{        Iso→isEquiv }\OtherTok{(} \OtherTok{(λ}\NormalTok{ x }\OtherTok{→}\NormalTok{ tt}\OtherTok{)} 
\NormalTok{                    , }\OtherTok{(} \OtherTok{(λ}\NormalTok{ a }\OtherTok{→}\NormalTok{ refl}\OtherTok{)} 
\NormalTok{                      , }\OtherTok{λ}\NormalTok{ b }\OtherTok{→}\NormalTok{ refl}\OtherTok{))}

\KeywordTok{open}\NormalTok{ ⇈Unit }\KeywordTok{public}

\KeywordTok{module}\NormalTok{ ⇈◃ }\OtherTok{\{}\NormalTok{ℓ0 ℓ1 ℓ2 κ0 κ1 κ2}\OtherTok{\}} \OtherTok{(}\NormalTok{p }\OtherTok{:}\NormalTok{ Poly ℓ0 κ0}\OtherTok{)} 
          \OtherTok{(}\NormalTok{q }\OtherTok{:}\NormalTok{ Poly ℓ1 κ1}\OtherTok{)} \OtherTok{(}\NormalTok{r }\OtherTok{:}\NormalTok{ Poly ℓ2 κ2}\OtherTok{)} \KeywordTok{where}

\NormalTok{    ⇈Distr }\OtherTok{:} \OtherTok{(}\NormalTok{p ⇈ }\OtherTok{(}\NormalTok{q ◃ r}\OtherTok{))}\NormalTok{ ⇆ }\OtherTok{((}\NormalTok{p ⇈ q}\OtherTok{)}\NormalTok{ ◃ }\OtherTok{(}\NormalTok{p ⇈ r}\OtherTok{))}
\NormalTok{    ⇈Distr }\OtherTok{=} \OtherTok{(} \OtherTok{(λ} \OtherTok{(}\NormalTok{a , h}\OtherTok{)} 
                  \OtherTok{→} \OtherTok{(} \OtherTok{(}\NormalTok{a , }\OtherTok{(λ}\NormalTok{ b }\OtherTok{→}\NormalTok{ fst }\OtherTok{(}\NormalTok{h b}\OtherTok{)))} 
\NormalTok{                    , }\OtherTok{λ}\NormalTok{ f }\OtherTok{→} \OtherTok{(}\NormalTok{a , }\OtherTok{(λ}\NormalTok{ b }\OtherTok{→}\NormalTok{ snd }\OtherTok{(}\NormalTok{h b}\OtherTok{)} \OtherTok{(}\NormalTok{f b}\OtherTok{)))} \OtherTok{))} 
\NormalTok{             , }\OtherTok{(λ} \OtherTok{(}\NormalTok{a , h}\OtherTok{)} \OtherTok{(}\NormalTok{f , g}\OtherTok{)}\NormalTok{ b }\OtherTok{→} \OtherTok{(}\NormalTok{f b , g b}\OtherTok{))} \OtherTok{)}
    
\NormalTok{    ⇈DistrCart }\OtherTok{:}\NormalTok{ isCartesian }\OtherTok{((}\NormalTok{p ⇈ q}\OtherTok{)}\NormalTok{ ◃ }\OtherTok{(}\NormalTok{p ⇈ r}\OtherTok{))}\NormalTok{ ⇈Distr}
\NormalTok{    ⇈DistrCart }\OtherTok{(}\NormalTok{a , h}\OtherTok{)} \OtherTok{=}
\NormalTok{        Iso→isEquiv }\OtherTok{(} \OtherTok{(λ}\NormalTok{ f }\OtherTok{→} \OtherTok{(} \OtherTok{(λ}\NormalTok{ b }\OtherTok{→}\NormalTok{ fst }\OtherTok{(}\NormalTok{f b}\OtherTok{))} 
\NormalTok{                             , }\OtherTok{(λ}\NormalTok{ b }\OtherTok{→}\NormalTok{ snd }\OtherTok{(}\NormalTok{f b}\OtherTok{))} \OtherTok{))}
\NormalTok{                    , }\OtherTok{(} \OtherTok{(λ} \OtherTok{(}\NormalTok{f , g}\OtherTok{)} \OtherTok{→}\NormalTok{ refl}\OtherTok{)} 
\NormalTok{                      , }\OtherTok{(λ}\NormalTok{ f }\OtherTok{→}\NormalTok{ refl}\OtherTok{)} \OtherTok{)} \OtherTok{)}

\KeywordTok{open}\NormalTok{ ⇈◃ }\KeywordTok{public}
\end{Highlighting}
\end{Shaded}

The putative distributive law induced by \(\upuparrows\):

\begin{Shaded}
\begin{Highlighting}[]
\NormalTok{distrLaw? }\OtherTok{:} \OtherTok{∀} \OtherTok{\{}\NormalTok{ℓ κ}\OtherTok{\}} \OtherTok{(}\NormalTok{u }\OtherTok{:}\NormalTok{ Poly ℓ κ}\OtherTok{)} \OtherTok{→} \OtherTok{(}\NormalTok{u ⇈ u}\OtherTok{)}\NormalTok{ ⇆ u}
            \OtherTok{→} \OtherTok{(}\NormalTok{u ◃ u}\OtherTok{)}\NormalTok{ ⇆ }\OtherTok{(}\NormalTok{u ◃ u}\OtherTok{)}
\NormalTok{distrLaw? u }\OtherTok{(}\NormalTok{π , π♯}\OtherTok{)} \OtherTok{=} 
    \OtherTok{(} \OtherTok{(λ} \OtherTok{(}\NormalTok{a , b}\OtherTok{)} \OtherTok{→}\NormalTok{ π }\OtherTok{(}\NormalTok{a , b}\OtherTok{)}\NormalTok{ , }\OtherTok{(λ}\NormalTok{ x }\OtherTok{→}\NormalTok{ a}\OtherTok{))} 
\NormalTok{    , }\OtherTok{λ} \OtherTok{(}\NormalTok{a , b}\OtherTok{)} \OtherTok{(}\NormalTok{f , x}\OtherTok{)} \OtherTok{→} \OtherTok{(}\NormalTok{x , }\OtherTok{(}\NormalTok{π♯ }\OtherTok{((}\NormalTok{a , b}\OtherTok{))}\NormalTok{ f x}\OtherTok{))} \OtherTok{)}
\end{Highlighting}
\end{Shaded}

Generalizing \(\upuparrows\) to an action of the twisted arrow category
of \(\mathbf{Poly}\) on \(\mathbf{Poly}\) and
\(\mathbf{Poly}^\mathbf{Cart}\):

\begin{Shaded}
\begin{Highlighting}[]
\OtherTok{\_}\NormalTok{⇈[}\OtherTok{\_}\NormalTok{][}\OtherTok{\_}\NormalTok{]}\OtherTok{\_} \OtherTok{:} \OtherTok{∀} \OtherTok{\{}\NormalTok{ℓ ℓ\textquotesingle{} ℓ\textquotesingle{}\textquotesingle{} κ κ\textquotesingle{} κ\textquotesingle{}\textquotesingle{}}\OtherTok{\}}
            \OtherTok{→} \OtherTok{(}\NormalTok{p }\OtherTok{:}\NormalTok{ Poly ℓ κ}\OtherTok{)} \OtherTok{(}\NormalTok{q }\OtherTok{:}\NormalTok{ Poly ℓ\textquotesingle{} κ\textquotesingle{}}\OtherTok{)}
            \OtherTok{→} \OtherTok{(}\NormalTok{p ⇆ q}\OtherTok{)} \OtherTok{→} \OtherTok{(}\NormalTok{r }\OtherTok{:}\NormalTok{ Poly ℓ\textquotesingle{}\textquotesingle{} κ\textquotesingle{}\textquotesingle{}}\OtherTok{)}
            \OtherTok{→}\NormalTok{ Poly }\OtherTok{(}\NormalTok{ℓ ⊔ κ ⊔ ℓ\textquotesingle{}\textquotesingle{}}\OtherTok{)} \OtherTok{(}\NormalTok{κ\textquotesingle{} ⊔ κ\textquotesingle{}\textquotesingle{}}\OtherTok{)}
\OtherTok{(}\NormalTok{A , B}\OtherTok{)}\NormalTok{ ⇈[ }\OtherTok{(}\NormalTok{C , D}\OtherTok{)}\NormalTok{ ][ }\OtherTok{(}\NormalTok{f , f♯}\OtherTok{)}\NormalTok{ ] }\OtherTok{(}\NormalTok{E , F}\OtherTok{)} \OtherTok{=}
   \OtherTok{(} \OtherTok{(}\NormalTok{Σ A }\OtherTok{(λ}\NormalTok{ a }\OtherTok{→}\NormalTok{ B a }\OtherTok{→}\NormalTok{ E}\OtherTok{))} 
\NormalTok{   , }\OtherTok{(λ} \OtherTok{(}\NormalTok{a , ε}\OtherTok{)} \OtherTok{→} \OtherTok{(}\NormalTok{d }\OtherTok{:}\NormalTok{ D }\OtherTok{(}\NormalTok{f a}\OtherTok{))} \OtherTok{→}\NormalTok{ F }\OtherTok{(}\NormalTok{ε }\OtherTok{(}\NormalTok{f♯ a d}\OtherTok{))))}

\KeywordTok{module}\NormalTok{ ⇈[]Functor }\OtherTok{\{}\NormalTok{ℓ0 ℓ1 ℓ2 ℓ3 ℓ4 ℓ5 κ0 κ1 κ2 κ3 κ4 κ5}\OtherTok{\}}
          \OtherTok{\{}\NormalTok{p }\OtherTok{:}\NormalTok{ Poly ℓ0 κ0}\OtherTok{\}} \OtherTok{\{}\NormalTok{p\textquotesingle{} }\OtherTok{:}\NormalTok{ Poly ℓ3 κ3}\OtherTok{\}}
          \OtherTok{(}\NormalTok{q }\OtherTok{:}\NormalTok{ Poly ℓ1 κ1}\OtherTok{)} \OtherTok{\{}\NormalTok{q\textquotesingle{} }\OtherTok{:}\NormalTok{ Poly ℓ4 κ4}\OtherTok{\}}
          \OtherTok{\{}\NormalTok{r }\OtherTok{:}\NormalTok{ Poly ℓ2 κ2}\OtherTok{\}} \OtherTok{(}\NormalTok{r\textquotesingle{} }\OtherTok{:}\NormalTok{ Poly ℓ5 κ5}\OtherTok{)}
          \OtherTok{(}\NormalTok{f }\OtherTok{:}\NormalTok{ p ⇆ q}\OtherTok{)} \OtherTok{(}\NormalTok{f\textquotesingle{} }\OtherTok{:}\NormalTok{ p\textquotesingle{} ⇆ q\textquotesingle{}}\OtherTok{)}
          \OtherTok{(}\NormalTok{g }\OtherTok{:}\NormalTok{ p ⇆ p\textquotesingle{}}\OtherTok{)} \OtherTok{(}\NormalTok{h }\OtherTok{:}\NormalTok{ q\textquotesingle{} ⇆ q}\OtherTok{)} \OtherTok{(}\NormalTok{k }\OtherTok{:}\NormalTok{ r ⇆ r\textquotesingle{}}\OtherTok{)}
          \OtherTok{(}\NormalTok{e }\OtherTok{:}\NormalTok{ EqLens q f }\OtherTok{(}\NormalTok{comp q g }\OtherTok{(}\NormalTok{comp q f\textquotesingle{} h}\OtherTok{)))} \KeywordTok{where}

\NormalTok{    ⇈[]Lens }\OtherTok{:} \OtherTok{(}\NormalTok{p ⇈[ q ][ f ] r}\OtherTok{)}\NormalTok{ ⇆ }\OtherTok{(}\NormalTok{p\textquotesingle{} ⇈[ q\textquotesingle{} ][ f\textquotesingle{} ] r\textquotesingle{}}\OtherTok{)}
\NormalTok{    ⇈[]Lens }\OtherTok{=} 
        \OtherTok{(} \OtherTok{(λ} \OtherTok{(}\NormalTok{a , γ}\OtherTok{)} \OtherTok{→} \OtherTok{(}\NormalTok{fst g a , }\OtherTok{λ}\NormalTok{ x }\OtherTok{→}\NormalTok{ fst k }\OtherTok{(}\NormalTok{γ }\OtherTok{(}\NormalTok{snd g a x}\OtherTok{))))} 
\NormalTok{        , }\OtherTok{λ} \OtherTok{(}\NormalTok{a , γ}\OtherTok{)}\NormalTok{ Ϝ x }\OtherTok{→}
\NormalTok{            snd k }\OtherTok{(}\NormalTok{γ }\OtherTok{(}\NormalTok{snd f a x}\OtherTok{))} 
               \OtherTok{(}\NormalTok{transp }\OtherTok{(λ}\NormalTok{ y }\OtherTok{→}\NormalTok{ snd r\textquotesingle{} }\OtherTok{(}\NormalTok{fst k }\OtherTok{(}\NormalTok{γ y}\OtherTok{)))} 
                       \OtherTok{(}\NormalTok{sym }\OtherTok{(}\NormalTok{snd e a x}\OtherTok{))} 
                       \OtherTok{(}\NormalTok{Ϝ }\OtherTok{(}\NormalTok{snd h }\OtherTok{(}\NormalTok{fst f\textquotesingle{} }\OtherTok{(}\NormalTok{fst g a}\OtherTok{))} 
                              \OtherTok{(}\NormalTok{transp }\OtherTok{(}\NormalTok{snd q}\OtherTok{)} \OtherTok{(}\NormalTok{fst e a}\OtherTok{)}\NormalTok{ x}\OtherTok{))))} \OtherTok{)}
\end{Highlighting}
\end{Shaded}

\begin{Shaded}
\begin{Highlighting}[]
\NormalTok{    ⇈[]LensCart }\OtherTok{:}\NormalTok{ isCartesian q h }\OtherTok{→}\NormalTok{ isCartesian r\textquotesingle{} k}
                  \OtherTok{→}\NormalTok{ isCartesian }\OtherTok{(}\NormalTok{p\textquotesingle{} ⇈[ q\textquotesingle{} ][ f\textquotesingle{} ] r\textquotesingle{}}\OtherTok{)}\NormalTok{ ⇈[]Lens}
\NormalTok{    ⇈[]LensCart ch ck }\OtherTok{(}\NormalTok{a , γ}\OtherTok{)} \OtherTok{=} 
\NormalTok{        compIsEquiv }
            \OtherTok{(}\NormalTok{PostCompEquiv }\OtherTok{(λ}\NormalTok{ x }\OtherTok{→}\NormalTok{ snd k }\OtherTok{(}\NormalTok{γ }\OtherTok{(}\NormalTok{snd f a x}\OtherTok{)))} 
                           \OtherTok{(λ}\NormalTok{ x }\OtherTok{→}\NormalTok{ ck }\OtherTok{(}\NormalTok{γ }\OtherTok{(}\NormalTok{snd f a x}\OtherTok{))))} 
            \OtherTok{(}\NormalTok{compIsEquiv }
                \OtherTok{(}\NormalTok{PostCompEquiv }
                    \OtherTok{(λ}\NormalTok{ x }\OtherTok{→}\NormalTok{ transp }\OtherTok{(λ}\NormalTok{ y }\OtherTok{→}\NormalTok{ snd r\textquotesingle{} }\OtherTok{(}\NormalTok{fst k }\OtherTok{(}\NormalTok{γ y}\OtherTok{)))} 
                                  \OtherTok{(}\NormalTok{sym }\OtherTok{(}\NormalTok{snd e a x}\OtherTok{)))} 
                    \OtherTok{(λ}\NormalTok{ x }\OtherTok{→}\NormalTok{ transpIsEquiv }\OtherTok{(}\NormalTok{sym }\OtherTok{(}\NormalTok{snd e a x}\OtherTok{))))} 
                \OtherTok{(}\NormalTok{compIsEquiv }
                    \OtherTok{(}\NormalTok{PreCompEquiv }\OtherTok{(}\NormalTok{transp }\OtherTok{(}\NormalTok{snd q}\OtherTok{)} \OtherTok{(}\NormalTok{fst e a}\OtherTok{))} 
                                  \OtherTok{(}\NormalTok{transpIsEquiv }\OtherTok{(}\NormalTok{fst e a}\OtherTok{)))} 
                    \OtherTok{(}\NormalTok{PreCompEquiv }\OtherTok{(λ}\NormalTok{ x }\OtherTok{→}\NormalTok{ snd h }\OtherTok{(}\NormalTok{fst f\textquotesingle{} }\OtherTok{(}\NormalTok{fst g a}\OtherTok{))}\NormalTok{ x}\OtherTok{)} 
                                  \OtherTok{(}\NormalTok{ch }\OtherTok{(}\NormalTok{fst f\textquotesingle{} }\OtherTok{(}\NormalTok{fst g a}\OtherTok{))))))}

\KeywordTok{open}\NormalTok{ ⇈[]Functor }\KeywordTok{public}
\end{Highlighting}
\end{Shaded}

Extension of properties noted above for \(\upuparrows\) to the
generalized \(\upuparrows\):

\begin{Shaded}
\begin{Highlighting}[]
\NormalTok{𝕪⇈[] }\OtherTok{:} \OtherTok{∀} \OtherTok{\{}\NormalTok{ℓ κ}\OtherTok{\}} \OtherTok{(}\NormalTok{p }\OtherTok{:}\NormalTok{ Poly ℓ κ}\OtherTok{)} \OtherTok{→} \OtherTok{(}\NormalTok{𝕪 ⇈[ 𝕪 ][ }\OtherTok{(}\NormalTok{id 𝕪}\OtherTok{)}\NormalTok{ ] p}\OtherTok{)}\NormalTok{ ⇆ p}
\NormalTok{𝕪⇈[] p }\OtherTok{=} \OtherTok{((λ} \OtherTok{(\_}\NormalTok{ , γ}\OtherTok{)} \OtherTok{→}\NormalTok{ γ tt}\OtherTok{)}\NormalTok{ , }\OtherTok{λ} \OtherTok{(\_}\NormalTok{ , γ}\OtherTok{)}\NormalTok{ Ϝ }\OtherTok{\_} \OtherTok{→}\NormalTok{ Ϝ}\OtherTok{)}

\NormalTok{⇈[]Curry }\OtherTok{:} \OtherTok{∀} \OtherTok{\{}\NormalTok{ℓ0 ℓ1 ℓ2 ℓ3 ℓ4 κ0 κ1 κ2 κ3 κ4}\OtherTok{\}}
           \OtherTok{→} \OtherTok{(}\NormalTok{p }\OtherTok{:}\NormalTok{ Poly ℓ0 κ0}\OtherTok{)} \OtherTok{(}\NormalTok{q }\OtherTok{:}\NormalTok{ Poly ℓ1 κ1}\OtherTok{)} 
           \OtherTok{→} \OtherTok{(}\NormalTok{r }\OtherTok{:}\NormalTok{ Poly ℓ2 κ2}\OtherTok{)} \OtherTok{(}\NormalTok{s }\OtherTok{:}\NormalTok{ Poly ℓ3 κ3}\OtherTok{)}
           \OtherTok{→} \OtherTok{(}\NormalTok{t }\OtherTok{:}\NormalTok{ Poly ℓ4 κ4}\OtherTok{)}
           \OtherTok{→} \OtherTok{(}\NormalTok{f }\OtherTok{:}\NormalTok{ p ⇆ q}\OtherTok{)} \OtherTok{(}\NormalTok{g }\OtherTok{:}\NormalTok{ r ⇆ s}\OtherTok{)}
           \OtherTok{→} \OtherTok{((}\NormalTok{p ◃ r}\OtherTok{)}\NormalTok{ ⇈[ q ◃ s ][ f ◃◃[ s ] g ] t}\OtherTok{)} 
\NormalTok{             ⇆ }\OtherTok{(}\NormalTok{p ⇈[ q ][ f ] }\OtherTok{(}\NormalTok{r ⇈[ s ][ g ] t}\OtherTok{))}
\NormalTok{⇈[]Curry p q r s t f g }\OtherTok{=} 
    \OtherTok{(} \OtherTok{(λ} \OtherTok{((}\NormalTok{a , h}\OtherTok{)}\NormalTok{ , k}\OtherTok{)} \OtherTok{→}\NormalTok{ a , }\OtherTok{(λ}\NormalTok{ b }\OtherTok{→} \OtherTok{(}\NormalTok{h b}\OtherTok{)}\NormalTok{ , }\OtherTok{(λ}\NormalTok{ d }\OtherTok{→}\NormalTok{ k }\OtherTok{(}\NormalTok{b , d}\OtherTok{))))} 
\NormalTok{    , }\OtherTok{λ} \OtherTok{((}\NormalTok{a , h}\OtherTok{)}\NormalTok{ , k}\OtherTok{)}\NormalTok{ Ϝ }\OtherTok{(}\NormalTok{b , d}\OtherTok{)} \OtherTok{→}\NormalTok{ Ϝ b d}\OtherTok{)}
\end{Highlighting}
\end{Shaded}

\begin{Shaded}
\begin{Highlighting}[]
\NormalTok{⇈[]𝕪 }\OtherTok{:} \OtherTok{∀} \OtherTok{\{}\NormalTok{ℓ0 κ0 ℓ1 κ1}\OtherTok{\}} \OtherTok{(}\NormalTok{p }\OtherTok{:}\NormalTok{ Poly ℓ0 κ0}\OtherTok{)} \OtherTok{(}\NormalTok{q }\OtherTok{:}\NormalTok{ Poly ℓ1 κ1}\OtherTok{)} 
       \OtherTok{→} \OtherTok{(}\NormalTok{f }\OtherTok{:}\NormalTok{ p ⇆ q}\OtherTok{)} \OtherTok{→} \OtherTok{(}\NormalTok{p ⇈[ q ][ f ] 𝕪}\OtherTok{)}\NormalTok{ ⇆ 𝕪}
\NormalTok{⇈[]𝕪 p q f }\OtherTok{=} \OtherTok{((λ} \OtherTok{\_} \OtherTok{→}\NormalTok{ tt}\OtherTok{)}\NormalTok{ , }\OtherTok{λ} \OtherTok{\_} \OtherTok{\_} \OtherTok{\_} \OtherTok{→}\NormalTok{ tt}\OtherTok{)}
      

\NormalTok{⇈[]Distr }\OtherTok{:} \OtherTok{∀} \OtherTok{\{}\NormalTok{ℓ0 ℓ1 ℓ2 ℓ3 ℓ4 κ0 κ1 κ2 κ3 κ4}\OtherTok{\}}
           \OtherTok{→} \OtherTok{(}\NormalTok{p }\OtherTok{:}\NormalTok{ Poly ℓ0 κ0}\OtherTok{)} \OtherTok{(}\NormalTok{q }\OtherTok{:}\NormalTok{ Poly ℓ1 κ1}\OtherTok{)} \OtherTok{(}\NormalTok{r }\OtherTok{:}\NormalTok{ Poly ℓ2 κ2}\OtherTok{)}
           \OtherTok{→} \OtherTok{(}\NormalTok{s }\OtherTok{:}\NormalTok{ Poly ℓ3 κ3}\OtherTok{)} \OtherTok{(}\NormalTok{t }\OtherTok{:}\NormalTok{ Poly ℓ4 κ4}\OtherTok{)}
           \OtherTok{→} \OtherTok{(}\NormalTok{f }\OtherTok{:}\NormalTok{ p ⇆ q}\OtherTok{)} \OtherTok{(}\NormalTok{g }\OtherTok{:}\NormalTok{ q ⇆ r}\OtherTok{)}
           \OtherTok{→} \OtherTok{(}\NormalTok{p ⇈[ r ][ comp r f g ] }\OtherTok{(}\NormalTok{s ◃ t}\OtherTok{))} 
\NormalTok{             ⇆ }\OtherTok{((}\NormalTok{p ⇈[ q ][ f ] s}\OtherTok{)}\NormalTok{ ◃ }\OtherTok{(}\NormalTok{q ⇈[ r ][ g ] t}\OtherTok{))}
\NormalTok{⇈[]Distr p q r s t }\OtherTok{(}\NormalTok{f , f♯}\OtherTok{)} \OtherTok{(}\NormalTok{g , g♯}\OtherTok{)} \OtherTok{=} 
    \OtherTok{(} \OtherTok{(λ} \OtherTok{(}\NormalTok{a , h}\OtherTok{)} \OtherTok{→} \OtherTok{(}\NormalTok{a , }\OtherTok{(λ}\NormalTok{ x }\OtherTok{→}\NormalTok{ fst }\OtherTok{(}\NormalTok{h x}\OtherTok{)))}\NormalTok{ , }\OtherTok{λ}\NormalTok{ k1 }\OtherTok{→}\NormalTok{ f a , }\OtherTok{λ}\NormalTok{ x }\OtherTok{→}\NormalTok{ snd }\OtherTok{(}\NormalTok{h }\OtherTok{(}\NormalTok{f♯ a x}\OtherTok{))} \OtherTok{(}\NormalTok{k1 x}\OtherTok{))} 
\NormalTok{    , }\OtherTok{λ} \OtherTok{(}\NormalTok{a , h}\OtherTok{)} \OtherTok{(}\NormalTok{k1 , k2}\OtherTok{)}\NormalTok{ d }\OtherTok{→} \OtherTok{(}\NormalTok{k1 }\OtherTok{(}\NormalTok{g♯ }\OtherTok{(}\NormalTok{f a}\OtherTok{)}\NormalTok{ d}\OtherTok{))}\NormalTok{ , k2 d }\OtherTok{)}
\end{Highlighting}
\end{Shaded}

Equality of distributors:

\begin{Shaded}
\begin{Highlighting}[]
\NormalTok{EqDistributor }\OtherTok{:} \OtherTok{∀} \OtherTok{\{}\NormalTok{ℓ0 ℓ1 ℓ2 ℓ3 κ0 κ1 κ2 κ3}\OtherTok{\}}
                \OtherTok{→} \OtherTok{(}\NormalTok{p }\OtherTok{:}\NormalTok{ Poly ℓ0 κ0}\OtherTok{)} \OtherTok{(}\NormalTok{q }\OtherTok{:}\NormalTok{ Poly ℓ1 κ1}\OtherTok{)}
                \OtherTok{→} \OtherTok{(}\NormalTok{r }\OtherTok{:}\NormalTok{ Poly ℓ2 κ2}\OtherTok{)} \OtherTok{(}\NormalTok{s }\OtherTok{:}\NormalTok{ Poly ℓ3 κ3}\OtherTok{)}
                \OtherTok{→} \OtherTok{(}\NormalTok{p ◃ r}\OtherTok{)}\NormalTok{ ⇆ }\OtherTok{(}\NormalTok{s ◃ q}\OtherTok{)} \OtherTok{→} \OtherTok{(}\NormalTok{p ◃ r}\OtherTok{)}\NormalTok{ ⇆ }\OtherTok{(}\NormalTok{s ◃ q}\OtherTok{)}
                \OtherTok{→}\NormalTok{ Type }\OtherTok{(}\NormalTok{ℓ0 ⊔ ℓ1 ⊔ ℓ2 ⊔ ℓ3 ⊔ κ0 ⊔ κ1 ⊔ κ2 ⊔ κ3}\OtherTok{)}
\NormalTok{EqDistributor p q r s }\OtherTok{(}\NormalTok{f , f♯}\OtherTok{)} \OtherTok{(}\NormalTok{g , g♯}\OtherTok{)} \OtherTok{=} 
    \OtherTok{(}\NormalTok{a }\OtherTok{:}\NormalTok{ fst p}\OtherTok{)} \OtherTok{(}\NormalTok{γ }\OtherTok{:}\NormalTok{ snd p a }\OtherTok{→}\NormalTok{ fst r}\OtherTok{)} 
    \OtherTok{→}\NormalTok{ Σ }\OtherTok{(}\NormalTok{fst }\OtherTok{(}\NormalTok{f }\OtherTok{(}\NormalTok{a , γ}\OtherTok{))}\NormalTok{ ≡ fst }\OtherTok{(}\NormalTok{g }\OtherTok{(}\NormalTok{a , γ}\OtherTok{)))} 
        \OtherTok{(λ}\NormalTok{ e1 }\OtherTok{→} \OtherTok{(}\NormalTok{x }\OtherTok{:}\NormalTok{ snd s }\OtherTok{(}\NormalTok{fst }\OtherTok{(}\NormalTok{f }\OtherTok{(}\NormalTok{a , γ}\OtherTok{))))}
                \OtherTok{→}\NormalTok{ Σ }\OtherTok{((}\NormalTok{snd }\OtherTok{(}\NormalTok{f }\OtherTok{(}\NormalTok{a , γ}\OtherTok{))}\NormalTok{ x}\OtherTok{)} 
\NormalTok{                    ≡ }\OtherTok{(}\NormalTok{snd }\OtherTok{(}\NormalTok{g }\OtherTok{(}\NormalTok{a , γ}\OtherTok{))} 
                           \OtherTok{(}\NormalTok{transp }\OtherTok{(}\NormalTok{snd s}\OtherTok{)}\NormalTok{ e1 x}\OtherTok{)))} 
                    \OtherTok{(λ}\NormalTok{ e2 }\OtherTok{→} \OtherTok{(}\NormalTok{y }\OtherTok{:}\NormalTok{ snd q }\OtherTok{(}\NormalTok{snd }\OtherTok{(}\NormalTok{f }\OtherTok{(}\NormalTok{a , γ}\OtherTok{))}\NormalTok{ x}\OtherTok{))} 
                            \OtherTok{→} \OtherTok{(}\NormalTok{f♯ }\OtherTok{(}\NormalTok{a , γ}\OtherTok{)} \OtherTok{(}\NormalTok{x , y}\OtherTok{))} 
\NormalTok{                              ≡ }\OtherTok{(}\NormalTok{g♯ }\OtherTok{(}\NormalTok{a , γ}\OtherTok{)} 
                                    \OtherTok{(} \OtherTok{(}\NormalTok{transp }\OtherTok{(}\NormalTok{snd s}\OtherTok{)}\NormalTok{ e1 x}\OtherTok{)} 
\NormalTok{                                    , }\OtherTok{(}\NormalTok{transp }\OtherTok{(}\NormalTok{snd q}\OtherTok{)}\NormalTok{ e2 y}\OtherTok{)))))}
\end{Highlighting}
\end{Shaded}

Converting from lenses out of \texttt{\_⇈{[}\_{]}{[}\_{]}\_} to
distributors:

\begin{Shaded}
\begin{Highlighting}[]
\NormalTok{⇈→Distributor }\OtherTok{:} \OtherTok{∀} \OtherTok{\{}\NormalTok{ℓ0 ℓ1 ℓ2 ℓ3 κ0 κ1 κ2 κ3}\OtherTok{\}}
                \OtherTok{→} \OtherTok{\{}\NormalTok{p }\OtherTok{:}\NormalTok{ Poly ℓ0 κ0}\OtherTok{\}} \OtherTok{(}\NormalTok{q }\OtherTok{:}\NormalTok{ Poly ℓ1 κ1}\OtherTok{)}
                \OtherTok{→} \OtherTok{(}\NormalTok{r }\OtherTok{:}\NormalTok{ Poly ℓ2 κ2}\OtherTok{)} \OtherTok{\{}\NormalTok{s }\OtherTok{:}\NormalTok{ Poly ℓ3 κ3}\OtherTok{\}}
                \OtherTok{→} \OtherTok{\{}\NormalTok{f }\OtherTok{:}\NormalTok{ p ⇆ q}\OtherTok{\}}
                \OtherTok{→} \OtherTok{(}\NormalTok{p ⇈[ q ][ f ] r}\OtherTok{)}\NormalTok{ ⇆ s}
                \OtherTok{→} \OtherTok{(}\NormalTok{p ◃ r}\OtherTok{)}\NormalTok{ ⇆ }\OtherTok{(}\NormalTok{s ◃ q}\OtherTok{)}
\NormalTok{⇈→Distributor q r }\OtherTok{\{}\NormalTok{f }\OtherTok{=} \OtherTok{(}\NormalTok{f , f♯}\OtherTok{)\}} \OtherTok{(}\NormalTok{g , g♯}\OtherTok{)} \OtherTok{=}
    \OtherTok{(} \OtherTok{(λ} \OtherTok{(}\NormalTok{a , h}\OtherTok{)} \OtherTok{→}\NormalTok{ g }\OtherTok{(}\NormalTok{a , h}\OtherTok{)}\NormalTok{ , }\OtherTok{λ}\NormalTok{ d\textquotesingle{} }\OtherTok{→}\NormalTok{ f a}\OtherTok{)} 
\NormalTok{    , }\OtherTok{λ} \OtherTok{(}\NormalTok{a , h}\OtherTok{)} \OtherTok{(}\NormalTok{d\textquotesingle{} , d}\OtherTok{)}
        \OtherTok{→}\NormalTok{ f♯ a d , g♯ }\OtherTok{(}\NormalTok{a , h}\OtherTok{)}\NormalTok{ d\textquotesingle{} d}\OtherTok{)}
\end{Highlighting}
\end{Shaded}

Functoriality of distributors:

\begin{Shaded}
\begin{Highlighting}[]
\KeywordTok{module}\NormalTok{ DistributorLens }\OtherTok{\{}\NormalTok{ℓ0 ℓ1 ℓ2 ℓ3 ℓ4 ℓ5 ℓ6 ℓ7}
\NormalTok{                        κ0 κ1 κ2 κ3 κ4 κ5 κ6 κ7}\OtherTok{\}}
                       \OtherTok{\{}\NormalTok{p }\OtherTok{:}\NormalTok{ Poly ℓ0 κ0}\OtherTok{\}} \OtherTok{\{}\NormalTok{p\textquotesingle{} }\OtherTok{:}\NormalTok{ Poly ℓ4 κ4}\OtherTok{\}}
                       \OtherTok{\{}\NormalTok{q }\OtherTok{:}\NormalTok{ Poly ℓ1 κ1}\OtherTok{\}} \OtherTok{(}\NormalTok{q\textquotesingle{} }\OtherTok{:}\NormalTok{ Poly ℓ5 κ5}\OtherTok{)}
                       \OtherTok{(}\NormalTok{r }\OtherTok{:}\NormalTok{ Poly ℓ2 κ2}\OtherTok{)} \OtherTok{\{}\NormalTok{r\textquotesingle{} }\OtherTok{:}\NormalTok{ Poly ℓ6 κ6}\OtherTok{\}}
                       \OtherTok{\{}\NormalTok{s }\OtherTok{:}\NormalTok{ Poly ℓ3 κ3}\OtherTok{\}} \OtherTok{(}\NormalTok{s\textquotesingle{} }\OtherTok{:}\NormalTok{ Poly ℓ7 κ7}\OtherTok{)}
                       \OtherTok{(}\NormalTok{g }\OtherTok{:}\NormalTok{ p\textquotesingle{} ⇆ p}\OtherTok{)} \OtherTok{(}\NormalTok{h }\OtherTok{:}\NormalTok{ q ⇆ q\textquotesingle{}}\OtherTok{)} 
                       \OtherTok{(}\NormalTok{k }\OtherTok{:}\NormalTok{ r\textquotesingle{} ⇆ r}\OtherTok{)} \OtherTok{(}\NormalTok{l }\OtherTok{:}\NormalTok{ s ⇆ s\textquotesingle{}}\OtherTok{)} \KeywordTok{where}

\NormalTok{    distrLens }\OtherTok{:} \OtherTok{(}\NormalTok{p ◃ r}\OtherTok{)}\NormalTok{ ⇆ }\OtherTok{(}\NormalTok{s ◃ q}\OtherTok{)} \OtherTok{→} \OtherTok{(}\NormalTok{p\textquotesingle{} ◃ r\textquotesingle{}}\OtherTok{)}\NormalTok{ ⇆ }\OtherTok{(}\NormalTok{s\textquotesingle{} ◃ q\textquotesingle{}}\OtherTok{)}
\NormalTok{    distrLens j }\OtherTok{=} 
\NormalTok{        comp }\OtherTok{(}\NormalTok{s\textquotesingle{} ◃ q\textquotesingle{}}\OtherTok{)} \OtherTok{(}\NormalTok{g ◃◃[ r ] k}\OtherTok{)} 
             \OtherTok{(}\NormalTok{comp }\OtherTok{((}\NormalTok{s\textquotesingle{} ◃ q\textquotesingle{}}\OtherTok{))}\NormalTok{ j }
                   \OtherTok{(}\NormalTok{l ◃◃[ q\textquotesingle{} ] h}\OtherTok{))}
\end{Highlighting}
\end{Shaded}

\begin{Shaded}
\begin{Highlighting}[]
\NormalTok{    ⇈→DistributorLens }\OtherTok{:} \OtherTok{\{}\NormalTok{f }\OtherTok{:}\NormalTok{ p ⇆ q}\OtherTok{\}} \OtherTok{→} \OtherTok{(}\NormalTok{p ⇈[ q ][ f ] r}\OtherTok{)}\NormalTok{ ⇆ s }
                        \OtherTok{→} \OtherTok{(}\NormalTok{p\textquotesingle{} ⇈[ q\textquotesingle{} ][ comp q\textquotesingle{} g }\OtherTok{(}\NormalTok{comp q\textquotesingle{} f h}\OtherTok{)}\NormalTok{ ] r\textquotesingle{}}\OtherTok{)}\NormalTok{ ⇆ s\textquotesingle{}}
\NormalTok{    ⇈→DistributorLens }\OtherTok{\{}\NormalTok{f }\OtherTok{=}\NormalTok{ f}\OtherTok{\}}\NormalTok{ j }\OtherTok{=} 
\NormalTok{        comp s\textquotesingle{} }\OtherTok{(}\NormalTok{⇈[]Lens q\textquotesingle{} r }\OtherTok{(}\NormalTok{comp q\textquotesingle{} g }\OtherTok{(}\NormalTok{comp q\textquotesingle{} f h}\OtherTok{))}\NormalTok{ f }
\NormalTok{                         g h k }\OtherTok{((λ}\NormalTok{ a }\OtherTok{→}\NormalTok{ refl}\OtherTok{)}\NormalTok{ , }\OtherTok{(λ}\NormalTok{ a d }\OtherTok{→}\NormalTok{ refl}\OtherTok{)))} 
             \OtherTok{(}\NormalTok{comp s\textquotesingle{} j l}\OtherTok{)}

\NormalTok{    ⇈→DistributorLens≡ }\OtherTok{:} \OtherTok{\{}\NormalTok{f }\OtherTok{:}\NormalTok{ p ⇆ q}\OtherTok{\}} \OtherTok{(}\NormalTok{j }\OtherTok{:} \OtherTok{(}\NormalTok{p ⇈[ q ][ f ] r}\OtherTok{)}\NormalTok{ ⇆ s}\OtherTok{)}
                         \OtherTok{→}\NormalTok{ distrLens }\OtherTok{(}\NormalTok{⇈→Distributor q r j}\OtherTok{)} 
\NormalTok{                           ≡ ⇈→Distributor q\textquotesingle{} r\textquotesingle{} }\OtherTok{(}\NormalTok{⇈→DistributorLens j}\OtherTok{)}
\NormalTok{    ⇈→DistributorLens≡ j }\OtherTok{=}\NormalTok{ refl}

\KeywordTok{open}\NormalTok{ DistributorLens }\KeywordTok{public}
\end{Highlighting}
\end{Shaded}

There are two distinct ways of composing distributors:

\begin{enumerate}
\def\labelenumi{\arabic{enumi}.}
\item
  Given distributors \(j1 : p \triangleleft s \leftrightarrows t \triangleleft q \) and
  \(j2 : q \triangleleft u \leftrightarrows v \triangleleft r \), we obtain a distributor
  \(p \triangleleft (s \triangleleft u) \leftrightarrows (t \triangleleft v) \triangleleft r\)
  as the composite \[
  p ◃ (s \triangleleft u) \simeq (p \triangleleft s) \triangleleft u \xrightarrow{j1 \triangleleft u} (t \triangleleft q) \triangleleft u \simeq t \triangleleft (q \triangleleft u) \xrightarrow{j2} t \triangleleft (v \triangleleft r) \simeq (t \triangleleft v) \triangleleft r
  \]
\end{enumerate}

\begin{Shaded}
\begin{Highlighting}[]
\KeywordTok{module}\NormalTok{ DistributorComp1 }\OtherTok{\{}\NormalTok{ℓ0 ℓ1 ℓ2 ℓ3 ℓ4 ℓ5 ℓ6 κ0 κ1 κ2 κ3 κ4 κ5 κ6}\OtherTok{\}}
                        \OtherTok{\{}\NormalTok{p }\OtherTok{:}\NormalTok{ Poly ℓ0 κ0}\OtherTok{\}} \OtherTok{\{}\NormalTok{q }\OtherTok{:}\NormalTok{ Poly ℓ1 κ1}\OtherTok{\}} \OtherTok{(}\NormalTok{r }\OtherTok{:}\NormalTok{ Poly ℓ2 κ2}\OtherTok{)}
                        \OtherTok{\{}\NormalTok{s }\OtherTok{:}\NormalTok{ Poly ℓ3 κ3}\OtherTok{\}} \OtherTok{\{}\NormalTok{t }\OtherTok{:}\NormalTok{ Poly ℓ4 κ4}\OtherTok{\}}
                        \OtherTok{(}\NormalTok{u }\OtherTok{:}\NormalTok{ Poly ℓ5 κ5}\OtherTok{)} \OtherTok{\{}\NormalTok{v }\OtherTok{:}\NormalTok{ Poly ℓ6 κ6}\OtherTok{\}} \KeywordTok{where}

\NormalTok{    distrComp1 }\OtherTok{:} \OtherTok{(}\NormalTok{p ◃ s}\OtherTok{)}\NormalTok{ ⇆ }\OtherTok{(}\NormalTok{t ◃ q}\OtherTok{)} \OtherTok{→} \OtherTok{(}\NormalTok{q ◃ u}\OtherTok{)}\NormalTok{ ⇆ }\OtherTok{(}\NormalTok{v ◃ r}\OtherTok{)}
                 \OtherTok{→} \OtherTok{(}\NormalTok{p ◃ }\OtherTok{(}\NormalTok{s ◃ u}\OtherTok{))}\NormalTok{ ⇆ }\OtherTok{((}\NormalTok{t ◃ v}\OtherTok{)}\NormalTok{ ◃ r}\OtherTok{)}
\NormalTok{    distrComp1 h k }\OtherTok{=} 
\NormalTok{        comp }\OtherTok{((}\NormalTok{t ◃ v}\OtherTok{)}\NormalTok{ ◃ r}\OtherTok{)} \OtherTok{(}\NormalTok{◃assoc⁻¹ p s u}\OtherTok{)} 
             \OtherTok{(}\NormalTok{comp }\OtherTok{((}\NormalTok{t ◃ v}\OtherTok{)}\NormalTok{ ◃ r}\OtherTok{)} \OtherTok{(}\NormalTok{h ◃◃[ u ] }\OtherTok{(}\NormalTok{id u}\OtherTok{))} 
                   \OtherTok{(}\NormalTok{comp }\OtherTok{((}\NormalTok{t ◃ v}\OtherTok{)}\NormalTok{ ◃ r}\OtherTok{)} \OtherTok{(}\NormalTok{◃assoc t q u}\OtherTok{)} 
                         \OtherTok{(}\NormalTok{comp }\OtherTok{((}\NormalTok{t ◃ v}\OtherTok{)}\NormalTok{ ◃ r}\OtherTok{)} \OtherTok{((}\NormalTok{id t}\OtherTok{)}\NormalTok{ ◃◃[ }\OtherTok{(}\NormalTok{v ◃ r}\OtherTok{)}\NormalTok{ ] k}\OtherTok{)} 
                               \OtherTok{(}\NormalTok{◃assoc⁻¹ t v r}\OtherTok{))))}
\end{Highlighting}
\end{Shaded}

The corresponding construction on morphisms
\texttt{(p\ ⇈{[}\ q\ {]}{[}\ f\ {]}\ s)\ ⇆\ t} and
\texttt{(q\ ⇈{[}\ r\ {]}{[}\ g\ {]}\ u)\ ⇆\ v} is to form the following
composite with the colaxator of \texttt{\_⇈{[}\_{]}{[}\_{]}\_}: \[
p {\upuparrows}[r][g \circ f] (s \triangleleft u) \leftrightarrows (p {\upuparrows}[q][f] s) \triangleleft (q {\upuparrows}[r][g] u) \leftrightarrows t \triangleleft v
\]

\begin{Shaded}
\begin{Highlighting}[]
\NormalTok{    ⇈→DistributorComp1 }\OtherTok{:} \OtherTok{\{}\NormalTok{f }\OtherTok{:}\NormalTok{ p ⇆ q}\OtherTok{\}} \OtherTok{\{}\NormalTok{g }\OtherTok{:}\NormalTok{ q ⇆ r}\OtherTok{\}} 
                         \OtherTok{→} \OtherTok{(}\NormalTok{p ⇈[ q ][ f ] s}\OtherTok{)}\NormalTok{ ⇆ t }
                         \OtherTok{→} \OtherTok{(}\NormalTok{q ⇈[ r ][ g ] u}\OtherTok{)}\NormalTok{ ⇆ v}
                         \OtherTok{→} \OtherTok{(}\NormalTok{p ⇈[ r ][ comp r f g ] }\OtherTok{(}\NormalTok{s ◃ u}\OtherTok{))}\NormalTok{ ⇆ }\OtherTok{(}\NormalTok{t ◃ v}\OtherTok{)}
\NormalTok{    ⇈→DistributorComp1 }\OtherTok{\{}\NormalTok{f }\OtherTok{=}\NormalTok{ f}\OtherTok{\}} \OtherTok{\{}\NormalTok{g }\OtherTok{=}\NormalTok{ g}\OtherTok{\}}\NormalTok{ h k }\OtherTok{=} 
\NormalTok{        comp }\OtherTok{(}\NormalTok{t ◃ v}\OtherTok{)} \OtherTok{(}\NormalTok{⇈[]Distr p q r s u f g}\OtherTok{)} 
             \OtherTok{(}\NormalTok{h ◃◃[ v ] k}\OtherTok{)}

\NormalTok{    ⇈→DistributorComp1≡ }\OtherTok{:} \OtherTok{\{}\NormalTok{f }\OtherTok{:}\NormalTok{ p ⇆ q}\OtherTok{\}} \OtherTok{\{}\NormalTok{g }\OtherTok{:}\NormalTok{ q ⇆ r}\OtherTok{\}} 
                          \OtherTok{(}\NormalTok{h }\OtherTok{:} \OtherTok{(}\NormalTok{p ⇈[ q ][ f ] s}\OtherTok{)}\NormalTok{ ⇆ t}\OtherTok{)}
                          \OtherTok{(}\NormalTok{k }\OtherTok{:} \OtherTok{(}\NormalTok{q ⇈[ r ][ g ] u}\OtherTok{)}\NormalTok{ ⇆ v}\OtherTok{)}
                          \OtherTok{→}\NormalTok{ distrComp1 }\OtherTok{(}\NormalTok{⇈→Distributor q s h}\OtherTok{)} \OtherTok{(}\NormalTok{⇈→Distributor r u k}\OtherTok{)}
\NormalTok{                            ≡ ⇈→Distributor r }\OtherTok{(}\NormalTok{s ◃ u}\OtherTok{)} \OtherTok{(}\NormalTok{⇈→DistributorComp1 h k}\OtherTok{)}
\NormalTok{    ⇈→DistributorComp1≡ h k }\OtherTok{=}\NormalTok{ refl}
    
\KeywordTok{open}\NormalTok{ DistributorComp1 }\KeywordTok{public}
\end{Highlighting}
\end{Shaded}

\begin{enumerate}
\def\labelenumi{\arabic{enumi}.}
\setcounter{enumi}{1}
\item
  Given distributors
  \(p \triangleleft u \leftrightarrows v \triangleleft q\) and
  \(r \triangleleft t \leftrightarrows u \triangleleft s\), we obtain a
  distributor
  \((p \triangleleft r) \triangleleft t \leftrightarrows v \triangleleft (q \triangleleft s)\)
  as the composite \[
  (p \triangleleft r) \triangleleft t \simeq p \triangleleft (r \triangleleft t) \leftrightarrows p \triangleleft (u \triangleleft s) \simeq (p \triangleleft u) \triangleleft s \leftrightarrows (v \triangleleft q) \triangleleft s \simeq v \triangleleft (q \triangleleft s)
  \]
\end{enumerate}

\begin{Shaded}
\begin{Highlighting}[]
\KeywordTok{module}\NormalTok{ DistributorComp2 }
           \OtherTok{\{}\NormalTok{ℓ0 ℓ1 ℓ2 ℓ3 ℓ4 ℓ5 ℓ6 κ0 κ1 κ2 κ3 κ4 κ5 κ6}\OtherTok{\}}
           \OtherTok{\{}\NormalTok{p }\OtherTok{:}\NormalTok{ Poly ℓ0 κ0}\OtherTok{\}} \OtherTok{\{}\NormalTok{q }\OtherTok{:}\NormalTok{ Poly ℓ1 κ1}\OtherTok{\}} 
           \OtherTok{\{}\NormalTok{r }\OtherTok{:}\NormalTok{ Poly ℓ2 κ2}\OtherTok{\}} \OtherTok{(}\NormalTok{s }\OtherTok{:}\NormalTok{ Poly ℓ3 κ3}\OtherTok{)}
           \OtherTok{(}\NormalTok{t }\OtherTok{:}\NormalTok{ Poly ℓ4 κ4}\OtherTok{)} \OtherTok{\{}\NormalTok{u }\OtherTok{:}\NormalTok{ Poly ℓ5 κ5}\OtherTok{\}} 
           \OtherTok{\{}\NormalTok{v }\OtherTok{:}\NormalTok{ Poly ℓ6 κ6}\OtherTok{\}} \KeywordTok{where} 

\NormalTok{    distrComp2 }\OtherTok{:} \OtherTok{(}\NormalTok{r ◃ t}\OtherTok{)}\NormalTok{ ⇆ }\OtherTok{(}\NormalTok{u ◃ s}\OtherTok{)} \OtherTok{→} \OtherTok{(}\NormalTok{p ◃ u}\OtherTok{)}\NormalTok{ ⇆ }\OtherTok{(}\NormalTok{v ◃ q}\OtherTok{)}
                 \OtherTok{→} \OtherTok{((}\NormalTok{p ◃ r}\OtherTok{)}\NormalTok{ ◃ t}\OtherTok{)}\NormalTok{ ⇆ }\OtherTok{(}\NormalTok{v ◃ }\OtherTok{(}\NormalTok{q ◃ s}\OtherTok{))}
\NormalTok{    distrComp2 h k }\OtherTok{=}
\NormalTok{        comp }\OtherTok{(}\NormalTok{v ◃ }\OtherTok{(}\NormalTok{q ◃ s}\OtherTok{))} \OtherTok{(}\NormalTok{◃assoc p r t}\OtherTok{)} 
             \OtherTok{(}\NormalTok{comp }\OtherTok{(}\NormalTok{v ◃ }\OtherTok{(}\NormalTok{q ◃ s}\OtherTok{))}  \OtherTok{((}\NormalTok{id p}\OtherTok{)}\NormalTok{ ◃◃[ u ◃ s ] h}\OtherTok{)} 
               \OtherTok{(}\NormalTok{comp }\OtherTok{(}\NormalTok{v ◃ }\OtherTok{(}\NormalTok{q ◃ s}\OtherTok{))} \OtherTok{(}\NormalTok{◃assoc⁻¹ p u s}\OtherTok{)} 
                     \OtherTok{(}\NormalTok{comp }\OtherTok{(}\NormalTok{v ◃ }\OtherTok{(}\NormalTok{q ◃ s}\OtherTok{))} \OtherTok{(}\NormalTok{k ◃◃[ s ] }\OtherTok{(}\NormalTok{id s}\OtherTok{))} 
                           \OtherTok{(}\NormalTok{◃assoc v q s}\OtherTok{))))}
\end{Highlighting}
\end{Shaded}

The corresponding construction on morphisms
\texttt{(p\ ⇈{[}\ q\ {]}{[}\ f\ {]}\ u)\ ⇆\ v} and
\texttt{(r\ ⇈{[}\ s\ {]}{[}\ g\ {]}\ t)\ ⇆\ u} is to form the following
composite with the morphism \texttt{⇈{[}{]}Curry} defined above: \[
(p \triangleleft r) {\upuparrows}[q \triangleleft s][f \triangleleft g] t \leftrightarrows p {\upuparrows}[q][f] (r {\upuparrows}[s][g] t) \leftrightarrows p {\upuparrows}[q][f] u \leftrightarrows v
\]

\begin{Shaded}
\begin{Highlighting}[]
\NormalTok{    ⇈→DistributorComp2 }\OtherTok{:} \OtherTok{\{}\NormalTok{f }\OtherTok{:}\NormalTok{ p ⇆ q}\OtherTok{\}} \OtherTok{\{}\NormalTok{g }\OtherTok{:}\NormalTok{ r ⇆ s}\OtherTok{\}}
        \OtherTok{→} \OtherTok{(}\NormalTok{r ⇈[ s ][ g ] t}\OtherTok{)}\NormalTok{ ⇆ u }\OtherTok{→} \OtherTok{(}\NormalTok{p ⇈[ q ][ f ] u}\OtherTok{)}\NormalTok{ ⇆ v}
        \OtherTok{→} \OtherTok{((}\NormalTok{p ◃ r}\OtherTok{)}\NormalTok{ ⇈[ }\OtherTok{(}\NormalTok{q ◃ s}\OtherTok{)}\NormalTok{ ][ f ◃◃[ s ] g ] t}\OtherTok{)}\NormalTok{ ⇆ v}
\NormalTok{    ⇈→DistributorComp2 }\OtherTok{\{}\NormalTok{f }\OtherTok{=}\NormalTok{ f}\OtherTok{\}} \OtherTok{\{}\NormalTok{g }\OtherTok{=}\NormalTok{ g}\OtherTok{\}}\NormalTok{ h k }\OtherTok{=}
\NormalTok{        comp v }\OtherTok{(}\NormalTok{⇈[]Curry p q r s t f g}\OtherTok{)} 
             \OtherTok{(}\NormalTok{comp v }\OtherTok{(}\NormalTok{⇈[]Lens q u f f }
                              \OtherTok{(}\NormalTok{id p}\OtherTok{)} \OtherTok{(}\NormalTok{id q}\OtherTok{)}\NormalTok{ h }
                              \OtherTok{(} \OtherTok{(λ}\NormalTok{ a }\OtherTok{→}\NormalTok{ refl}\OtherTok{)} 
\NormalTok{                              , }\OtherTok{(λ}\NormalTok{ a d }\OtherTok{→}\NormalTok{ refl}\OtherTok{)))} 
\NormalTok{                   k}\OtherTok{)}
    
\NormalTok{    ⇈→DistributorComp2≡ }\OtherTok{:} \OtherTok{\{}\NormalTok{f }\OtherTok{:}\NormalTok{ p ⇆ q}\OtherTok{\}} \OtherTok{\{}\NormalTok{g }\OtherTok{:}\NormalTok{ r ⇆ s}\OtherTok{\}}
        \OtherTok{→} \OtherTok{(}\NormalTok{h }\OtherTok{:} \OtherTok{(}\NormalTok{r ⇈[ s ][ g ] t}\OtherTok{)}\NormalTok{ ⇆ u}\OtherTok{)} \OtherTok{(}\NormalTok{k }\OtherTok{:} \OtherTok{(}\NormalTok{p ⇈[ q ][ f ] u}\OtherTok{)}\NormalTok{ ⇆ v}\OtherTok{)}
        \OtherTok{→} \OtherTok{(}\NormalTok{distrComp2 }\OtherTok{(}\NormalTok{⇈→Distributor s t h}\OtherTok{)} 
                      \OtherTok{(}\NormalTok{⇈→Distributor q u k}\OtherTok{))} 
\NormalTok{          ≡ ⇈→Distributor }\OtherTok{(}\NormalTok{q ◃ s}\OtherTok{)}\NormalTok{ t }
                          \OtherTok{(}\NormalTok{⇈→DistributorComp2 h k}\OtherTok{)}
\NormalTok{    ⇈→DistributorComp2≡ h k }\OtherTok{=}\NormalTok{ refl}

\KeywordTok{open}\NormalTok{ DistributorComp2 }\KeywordTok{public}
\end{Highlighting}
\end{Shaded}

Likewise, there are two corresponding notions of ``identity
distributor'' on a polynomial \texttt{p}, the first of which is given by
the following composition of unitors for \texttt{◃}: \[
p \triangleleft y \simeq p \simeq y \triangleleft p
\] and the second of which is given by the inverse such composition \[
y \triangleleft p \simeq p \simeq p \triangleleft y
\]

\begin{Shaded}
\begin{Highlighting}[]
\KeywordTok{module}\NormalTok{ DistributorId }\OtherTok{\{}\NormalTok{ℓ κ}\OtherTok{\}} \OtherTok{(}\NormalTok{p }\OtherTok{:}\NormalTok{ Poly ℓ κ}\OtherTok{)} \KeywordTok{where}

\NormalTok{    distrId1 }\OtherTok{:} \OtherTok{(}\NormalTok{p ◃ 𝕪}\OtherTok{)}\NormalTok{ ⇆ }\OtherTok{(}\NormalTok{𝕪 ◃ p}\OtherTok{)}
\NormalTok{    distrId1 }\OtherTok{=}\NormalTok{ comp }\OtherTok{(}\NormalTok{𝕪 ◃ p}\OtherTok{)} \OtherTok{(}\NormalTok{◃unitr p}\OtherTok{)} \OtherTok{(}\NormalTok{◃unitl⁻¹ p}\OtherTok{)}

\NormalTok{    distrId2 }\OtherTok{:} \OtherTok{(}\NormalTok{𝕪 ◃ p}\OtherTok{)}\NormalTok{ ⇆ }\OtherTok{(}\NormalTok{p ◃ 𝕪}\OtherTok{)}
\NormalTok{    distrId2 }\OtherTok{=}\NormalTok{ comp }\OtherTok{(}\NormalTok{p ◃ 𝕪}\OtherTok{)} \OtherTok{(}\NormalTok{◃unitl p}\OtherTok{)} \OtherTok{(}\NormalTok{◃unitr⁻¹ p}\OtherTok{)}
\end{Highlighting}
\end{Shaded}

The corresponding morphisms
\texttt{p\ ⇈{[}\ p\ {]}{[}\ id\ p\ {]}\ 𝕪\ ⇆\ 𝕪} and
\texttt{𝕪\ ⇈{[}\ 𝕪\ {]}{[}\ id\ 𝕪\ {]}\ p\ ⇆\ p} are precisely the maps
\texttt{⇈{[}{]}𝕪} and \texttt{𝕪⇈{[}{]}} defined above, respectively:

\begin{Shaded}
\begin{Highlighting}[]
\NormalTok{    ⇈→DistributorId1≡ }\OtherTok{:}\NormalTok{ distrId1 ≡ ⇈→Distributor p 𝕪 }\OtherTok{(}\NormalTok{⇈[]𝕪 p p }\OtherTok{(}\NormalTok{id p}\OtherTok{))}
\NormalTok{    ⇈→DistributorId1≡ }\OtherTok{=}\NormalTok{ refl}

\NormalTok{    ⇈→DistributorId2≡ }\OtherTok{:}\NormalTok{ distrId2 ≡ ⇈→Distributor 𝕪 p }\OtherTok{(}\NormalTok{𝕪⇈[] p}\OtherTok{)}
\NormalTok{    ⇈→DistributorId2≡ }\OtherTok{=}\NormalTok{ refl}

\KeywordTok{open}\NormalTok{ DistributorId }\KeywordTok{public}
\end{Highlighting}
\end{Shaded}

Proof of Theorem 4.2:

\begin{Shaded}
\begin{Highlighting}[]
\NormalTok{ap⇈→Distributor }\OtherTok{:} \OtherTok{∀} \OtherTok{\{}\NormalTok{ℓ0 ℓ1 ℓ2 ℓ3 κ0 κ1 κ2 κ3}\OtherTok{\}}
                  \OtherTok{→} \OtherTok{(}\NormalTok{p }\OtherTok{:}\NormalTok{ Poly ℓ0 κ0}\OtherTok{)} \OtherTok{(}\NormalTok{q }\OtherTok{:}\NormalTok{ Poly ℓ1 κ1}\OtherTok{)}
                  \OtherTok{→} \OtherTok{(}\NormalTok{r }\OtherTok{:}\NormalTok{ Poly ℓ2 κ2}\OtherTok{)} \OtherTok{(}\NormalTok{s }\OtherTok{:}\NormalTok{ Poly ℓ3 κ3}\OtherTok{)}
                  \OtherTok{→} \OtherTok{(}\NormalTok{f }\OtherTok{:}\NormalTok{ p ⇆ q}\OtherTok{)}
                  \OtherTok{→} \OtherTok{(}\NormalTok{h k }\OtherTok{:} \OtherTok{(}\NormalTok{p ⇈[ q ][ f ] r}\OtherTok{)}\NormalTok{ ⇆ s}\OtherTok{)}
                  \OtherTok{→}\NormalTok{ EqLens s h k }
                  \OtherTok{→}\NormalTok{ EqDistributor p q r s}
                        \OtherTok{(}\NormalTok{⇈→Distributor q r h}\OtherTok{)}
                        \OtherTok{(}\NormalTok{⇈→Distributor q r k}\OtherTok{)}
\NormalTok{ap⇈→Distributor p q r s f h k }\OtherTok{(}\NormalTok{e , e♯}\OtherTok{)}\NormalTok{ a γ }\OtherTok{=} 
    \OtherTok{(}\NormalTok{ e }\OtherTok{(}\NormalTok{a , γ}\OtherTok{)} 
\NormalTok{    , }\OtherTok{λ}\NormalTok{ x }\OtherTok{→} \OtherTok{(}\NormalTok{ refl }
\NormalTok{            , }\OtherTok{(λ}\NormalTok{ y }\OtherTok{→}\NormalTok{ pairEq refl }
                        \OtherTok{(}\NormalTok{coAp }\OtherTok{(}\NormalTok{e♯ }\OtherTok{(}\NormalTok{a , γ}\OtherTok{)}\NormalTok{ x}\OtherTok{)}\NormalTok{ y}\OtherTok{))} \OtherTok{)} \OtherTok{)}

\KeywordTok{module}\NormalTok{ DistrLaw }\OtherTok{\{}\NormalTok{ℓ κ}\OtherTok{\}} \OtherTok{(}\NormalTok{𝔲 }\OtherTok{:}\NormalTok{ Poly ℓ κ}\OtherTok{)} \OtherTok{(}\NormalTok{univ }\OtherTok{:}\NormalTok{ isUnivalent 𝔲}\OtherTok{)}
                \OtherTok{(}\NormalTok{η }\OtherTok{:}\NormalTok{ 𝕪 ⇆ 𝔲}\OtherTok{)} \OtherTok{(}\NormalTok{cη }\OtherTok{:}\NormalTok{ isCartesian 𝔲 η}\OtherTok{)}
                \OtherTok{(}\NormalTok{σ }\OtherTok{:} \OtherTok{(}\NormalTok{𝔲 ◃ 𝔲}\OtherTok{)}\NormalTok{ ⇆ 𝔲}\OtherTok{)} \OtherTok{(}\NormalTok{cσ }\OtherTok{:}\NormalTok{ isCartesian 𝔲 σ}\OtherTok{)}
                \OtherTok{(}\NormalTok{π }\OtherTok{:} \OtherTok{(}\NormalTok{𝔲 ⇈ 𝔲}\OtherTok{)}\NormalTok{ ⇆ 𝔲}\OtherTok{)} \OtherTok{(}\NormalTok{cπ }\OtherTok{:}\NormalTok{ isCartesian 𝔲 π}\OtherTok{)} \KeywordTok{where}
    
\NormalTok{    distrLaw1 }\OtherTok{:}\NormalTok{ EqDistributor 𝔲 𝔲 }\OtherTok{(}\NormalTok{𝔲 ◃ 𝔲}\OtherTok{)}\NormalTok{ 𝔲}
                    \OtherTok{(}\NormalTok{distrLens 𝔲 }\OtherTok{(}\NormalTok{𝔲 ◃ 𝔲}\OtherTok{)}\NormalTok{ 𝔲 }\OtherTok{(}\NormalTok{id 𝔲}\OtherTok{)} \OtherTok{(}\NormalTok{id 𝔲}\OtherTok{)} \OtherTok{(}\NormalTok{id }\OtherTok{(}\NormalTok{𝔲 ◃ 𝔲}\OtherTok{))}\NormalTok{ σ }
                               \OtherTok{(}\NormalTok{distrComp1 𝔲 𝔲 }\OtherTok{(}\NormalTok{distrLaw? 𝔲 π}\OtherTok{)} 
                                               \OtherTok{(}\NormalTok{distrLaw? 𝔲 π}\OtherTok{)))} 
                    \OtherTok{(}\NormalTok{distrLens 𝔲 𝔲 𝔲 }\OtherTok{(}\NormalTok{id 𝔲}\OtherTok{)} \OtherTok{(}\NormalTok{id 𝔲}\OtherTok{)}\NormalTok{ σ }\OtherTok{(}\NormalTok{id 𝔲}\OtherTok{)} 
                               \OtherTok{(}\NormalTok{distrLaw? 𝔲 π}\OtherTok{))}
\NormalTok{    distrLaw1 }\OtherTok{=}\NormalTok{ ap⇈→Distributor 𝔲 𝔲 }\OtherTok{(}\NormalTok{𝔲 ◃ 𝔲}\OtherTok{)}\NormalTok{ 𝔲 }\OtherTok{(}\NormalTok{id 𝔲}\OtherTok{)}
                    \OtherTok{(}\NormalTok{comp 𝔲 }\OtherTok{(}\NormalTok{comp }\OtherTok{(}\NormalTok{𝔲 ◃ 𝔲}\OtherTok{)} \OtherTok{(}\NormalTok{⇈Distr 𝔲 𝔲 𝔲}\OtherTok{)} \OtherTok{(}\NormalTok{π ◃◃[ 𝔲 ] π}\OtherTok{))}\NormalTok{ σ}\OtherTok{)}
                    \OtherTok{(}\NormalTok{comp 𝔲 }\OtherTok{(}\NormalTok{⇈[]Lens 𝔲 𝔲 }\OtherTok{(}\NormalTok{id 𝔲}\OtherTok{)} \OtherTok{(}\NormalTok{id 𝔲}\OtherTok{)} \OtherTok{(}\NormalTok{id 𝔲}\OtherTok{)} \OtherTok{(}\NormalTok{id 𝔲}\OtherTok{)}\NormalTok{ σ }
                                     \OtherTok{((λ}\NormalTok{ a }\OtherTok{→}\NormalTok{ refl}\OtherTok{)}\NormalTok{ , }\OtherTok{(λ}\NormalTok{ a d }\OtherTok{→}\NormalTok{ refl}\OtherTok{)))}\NormalTok{ π}\OtherTok{)}
                    \OtherTok{(}\NormalTok{univ }\OtherTok{(}\NormalTok{compCartesian 𝔲 }
                                \OtherTok{(}\NormalTok{compCartesian }\OtherTok{(}\NormalTok{𝔲 ◃ 𝔲}\OtherTok{)} 
                                    \OtherTok{(}\NormalTok{⇈DistrCart 𝔲 𝔲 𝔲}\OtherTok{)} 
                                    \OtherTok{(}\NormalTok{◃◃Cart 𝔲 𝔲 cπ cπ}\OtherTok{))} 
\NormalTok{                                cσ}\OtherTok{)} 
                          \OtherTok{(}\NormalTok{compCartesian 𝔲 }
                            \OtherTok{(}\NormalTok{⇈[]LensCart 𝔲 𝔲 }\OtherTok{(}\NormalTok{id 𝔲}\OtherTok{)} \OtherTok{(}\NormalTok{id 𝔲}\OtherTok{)} \OtherTok{(}\NormalTok{id 𝔲}\OtherTok{)} \OtherTok{(}\NormalTok{id 𝔲}\OtherTok{)}\NormalTok{ σ }
                                \OtherTok{((λ}\NormalTok{ a }\OtherTok{→}\NormalTok{ refl}\OtherTok{)}\NormalTok{ , }\OtherTok{(λ}\NormalTok{ a d }\OtherTok{→}\NormalTok{ refl}\OtherTok{))} 
                                \OtherTok{(}\NormalTok{idCart 𝔲}\OtherTok{)}\NormalTok{ cσ}\OtherTok{)} 
\NormalTok{                            cπ}\OtherTok{))}
    
\NormalTok{    distrLaw2 }\OtherTok{:}\NormalTok{ EqDistributor }\OtherTok{(}\NormalTok{𝔲 ◃ 𝔲}\OtherTok{)}\NormalTok{ 𝔲 𝔲 𝔲}
                    \OtherTok{(}\NormalTok{distrLens 𝔲 𝔲 𝔲 }\OtherTok{(}\NormalTok{id }\OtherTok{(}\NormalTok{𝔲 ◃ 𝔲}\OtherTok{))}\NormalTok{ σ }\OtherTok{(}\NormalTok{id 𝔲}\OtherTok{)} \OtherTok{(}\NormalTok{id 𝔲}\OtherTok{)} 
                               \OtherTok{(}\NormalTok{distrComp2 𝔲 𝔲 }\OtherTok{(}\NormalTok{distrLaw? 𝔲 π}\OtherTok{)} 
                                               \OtherTok{(}\NormalTok{distrLaw? 𝔲 π}\OtherTok{)))} 
                    \OtherTok{(}\NormalTok{distrLens 𝔲 𝔲 𝔲 σ }\OtherTok{(}\NormalTok{id 𝔲}\OtherTok{)} \OtherTok{(}\NormalTok{id 𝔲}\OtherTok{)} \OtherTok{(}\NormalTok{id 𝔲}\OtherTok{)} 
                               \OtherTok{(}\NormalTok{distrLaw? 𝔲 π}\OtherTok{))}
\NormalTok{    distrLaw2 }\OtherTok{=}\NormalTok{ ap⇈→Distributor }\OtherTok{(}\NormalTok{𝔲 ◃ 𝔲}\OtherTok{)}\NormalTok{ 𝔲 𝔲 𝔲 σ}
                    \OtherTok{(}\NormalTok{comp 𝔲 }
                        \OtherTok{(}\NormalTok{comp }\OtherTok{(}\NormalTok{𝔲 ⇈ 𝔲}\OtherTok{)} 
                            \OtherTok{(}\NormalTok{comp }\OtherTok{(}\NormalTok{𝔲 ⇈ }\OtherTok{(}\NormalTok{𝔲 ⇈ 𝔲}\OtherTok{))} 
                                \OtherTok{(}\NormalTok{⇈[]Lens 𝔲 𝔲 σ }\OtherTok{(}\NormalTok{id }\OtherTok{(}\NormalTok{𝔲 ◃ 𝔲}\OtherTok{))} 
                                    \OtherTok{(}\NormalTok{id }\OtherTok{(}\NormalTok{𝔲 ◃ 𝔲}\OtherTok{))}\NormalTok{ σ }\OtherTok{(}\NormalTok{id 𝔲}\OtherTok{)} 
                                    \OtherTok{((λ}\NormalTok{ a }\OtherTok{→}\NormalTok{ refl}\OtherTok{)}\NormalTok{ , }\OtherTok{(λ}\NormalTok{ a d }\OtherTok{→}\NormalTok{ refl}\OtherTok{)))} 
                                \OtherTok{(}\NormalTok{⇈Curry 𝔲 𝔲 𝔲}\OtherTok{))} 
                            \OtherTok{(}\NormalTok{⇈Lens 𝔲 𝔲 }\OtherTok{(}\NormalTok{id 𝔲}\OtherTok{)} \OtherTok{(}\NormalTok{id 𝔲}\OtherTok{)} 
                                   \OtherTok{((λ}\NormalTok{ a }\OtherTok{→}\NormalTok{ refl}\OtherTok{)}\NormalTok{ , }\OtherTok{(λ}\NormalTok{ a d }\OtherTok{→}\NormalTok{ refl}\OtherTok{))} 
\NormalTok{                                   π}\OtherTok{))} 
\NormalTok{                        π}\OtherTok{)}
                    \OtherTok{(}\NormalTok{comp 𝔲 }\OtherTok{(}\NormalTok{⇈[]Lens 𝔲 𝔲 σ }\OtherTok{(}\NormalTok{id 𝔲}\OtherTok{)}\NormalTok{ σ }\OtherTok{(}\NormalTok{id 𝔲}\OtherTok{)} \OtherTok{(}\NormalTok{id 𝔲}\OtherTok{)} 
                                     \OtherTok{((λ}\NormalTok{ a }\OtherTok{→}\NormalTok{ refl}\OtherTok{)}\NormalTok{ , }\OtherTok{(λ}\NormalTok{ a d }\OtherTok{→}\NormalTok{ refl}\OtherTok{)))} 
\NormalTok{                            π}\OtherTok{)}
                    \OtherTok{(}\NormalTok{univ }\OtherTok{(}\NormalTok{compCartesian 𝔲 }
                            \OtherTok{(}\NormalTok{compCartesian }\OtherTok{(}\NormalTok{𝔲 ⇈ 𝔲}\OtherTok{)} 
                                \OtherTok{(}\NormalTok{compCartesian }\OtherTok{(}\NormalTok{𝔲 ⇈ }\OtherTok{(}\NormalTok{𝔲 ⇈ 𝔲}\OtherTok{))} 
                                    \OtherTok{(}\NormalTok{⇈[]LensCart 𝔲 𝔲 σ }\OtherTok{(}\NormalTok{id }\OtherTok{(}\NormalTok{𝔲 ◃ 𝔲}\OtherTok{))} 
                                        \OtherTok{(}\NormalTok{id }\OtherTok{(}\NormalTok{𝔲 ◃ 𝔲}\OtherTok{))}\NormalTok{ σ }\OtherTok{(}\NormalTok{id 𝔲}\OtherTok{)} 
                                        \OtherTok{((λ}\NormalTok{ a }\OtherTok{→}\NormalTok{ refl}\OtherTok{)}\NormalTok{ , }\OtherTok{(λ}\NormalTok{ a d }\OtherTok{→}\NormalTok{ refl}\OtherTok{))} 
\NormalTok{                                        cσ }\OtherTok{(}\NormalTok{idCart 𝔲}\OtherTok{))} 
                                    \OtherTok{(}\NormalTok{⇈CurryCart 𝔲 𝔲 𝔲}\OtherTok{))} 
                                \OtherTok{(}\NormalTok{⇈[]LensCart 𝔲 𝔲 }\OtherTok{(}\NormalTok{id 𝔲}\OtherTok{)} \OtherTok{(}\NormalTok{id 𝔲}\OtherTok{)} \OtherTok{(}\NormalTok{id 𝔲}\OtherTok{)} \OtherTok{(}\NormalTok{id 𝔲}\OtherTok{)}\NormalTok{ π}
                                             \OtherTok{((λ}\NormalTok{ a }\OtherTok{→}\NormalTok{ refl}\OtherTok{)}\NormalTok{ , }\OtherTok{(λ}\NormalTok{ a d }\OtherTok{→}\NormalTok{ refl}\OtherTok{))} 
                                             \OtherTok{(}\NormalTok{idCart 𝔲}\OtherTok{)}\NormalTok{ cπ}\OtherTok{))} 
\NormalTok{                            cπ}\OtherTok{)}
                          \OtherTok{(}\NormalTok{compCartesian 𝔲 }
                            \OtherTok{(}\NormalTok{⇈[]LensCart 𝔲 𝔲 σ }\OtherTok{(}\NormalTok{id 𝔲}\OtherTok{)}\NormalTok{ σ }\OtherTok{(}\NormalTok{id 𝔲}\OtherTok{)} \OtherTok{(}\NormalTok{id 𝔲}\OtherTok{)} 
                                \OtherTok{((λ}\NormalTok{ a }\OtherTok{→}\NormalTok{ refl}\OtherTok{)}\NormalTok{ , }\OtherTok{(λ}\NormalTok{ a d }\OtherTok{→}\NormalTok{ refl}\OtherTok{))} 
                                \OtherTok{(}\NormalTok{idCart 𝔲}\OtherTok{)} \OtherTok{(}\NormalTok{idCart 𝔲}\OtherTok{))} 
\NormalTok{                            cπ}\OtherTok{))}
    
\NormalTok{    distrLaw3 }\OtherTok{:}\NormalTok{ EqDistributor 𝔲 𝔲 𝕪 𝔲 }
                    \OtherTok{(}\NormalTok{distrLens 𝔲 𝕪 𝔲 }\OtherTok{(}\NormalTok{id 𝔲}\OtherTok{)} \OtherTok{(}\NormalTok{id 𝔲}\OtherTok{)} \OtherTok{(}\NormalTok{id 𝕪}\OtherTok{)}\NormalTok{ η }\OtherTok{(}\NormalTok{distrId1 𝔲}\OtherTok{))} 
                    \OtherTok{(}\NormalTok{distrLens 𝔲 𝔲 𝔲 }\OtherTok{(}\NormalTok{id 𝔲}\OtherTok{)} \OtherTok{(}\NormalTok{id 𝔲}\OtherTok{)}\NormalTok{ η }\OtherTok{(}\NormalTok{id 𝔲}\OtherTok{)} \OtherTok{(}\NormalTok{distrLaw? 𝔲 π}\OtherTok{))}
\NormalTok{    distrLaw3 }\OtherTok{=} 
\NormalTok{        ap⇈→Distributor 𝔲 𝔲 𝕪 𝔲 }\OtherTok{(}\NormalTok{id 𝔲}\OtherTok{)}
            \OtherTok{(}\NormalTok{comp 𝔲 }\OtherTok{(}\NormalTok{⇈𝕪 𝔲}\OtherTok{)}\NormalTok{ η}\OtherTok{)} 
            \OtherTok{(}\NormalTok{comp 𝔲 }\OtherTok{(}\NormalTok{⇈Lens 𝔲 𝔲 }\OtherTok{(}\NormalTok{id 𝔲}\OtherTok{)} \OtherTok{(}\NormalTok{id 𝔲}\OtherTok{)} \OtherTok{((λ}\NormalTok{ a }\OtherTok{→}\NormalTok{ refl}\OtherTok{)}\NormalTok{ , }\OtherTok{(λ}\NormalTok{ a d }\OtherTok{→}\NormalTok{ refl}\OtherTok{))}\NormalTok{ η}\OtherTok{)}\NormalTok{ π}\OtherTok{)}
            \OtherTok{(}\NormalTok{univ }\OtherTok{(}\NormalTok{compCartesian 𝔲 }\OtherTok{(}\NormalTok{⇈𝕪Cart 𝔲}\OtherTok{)}\NormalTok{ cη}\OtherTok{)} 
                  \OtherTok{(}\NormalTok{compCartesian 𝔲 }
                    \OtherTok{(}\NormalTok{⇈[]LensCart 𝔲 𝔲 }\OtherTok{(}\NormalTok{id 𝔲}\OtherTok{)} \OtherTok{(}\NormalTok{id 𝔲}\OtherTok{)} \OtherTok{(}\NormalTok{id 𝔲}\OtherTok{)} \OtherTok{(}\NormalTok{id 𝔲}\OtherTok{)}\NormalTok{ η }
                                 \OtherTok{((λ}\NormalTok{ a }\OtherTok{→}\NormalTok{ refl}\OtherTok{)}\NormalTok{ , }\OtherTok{(λ}\NormalTok{ a d }\OtherTok{→}\NormalTok{ refl}\OtherTok{))} 
                                 \OtherTok{(}\NormalTok{idCart 𝔲}\OtherTok{)}\NormalTok{ cη}\OtherTok{)} 
\NormalTok{                    cπ}\OtherTok{))}
    
\NormalTok{    distrLaw4 }\OtherTok{:}\NormalTok{ EqDistributor 𝕪 𝔲 𝔲 𝔲}
                    \OtherTok{(}\NormalTok{distrLens 𝔲 𝔲 𝔲 }\OtherTok{(}\NormalTok{id 𝕪}\OtherTok{)}\NormalTok{ η }\OtherTok{(}\NormalTok{id 𝔲}\OtherTok{)} \OtherTok{(}\NormalTok{id 𝔲}\OtherTok{)} \OtherTok{(}\NormalTok{distrId2 𝔲}\OtherTok{))} 
                    \OtherTok{(}\NormalTok{distrLens 𝔲 𝔲 𝔲 η }\OtherTok{(}\NormalTok{id 𝔲}\OtherTok{)} \OtherTok{(}\NormalTok{id 𝔲}\OtherTok{)} \OtherTok{(}\NormalTok{id 𝔲}\OtherTok{)} \OtherTok{(}\NormalTok{distrLaw? 𝔲 π}\OtherTok{))}
\NormalTok{    distrLaw4 }\OtherTok{=}
\NormalTok{        ap⇈→Distributor 𝕪 𝔲 𝔲 𝔲 η }
            \OtherTok{(}\NormalTok{comp 𝔲 }\OtherTok{(}\NormalTok{⇈[]Lens 𝔲 𝔲 η }\OtherTok{(}\NormalTok{id 𝕪}\OtherTok{)} \OtherTok{(}\NormalTok{id 𝕪}\OtherTok{)}\NormalTok{ η }\OtherTok{(}\NormalTok{id 𝔲}\OtherTok{)} 
                             \OtherTok{((λ}\NormalTok{ a }\OtherTok{→}\NormalTok{ refl}\OtherTok{)}\NormalTok{ , }\OtherTok{(λ}\NormalTok{ a d }\OtherTok{→}\NormalTok{ refl}\OtherTok{)))} 
                    \OtherTok{(}\NormalTok{𝕪⇈ 𝔲}\OtherTok{))}
            \OtherTok{(}\NormalTok{comp 𝔲 }\OtherTok{(}\NormalTok{⇈[]Lens 𝔲 𝔲 η }\OtherTok{(}\NormalTok{id 𝔲}\OtherTok{)}\NormalTok{ η }\OtherTok{(}\NormalTok{id 𝔲}\OtherTok{)} \OtherTok{(}\NormalTok{id 𝔲}\OtherTok{)}
                             \OtherTok{((λ}\NormalTok{ a }\OtherTok{→}\NormalTok{ refl}\OtherTok{)}\NormalTok{ , }\OtherTok{(λ}\NormalTok{ a d }\OtherTok{→}\NormalTok{ refl}\OtherTok{)))} 
\NormalTok{                    π}\OtherTok{)} 
            \OtherTok{(}\NormalTok{univ }\OtherTok{(}\NormalTok{compCartesian 𝔲 }
                    \OtherTok{(}\NormalTok{⇈[]LensCart 𝔲 𝔲 η }\OtherTok{(}\NormalTok{id 𝕪}\OtherTok{)} \OtherTok{(}\NormalTok{id 𝕪}\OtherTok{)}\NormalTok{ η }\OtherTok{(}\NormalTok{id 𝔲}\OtherTok{)} 
                                 \OtherTok{((λ}\NormalTok{ a }\OtherTok{→}\NormalTok{ refl}\OtherTok{)}\NormalTok{ , }\OtherTok{(λ}\NormalTok{ a d }\OtherTok{→}\NormalTok{ refl}\OtherTok{))} 
\NormalTok{                                 cη }\OtherTok{(}\NormalTok{idCart 𝔲}\OtherTok{))} 
                    \OtherTok{(}\NormalTok{𝕪⇈Cart 𝔲}\OtherTok{))} 
                  \OtherTok{(}\NormalTok{compCartesian 𝔲 }
                    \OtherTok{(}\NormalTok{⇈[]LensCart 𝔲 𝔲 η }\OtherTok{(}\NormalTok{id 𝔲}\OtherTok{)}\NormalTok{ η }\OtherTok{(}\NormalTok{id 𝔲}\OtherTok{)} \OtherTok{(}\NormalTok{id 𝔲}\OtherTok{)} 
                                 \OtherTok{((λ}\NormalTok{ a }\OtherTok{→}\NormalTok{ refl}\OtherTok{)}\NormalTok{ , }\OtherTok{(λ}\NormalTok{ a d }\OtherTok{→}\NormalTok{ refl}\OtherTok{))} 
                                 \OtherTok{(}\NormalTok{idCart 𝔲}\OtherTok{)} \OtherTok{(}\NormalTok{idCart 𝔲}\OtherTok{))} 
\NormalTok{                    cπ}\OtherTok{))}
\end{Highlighting}
\end{Shaded}

\end{document}